\documentclass[11pt,a4paper]{article}

\usepackage{amsmath,amssymb,tikz,hyperref,empheq}
\usepackage{graphicx}
 \usepackage{verbatim}
 \allowdisplaybreaks
\def\be{\begin{equation}}
\def\ee{\end{equation}}
\def\ba#1\ea{\begin{align}#1\end{align}}
\def\bg#1\eg{\begin{gather}#1\end{gather}}
\def\bm#1\em{\begin{multline}#1\end{multline}}
\def\bmd#1\emd{\begin{multlined}#1\end{multlined}}

\def\a{\alpha}
\def\b{\beta}

\def\e{\epsilon}

\def\({\left(}
\def\){\right)}
\def\[{\left[}
\def\]{\right]}

\def \be {\begin{equation}}
\def \ee {\end{equation}}
\def \ba {\begin{array}}
\def \ea {\end{array}}
\def \bea{\begin{eqnarray}}
\def \eea{\end{eqnarray}}

\def \a {\alpha}
\def \b {\beta}

\def \e {\epsilon}

\def\bea{\begin{eqnarray}}
\def\eea{\end{eqnarray}}

\newcommand{\bit}{\begin{itemize}}  \newcommand{\eit}{\end{itemize}}
\newcommand{\ben}{\begin{enumerate}}  \newcommand{\een}{\end{enumerate}}

\long\def\symbolfootnote[#1]#2{\begingroup%
\def\thefootnote{\fnsymbol{footnote}}\footnote[#1]{#2}\endgroup}

\newcommand{\sysu}{{\it School of Physics and Astronomy, Sun Yat-Sen University, 2 Daxue Road, Zhuhai 519082, China}}

\begin{document}
\thispagestyle{empty}
\begin{center}

~\vspace{20pt}

{\Large\bf Ghost Problem, Spectrum Identities and Various Constraints on Brane-localized Gravity}

\vspace{25pt}

 Rong-Xin Miao ${}$\symbolfootnote[1]{Email:~\sf
  miaorx@mail.sysu.edu.cn}

\vspace{10pt}${}$\sysu

\vspace{2cm}

\begin{abstract}
This paper investigates the brane-localized interactions, including DGP gravity and higher derivative (HD) gravity localized on the brane. We derive the effective action on the brane, which suggests the brane-localized HD gravity suffers the ghost problem generally. Besides, we obtain novel algebraic identities of the mass spectrum, which reveal the global nature and can characterize the phase transformation of the mass spectrum. We get a powerful ghost-free condition from the spectrum identities, which rules out one type of brane-localized HD gravity. We further prove the mass spectrum is real and non-negative $m^2\ge 0$ under the ghost-free condition. 

Furthermore, we discuss various constraints on parameters of brane-localized gravity in AdS/BCFT and wedge holography, respectively. They include the ghost-free condition of Kaluza-Klein and brane-bending modes, the positive definiteness of boundary central charges, and entanglement entropy. The ghost-free condition imposes strict constraint, which requires non-negative couplings for pure DGP gravity and Gauss-Bonnet gravity on the brane. It also rules out one class of brane-localized HD gravity. Thus, such HD gravity should be understood as a low-energy effective theory on the brane under the ghost energy scale. Finally, we briefly discuss the applications of our results.
\end{abstract}

\end{center}

\newpage
\setcounter{footnote}{0}
\setcounter{page}{1}

\tableofcontents

\section{Introduction}

Higher derivative (HD) gravity is interesting in many aspects. First, string theory predicts there are HD corrections to the effective action of gravity \cite{Gross:1986mw, Gross:1986iv, Fradkin:1984pq}. Second, unlike Einstein's gravity, HD gravity is generally renormalizable \cite{Stelle:1976gc, Buchbinder:1992rb}. Third, HD gravity can describe more general strongly coupled CFTs in the AdS/CFT correspondence \cite{Buchel:2009sk, Gregory:2009fj, Nojiri:1999mh}. Fourth, as a modified gravity theory, HD gravity has important applications in cosmology \cite{DeFelice:2010aj, Nojiri:2010wj, Nojiri:2017ncd}. Despite so many advantages, HD gravity suffers the ghost problem generally \cite{Stelle:1977ry}. See Appendix A for an example. Inspired by DGP gravity \cite{Dvali:2000hr}, this paper investigates HD gravity localized on the brane. Like the bulk case, the brane-localized HD gravity also suffers from the ghost problem. For simplicity, we mainly focus on the AdS/BCFT correspondence \cite{Takayanagi:2011zk, Fujita:2011fp, Nozaki:2012qd, Miao:2018qkc, Miao:2017gyt, Chu:2017aab, Chu:2021mvq}, which is closely related to braneworld theory \cite{Randall:1999ee, Randall:1999vf, Karch:2000ct} and has a wind range of applications. In particular, it plays a vital role in recent breakthroughs in resolving black hole information problems \cite{Penington:2019npb, Almheiri:2019psf, Almheiri:2019hni}. See also \cite{Almheiri:2019yqk,Almheiri:2019psy,Chen:2020uac,Chen:2020hmv,Ling:2020laa,Geng:2020qvw,Krishnan:2020fer,Yadav:2022mnv,Emparan:2023dxm,Kawabata:2021hac,Chou:2021boq,Alishahiha:2020qza,Hu:2022ymx,Hu:2022zgy,Miao:2022mdx,Miao:2023unv,Li:2023fly,Jeong:2023hrb,Yu:2023whl,Chang:2023gkt,Tong:2023nvi,Ghodrati:2022hbb,Lee:2022efh,Aguilar-Gutierrez:2023kfn,Aguilar-Gutierrez:2023tic,dS wedge} for some related works.  Besides, it is a powerful tool for studying the boundary effect, such as the Casimir effect \cite{Miao:2017aba}, anomalous transport \cite{Chu:2018ntx, Miao:2022oas}, and so on. The main motivation of this paper is to formulate AdS/BCFT with brane-localized gravity and investigate various self-consistent conditions. For simplicity, we focus on DGP gravity and brane-localized curvature-squared gravity.

Let us summarize our main results. See (\ref{gravityaction}) for a glance at the action of brane-localized gravity, where $\lambda, \a_1, \a_2, \a_3$ denote the DGP, Gauss-Bonnet (GB) and other curvature-squared couplings, respectively. We work out the squared gravitational effective action on the brane, which includes HD terms and thus ghosts for $\a_2\ne 0$. Furthermore, we obtain novel algebraic identities of the mass spectrum on the brane in AdS$_{d+1}$/BCFT$_d$
\begin{eqnarray}\label{gravity: SI DGP0}
&&\sum_{m} \frac{H_m^2(\rho)}{\langle H_m, H_m \rangle} = \frac{1}{2\Big(\lambda +4(d-3) \a_1 \text{sech}^2(\rho )\Big)}, \ \ \ \ \ \ \ \ \ \ \ \ \text{for }\alpha_2=0, \\ \label{gravity: SI HD0}
&&\sum_{m} \frac{H_m^2(\rho)}{\langle H_m, H_m \rangle} = 0, \ \ \ \ \sum_{m} \frac{H_m^2(\rho) m^2}{\langle H_m, H_m \rangle}=\frac{\cosh^2(\rho)}{2\alpha_2}, \ \ \ \ \ \ \ \text{for } \alpha_2\ne0,
\end{eqnarray} 
where the sum is over all gravitational Kaluza-Klein (KK) modes $H_m$, $\langle H_m, H_m \rangle$ is the inner product, $m$ denotes the mass, and $\rho$ labels the brane location. 
Note that $\alpha_3$ is related to the scalar mode of brane-localized curvature-squared gravity $\hat{R}^2$, where $\hat{R}$ is defined in (\ref{background curvature}). Thus, it does not appear in the spectrum identities (\ref{gravity: SI DGP0},\ref{gravity: SI HD0}) for gravitational KK modes. 
Interestingly, the above spectrum identities reveal the global nature of the mass spectrum and can describe the phase transformation. Comparing (\ref{gravity: SI DGP0}) with (\ref{gravity: SI HD0}), we observe a phase transition when turning on the HD coupling $\alpha_2$. 
That is reasonable since an additional KK mode (ghost mode) emerges by turning on $\alpha_2$, which changes the mass spectrum discontinuously. From the spectrum identities, we derive a ghost-free condition, i.e., $\Big(\lambda +4(d-3) \a_1 \text{sech}^2(\rho )\Big)\ge 0$ and $\a_2=0$, which is consistent with the effective action that implies a ghost for $\a_2\ne 0$. Remarkably, the mass spectrum is always real and positive $m^2>0$ under this ghost-free condition. Furthermore, we study various constraints on the parameters of brane-localized gravity, which include the ghost-free condition of KK and brane-bending modes, the positive definiteness of boundary central charges, and entanglement entropy. See Table. \ref{sect4:Constraint in AdSBCFT} of sect.4.3 and Table. \ref{table:Constraint in wedge holography} of sect.5 for the summary of various constraints in AdS/BCFT and wedge holography, respectively. Take the pure DGP gravity with $\a_i=0$ as an example; the strongest constraint comes from the ghost-free condition of KK modes, which requires $\lambda \ge 0$. As for the general cases, please refer to (\ref{sect4: constraint HD0}-\ref{sect4: constraint HD3}) and (\ref{sect5: constraint HD0}-\ref{sect5: constraint HD3}) for AdS/BCFT and wedge holography, respectively.

The paper is organized as follows. To warm up, we study a toy model of a brane-localized HD scalar in section 2. We show the HD interaction always results in a ghost on the brane. Besides, we prove novel algebraic identities of the mass spectrum and verify them with numerical and perturbation calculations. Section 3 generalizes the discussions to brane-localized HD gravity in AdS/BCFT. We analyze carefully the spectrum identities and derive the effective actions of KK and brane-bending modes. Section 4 and Section 5 discuss various constraints from AdS/BCFT and wedge holography, respectively. Finally, we conclude with some open problems in section 6. 

Abbreviations: This paper labels ``higher-derivative", ``Dvali-Gabadadze-Porrati," ``Gauss-Bonnet," and ``Kaluza-Klein"  by  ``HD," ``DGP," ``GB," and ``KK," respectively.

\section{A toy model}

To warm up, we first study a toy model of a brane-localized HD scalar whose qualitative characteristics are similar to those of gravity. In sect.2.1, we analyze the mass spectrum of the toy model in a flat space and work out the effective action on the brane. The effective action includes HD terms and implies a ghost for the non-zero HD coupling. In sect.2.2, we prove novel spectrum identities on the brane and verify them in the large parameter limits. The spectrum identities lead to a powerful ghost-free condition. Interestingly, the mass spectrum is automatically real and positive $m^2>0$ under the ghost-free condition. In sect.2.3, we generalize the discussions to AdS space.

The action of toy model is given by
 \begin{eqnarray}\label{actionscalar}
I=\int_N d^{d+1}x \sqrt{|g|} \left(-\frac{1}{2} \nabla_{\mu} \phi \nabla^{\mu} \phi \right)+\int_Q d^dy \sqrt{|h|} \left(-\frac{\lambda}{2} D_{i} \phi D^{i} \phi -\frac{\alpha}{2} D_{i} \phi \Box D^{i} \phi \right),
\end{eqnarray}
where $N$ denotes the bulk space, $Q$ is the brane, $\lambda, \alpha$ are parameters, $\Box=D^i D_i$, $\nabla_{\mu}$ and $D_i$ are covariant derivatives with respect to the bulk metric $g_{\mu\nu}$ and boundary metric $h_{ij}$, respectively. The first boundary term $-\frac{\lambda}{2} D_{i} \phi D^{i} \phi$ mimics DGP gravity and the second boundary term $ -\frac{\alpha}{2} D_{i} \phi \Box D^{i} \phi $ mimics brane-localized HD gravity. From the above action, we derive the equation of motion (EOM) in bulk
 \begin{eqnarray}\label{EOMscalar}
\nabla^{\mu} \nabla_{\mu} \phi=0,
\end{eqnarray}
and Neumann boundary condition (NBC) on the brane $Q$
\begin{eqnarray}\label{NBCscalar}
n^{\mu}\nabla_{\mu} \phi&=&\lambda \Box \phi+\alpha D^i \Box D_i \phi\nonumber\\
&=& \lambda \Box \phi+\alpha \Box^2 \phi+ \alpha D^i(R_{ij} D^j\phi),
\end{eqnarray}
where $n^{\mu}$ and $R_{ij}$ are the outpointing normal vector and intrinsic Ricci tensor on the brane. 

\subsection{Flat space}
\begin{figure}[t]
\centering
\includegraphics[width=8cm]{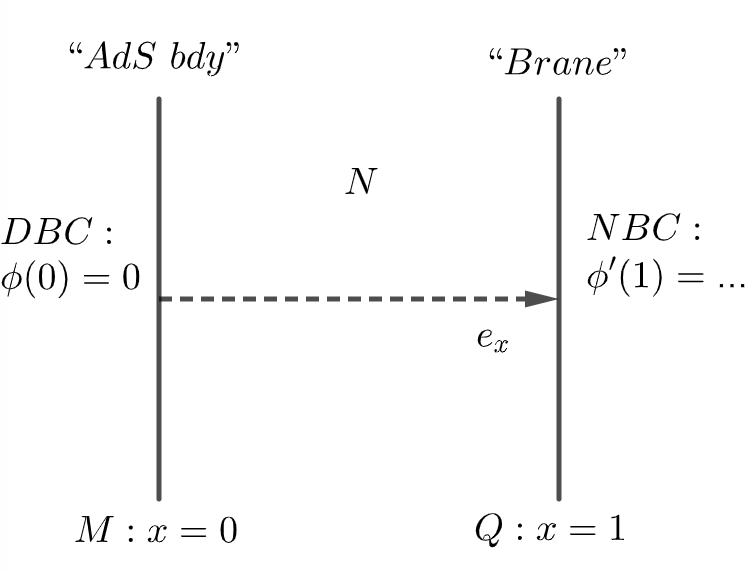}
\caption{Geometry of toy model. $N$ denotes the flat space in bulk, $M$ and $Q$ are the boundaries at $x=0$ and $x=1$ mimicking the AdS boundary and the brane, respectively.  Similar to AdS/BCFT, we impose DBC on $M$ while NBC on $Q$.}
\label{toymodel}
\end{figure}

Let us first study the case of flat space, which simplifies the calculations greatly.  The case of AdS is similar and will be discussed in sect. 2.3. The bulk metric in Gauss normal coordinates is given by
\begin{eqnarray}\label{scalarflatmetric}
ds^2=dx^2+\eta_{ij} dy^i dy^j, \ \ \ 0\le x \le 1,
\end{eqnarray}
where $\eta_{ij}$ denotes the Minkowski metric on boundaries, the ``brane" locates at $x=1$, and $x=0$ mimics the AdS boundary in AdS/BCFT. Similar to AdS/BCFT, we choose Dirichlet boundary condition (DBC) on the left boundary
\begin{eqnarray}\label{DBCscalar}
\phi|_{x=0}=0,
\end{eqnarray}
while imposing NBC (\ref{NBCscalar}) on the right boundary (brane) $x=1$.  See  Fig. \ref{toymodel} for the schematic geometry. 

Taking the ansatz 
\begin{eqnarray}\label{bulkscalar}
\phi=X(x) Y(y),
\end{eqnarray}
and separating variables of EOM (\ref{EOMscalar}), we get
 \begin{eqnarray}\label{EOMX}
&& X''+ m^2 X=0,\\
&& \Box Y- m^2 Y=0, \label{EOMY}
\end{eqnarray}
where $m$ labels the mass of the Kaluza-Klein (KK) model on the brane. For the ansatz (\ref{bulkscalar}), the BCs (\ref{NBCscalar},\ref{DBCscalar}) become
\begin{eqnarray}\label{scalarDBC1}
&&\text{DBC}: \ X(0)=0, \\
&&\text{NBC}: \ X'(1)=(\lambda m^2+\alpha m^4) X(1), \label{scalarNBC1}
\end{eqnarray}
where we have used (\ref{EOMY}) in the derivation of (\ref{scalarNBC1}). From (\ref{EOMX}, \ref{scalarDBC1}), we solve
\begin{equation}\label{solutionX}
X(x)=\begin{cases}
\ a_1 x,&\
\text{$m=0$} ,\\
\ a_1 \sin( \sqrt{m^2} x),&\
\text{$m\ne 0$}.
\end{cases}
\end{equation}
which together with NBC (\ref{scalarNBC1}) yields a constraint for the mass spectrum
\begin{eqnarray}\label{massscalar}
\sqrt{m^2} \sin (\sqrt{m^2}) \left(\lambda +\alpha m^2\right)-\cos (\sqrt{m^2})=0,
\end{eqnarray}
where $m^2$ can be complex generally. So far, we do not impose any condition. Some comments are in order. 
\begin{itemize}
  \item The massless mode $X=a_1 x$ does not lie in the mass spectrum since it does not satisfy NBC (\ref{scalarNBC1}). One can check that, indeed, $m^2=0$ is not a solution to the mass spectrum (\ref{massscalar}). It is similar to AdS/BCFT.
 \item For large and real $\sqrt{m^2}$, the first term of (\ref{massscalar}) dominates, which yields $\sin(\sqrt{m^2}) \to 0$ and thus
\begin{eqnarray}\label{massscalar1}
\lim_{m^2\to \infty} m^2 \to (q \pi)^2, 
\end{eqnarray}
where $q$ is a positive integer. For large and imaginary $\sqrt{m^2}$, (\ref{massscalar}) gives $m^2\to -\lambda/\alpha $, which is possible only if $\lambda/\alpha $ is a large positive number.   
 \item For $\lambda=0$ and $\alpha>0$, there is a tachyon mode with $m^2<0$. The reason is as follows. 
From BC (\ref{massscalar}) with $\lambda=0$ and $m^2<0$ , we derive 
\begin{eqnarray}\label{positivealphanegativemass}
\alpha= \frac{\cot \left(\sqrt{m^2}\right)}{(m^2)^{3/2}}=\frac{\coth \left(\sqrt{-m^2}\right)}{(-m^2)^{3/2}} > 0,
\end{eqnarray}
which implies there could be a tachyon mode with $m^2<0$ only if $\alpha>0$. Furthermore, since $\frac{\coth \left(\sqrt{-m^2}\right)}{(-m^2)^{3/2}}$ is a monotonous function and takes values from $0^+$ to $\infty$, there is one and only one tachyon mode for any given positive $\alpha$.  Following a similar approach, one can show that there is a tachyon mode for $\lambda<0$ and $\alpha=0$. 
 \item In comparison,  for $\alpha<0$ and $\lambda=0$, there is a pair of modes with complex conjugate $m^2$, which implies brane-localized HD scalar is unstable.  See Table. \ref{table1scalarspectrum} below. There is no complex-mass modes for $\lambda>0$ and $\alpha=0$, which suggests the DGP-like scalar term is well-defined. The proof is as follows. From NBC (\ref{massscalar}) with $\alpha=0$  we have
\begin{eqnarray}\label{sect2:prooflambdanocomplexmode}
\lambda= \frac{\cot \left(\sqrt{m^2}\right)}{\sqrt{m^2}}.
\end{eqnarray}
Labeling $\sqrt{m^2}=a+ b\  i$ and $m^2=a^2-b^2+2ab\  i$ with real $a,b$,  we derive the imaginary part of (\ref{sect2:prooflambdanocomplexmode})
\begin{eqnarray}\label{sect2:prooflambdanocomplexmode1}
\frac{b \sin (2 a)+a \sinh (2 b)}{\left(a^2+b^2\right) (\cos (2 a)-\cosh (2 b))}=0.
\end{eqnarray}
Clearly, $ab=0$ is a solution to (\ref{sect2:prooflambdanocomplexmode1}), which means the $m^2=a^2-b^2+2ab\  i$ has no imaginary part. There are no other solutions. That is because, for $ab\ne 0$, (\ref{sect2:prooflambdanocomplexmode1}) is equivalent to
\begin{eqnarray}\label{sect2:prooflambdanocomplexmode2}
\frac{\sin (2 a)}{a}+\frac{\sinh (2 b)}{b}=0,
\end{eqnarray}
which has no solution since $\frac{\sinh (2 b)}{b}>2$ while $|\frac{\sin (2 a)}{a}|<2$ for $ab\ne 0$. Now we finish the proof that there are no complex $m^2$ modes for $\lambda\ne 0, \alpha=0$.
\item  For large $\alpha$, we can solve perturbatively the mass spectrum
\begin{eqnarray}\label{sect2:masslargealpha}
m^2: \frac{-1}{\sqrt{\alpha }} \Big(1+ \frac{1+3 \lambda }{6 \sqrt{\alpha }}+O(\frac{1}{\alpha})\Big), \ \frac{1}{\sqrt{\alpha }} \Big(1- \frac{1+3 \lambda }{6 \sqrt{\alpha }}+O(\frac{1}{\alpha})\Big), \  \pi ^2 q^2\Big(1+ \frac{2}{\pi ^4 q^4 \alpha}+O(\frac{1}{\alpha^2}) \Big),
\end{eqnarray} 
where $q=1,2,...$ are integers. One can check directly (\ref{sect2:masslargealpha}) obeys NBC (\ref{massscalar}). Besides, it clearly shows $m^2$ can be negative for $\alpha>0$ and complex for $\alpha<0$. 
\item Finally, we list the mass spectrum in Table.\ref{table1scalarspectrum} for finite $\alpha$. It shows positive $\alpha$ yields a tachyon mode with $m^2<0$, while negative $\alpha$ produces a pair of complex conjugate $m^2$. Besides, the large mass limit (\ref{massscalar1}) is a good approximation for even small $q$.
\begin{table}[ht]
\caption{Mass spectrum for HD scalar with $\lambda=0, \alpha=0.1, -0.1$}
\begin{center}
    \begin{tabular}{| c | c | c | c |  c | c | c | c| c| c|c| }
    \hline
     &$-1$&0& $1$ & $2$ & 3 &4 &5  \\ \hline
  $m^2$ for $\alpha=0.1$  &{\color{blue}-4.722} & 1.786 & 11.508 & 39.977 &89.051 & 158.040& 246.821\\ \hline
  $m^2$ for $\alpha=-0.1$ &$\times $ &{\color{blue}$2.089 \pm 3.157 i$}& 7.008 & 38.964 & 88.601& 157.787&246.659\\ \hline
  large mass limit (\ref{massscalar1}) &$\times$ &$\times$& 9.870 & 39.478 & 88.826& 157.914&246.740\\ \hline
    \end{tabular}
\end{center}
\label{table1scalarspectrum}
\end{table}

\end{itemize}

Now let us study the effective action on the brane. To do so, we expand the general solution in powers of KK modes 
 \begin{eqnarray}\label{scalar-power}
\phi= \sum_{m} X_m(x) Y_m(y)
\end{eqnarray}
where $\sum_{m}$ denotes the sum over the mass spectrum (\ref{massscalar}), and $X_m$ obeys EOM (\ref{EOMX}). From the solution (\ref{solutionX}) and mass constraint (\ref{massscalar}), we can derive an orthogonal relation
\begin{eqnarray}\label{scalarorthogonal}
\int_0^1 dx X_m(x) X_{m'}(x) +\lambda X_m(1) X_{m'}(1)+ \alpha (m^2+ m'^2)X_m(1) X_{m'}(1) = c_m \delta_{m, m'},
\end{eqnarray}
where 
\begin{eqnarray}\label{cmorthogonal}
c_m=\int_0^1 dx X^2_m(x) +\lambda X^2_m(1) + 2\alpha m^2 X^2_m(1),
\end{eqnarray}
is a constant depending on the normalization of $X_m(x)$. We do not choose the normalization $c_m=1$ as usual since $c_m$ can also be negative and zero for some choices of parameters $\lambda, \alpha$. As shown below, $c_m\ge 0$ is the ghost-free condition and imposes constraints on the parameters. We remark that (\ref{cmorthogonal}) can be interpreted as an inner product, i.e., $c_m=\langle X_m, X_m \rangle$. Thus it is natural that it should be non-negative. 

Substituting (\ref{scalar-power}) into the action (\ref{actionscalar}), we get
 \begin{eqnarray}\label{action-scalar1}
I&=&-\frac{1}{2}\sum_{m,m'}\int_{0}^{1}dx \int_Q d^dy \Big{[} X'_{m}(x)X'_{m'}(x) Y_m(y) Y_{m'}(y)+X_{m}(x)X_{m'}(x) D_i Y_m(y) D^i Y_{m'}(y) \Big{]}\nonumber\\
&& -\frac{1}{2}\sum_{m,m'}\int_Q d^dy \Big{[}X_m(1) X_{m'}(1)\left( \lambda D_{i} Y_m D^{i} Y_{m'} +\alpha D_{i} Y_m \Box D^{i} Y_{m'} \right) \Big{]}.
\end{eqnarray}
Integrating by parts and using EOM (\ref{EOMX}) together with BCs (\ref{scalarDBC1},\ref{scalarNBC1}), we obtain  \begin{eqnarray}\label{action-scalar2}
I=&&-\frac{1}{2}\sum_{m,m'}\int_{0}^{1}dx  X_{m}(x)X_{m'}(x) \int_Q d^dy \Big{[} m^2Y_m(y) Y_{m'}(y)+ D_i Y_m(y) D^i Y_{m'}(y) \Big{]}\nonumber\\
&&-\frac{1}{2}\sum_{m,m'} X_m(1) X_{m'}(1) \int_Q d^dy \Big{[}(\lambda m^2+ \alpha m^4 ) Y_m Y_{m'}+\lambda D_{i} Y_m D^{i} Y_{m'} +\alpha D_{i} Y_m \Box D^{i} Y_{m'} \Big{]}.\nonumber\\
\end{eqnarray}
By applying the orthogonal relation (\ref{scalarorthogonal}) and dropping some total derivatives on brane $Q$, we finally obtain the effective action on the brane
 \begin{eqnarray}\label{action-scalar3}
I=\frac{1}{2}\sum_{m,m'} \int_Q d^dy Y_m \Big{[} c_m \delta_{m,m'}(\Box -m^2) + \alpha  X_m(1) X_{m'}(1) (\Box -m^2) (\Box -m'^2)   \Big{]} Y_{m'},
\end{eqnarray}
which takes an elegant expression. Due to the HD coupling $\alpha$, there are cross terms between different modes.  

Now, we are ready to discuss the ghost problem. For simplicity, we focus on the case $\lambda=0$ below. In the following subsection, we discuss the ghost problem for general parameters by applying a different method.  

{\bf{Case I}: single mode  }

First, we consider a single mode excitation $Y_m$ and set $Y_{m'}=0$ for $m'\ne m$. 
Then the action (\ref{action-scalar3}) becomes 
 \begin{eqnarray}\label{action-scalar6}
I_m&=&\frac{1}{2} \int_Q d^dy \ Y_m (\Box -m^2)\Big[c_m+\alpha X_m^2(1) (\Box -m^2) \Big] Y_m\nonumber\\
&\simeq&\frac{c_m}{2} \int_Q d^dy \ Y_m (\Box -m^2)Y_m,
\end{eqnarray}
where we have used the trick of Appendix A to transform the fourth-order action into an effective second-order action, that we have used $(\Box-m^2)Y_m=0$ in the bracket. To have the correct action sign, we require $c_m\ge 0$ for all modes. We remark that $c_m$ (\ref{cmorthogonal}) is positive for the minimal theory without boundary action, i.e., $\lambda=\alpha=0$. And the requirement $c_m\ge 0$ imposes a constraint for the boundary couplings $\lambda$ and $\alpha$. $c_m=0$ is the critical case, corresponding to the so-called critical gravity \cite{Lu:2011zk}. Note that the first line of (\ref{action-scalar6}) seems suggest an additional massive mode with mass $m_a^2=m^2-\frac{c_m}{\alpha X^2_m(1)}$. We numerically find $m_a^2$ does not obey the NBC (\ref{massscalar}) and thus is a fake mode generally. Recall that we have a tachyon mode with $m^2<0$ for $\alpha>0, \lambda=0$. Now we show this tachyon mode is a ghost. Since we focus on the real scalar, we rewrite (\ref{solutionX}) as
 \begin{eqnarray}\label{sect2:negativemXm}
X_m=\sinh(|m| x)
\end{eqnarray}
for $m^2=-|m|^2<0$. From (\ref{positivealphanegativemass}) and (\ref{cmorthogonal}), we derive a negative inner product
 \begin{eqnarray}\label{sect2:negativemcm}
c_m=-\frac{1}{2}-\frac{3 \sinh (2 \left| m\right| )}{4 \left| m\right| } <0,
\end{eqnarray}
which shows the tachyon mode with $m^2<0$ is a ghost. In conclusion, the case with $\alpha>0$ includes a ghost mode.

{\bf{Case II}: complex conjugate modes}

Let us go on to discuss the case of $\alpha<0$. For $\alpha<0$, there are modes with complex conjugate mass. See again Table. \ref{table1scalarspectrum}. It turns out one real combination of the complex conjugate modes is a ghost. To see this, let us calculate the inner products of the following real combinations
 \begin{eqnarray}\label{sect2:cm complex conjugate 1}
&&\langle \frac{X_m+X_{m*}}{2} , \frac{X_m+X_{m*}}{2} \rangle = \frac{c_m+ c_{m*}}{4}\\
&&\langle \frac{X_m-X_{m*}}{2i} , \frac{X_m-X_{m*}}{2i} \rangle = -\frac{c_m+ c_{m*}}{4}, \label{sect2:cm complex conjugate 2}
\end{eqnarray}
where we have used orthogonal condition $\langle X_m, X_{m*}\rangle=0$.  Since the above inner products differ by a minus sign, one of them must be negative, which leads to a ghost. One can also see this from the effective action. By applying (\ref{action-scalar6}), we obtain the total action of the complex conjugate pair 
 \begin{eqnarray}\label{sect2:complexconjugateaction}
I_m+I_{m*} &\simeq&\frac{c_m}{2} \int_Q d^dy \ Y_m (\Box -m^2)Y_m + \frac{c_{m*}}{2} \int_Q d^dy \ Y_{m*} (\Box -{m*}^2)Y_{m*},\nonumber\\
&\simeq& \int_Q d^dy \Big( \frac{c_m+c_{m*}}{2} (Y_r \Box Y_r-Y_i \Box Y_i ) -2 \frac{c_m-c_{m*}}{2i} Y_r \Box Y_i +...\Big),
\end{eqnarray}
where $...$ label the non-kinetic energy term, $Y_r$ and $Y_i$ denote the real and imaginary parts of $Y_m$, respectively. The kinetic energy terms of $Y_r$ and $Y_i$ differ by a minus sign, implying a ghost. Note that the cross-term does not matter. Rewrite (\ref{sect2:complexconjugateaction}) into the matrix form $Y^T M \Box Y$, where $Y^T=(Y_r, Y_i)$ and $M$ is a $2\times 2$ matrix. The determinant of $M$ is negative,
 \begin{eqnarray}\label{sect2:complexconjugateaction1}
|M|=-(\frac{c_m+c_{m*}}{2})^2-( \frac{c_m-c_{m*}}{2i} )^2<0.
\end{eqnarray}
As a result, the action (\ref{sect2:complexconjugateaction}) includes a ghost. To summarize, we have ghosts for both $\alpha>0$ and $\alpha<0$.

{\bf{Case III}: all non-ghost modes}

Let us discuss the effective action (\ref{action-scalar3}) summing over all the non-ghost modes with $c_m>0$, which has interesting properties. We comment on the sum of all modes at the end. For $\alpha>0$, all the modes with $m^2>0$ are non-ghost fields. As for $\alpha<0$, non-ghost modes include one of the complex conjugate pairs.

Define $M_{m,m'}=\Big{[} c_m \delta_{m,m'}(\Box -m^2) + \alpha X_m(1) X_{m'}(1) (\Box -m^2) (\Box -m'^2) \Big{]}$ in the effective action (\ref{action-scalar3}) and calculate its determinant, we get
 \begin{eqnarray}\label{detM}
|M| \sim (\Box-m_g^2)\prod_{m\ne m_g} (\Box-m^2),
\end{eqnarray}
where $\prod_{m\ne m_g}$ denotes the product overall non-ghost modes in the mass spectrum (\ref{massscalar}) and 
 \begin{eqnarray}\label{massghost}
m_g^2=\frac{-\frac{1}{\alpha}+\sum_{m\ne m_g} \frac{X_{m}^2(1) m^2}{c_{m}} }{\sum_{m\ne m_g} \frac{X_{m}^2(1)}{c_{m}} }.
\end{eqnarray}
Eq.(\ref{detM}) suggests that apart from the non-ghost modes, there is an additional mode with mass $m_g$ on the brane. This extra mode depends on the HD coupling $\alpha$ and thus can be identified with the ghost from the HD interaction. We choose $Y_m=Y_g X_m(1)/c_m$ to see this. Then the effective action (\ref{action-scalar3}) becomes 
 \begin{eqnarray}\label{action-scalar4}
I_g=\frac{1}{2}\sum_{m'\ne m_g}\alpha \frac{X^2_{m'}(1)}{c_{m'}}\int_Q d^dy \ Y_g \sum_{m\ne m_g}\frac{X^2_m(1)}{c_m}(\Box -m^2) (\Box -m_g^2) Y_g.
\end{eqnarray}
To get a second-order effective action for the mode obeying $(\Box-m_g^2) Y_g=0$, we replace $(\Box-m^2)$ with $(m_g^2-m^2)$. Then, the above action can be further simplified as
 \begin{eqnarray}\label{action-scalar5}
I_g\simeq-\frac{1}{2}\sum_{m'\ne m_g}\frac{X^2_{m'}(1)}{c_{m'}}\int_Q d^dy \ Y_g (\Box -m_g^2) Y_g.
\end{eqnarray}
Since we sum over the non-ghost modes, we have $c_{m'}>0$. As a result, the action (\ref{action-scalar5}) has a wrong sign, which suggests the extra mode with mass $m_g$ is a ghost. One may wonder if this extra mode is the ghost discussed in Case I and Case II (the blue modes of Table. \ref{table1scalarspectrum}). As we will show below, the answer is yes. Remarkably, although we consider only the sum over non-ghost modes in the effective action (\ref{action-scalar3}), the final result contains the ghost information. It implies the mass spectrum (\ref{massscalar}) has novel global properties.

Let us calculate the mass (\ref{massghost}) of the extra mode. Recall that for large mass, we have $m^2\to (q \pi)^2$ (\ref{massscalar1}), which yields
 \begin{eqnarray}\label{BCcheck1}
 c_m\to \frac{1}{2}, \ \frac{X_{m}^2(1) m^2}{c_{m}}\to \frac{2}{\pi ^4 \alpha ^2 q^4}, \ \frac{X_{m}^2(1)}{c_{m}} \to \frac{2}{\pi ^6 \alpha ^2 q^6},
 \end{eqnarray}
 where we have used (\ref{solutionX}) with $a_1=1$ and (\ref{cmorthogonal}) in the above derivations. For the large mass, the $\alpha$ terms dominate the mass spectrum (\ref{massscalar}). As a result, only parameter $\alpha$ appears in the above limits \footnote{Note that (\ref{BCcheck1}) applies to the case $\alpha\ne 0$. Note also that $m^2\to (q \pi)^2$ is not necessarily the mass of the q-th mode. It may be the (q-1)-th mode.}. Note that the above equations are good approximates for even small integer $q$. Then, we could take the numerical results for the first few non-ghost modes (see table.\ref{table1scalarspectrum}) and the good approximate (\ref{BCcheck1}) for the other modes. In this way, we derive the mass (\ref{massghost})
\begin{equation}\label{sect.2.1:massghost}
m_g^2 \approx \begin{cases}
-4.722, &\ \text{for $\lambda=0, \alpha=0.1$},\\
2.089 \mp 3.157 i, &\ \text{for $\lambda=0, \alpha=-0.1$}
\end{cases}
\end{equation}
which agrees with the ghost mass of Table. \ref{table1scalarspectrum} with high numerical accuracy.  It strongly supports that the extra mode implied by the effective action is the ghost mode contained in the mass spectrum (\ref{massscalar}). We also checked some examples with non-zero $\lambda$ and find the ghost mass  (\ref{massghost}) derived from the effective action indeed obeys NBC  (\ref{massscalar}). Finally, we want to mention that if one sums over all the modes (including both ghost and non-ghost modes) in the effective action, one also gets an extra mode with the mass formally given by (\ref{massghost}).  We numerically find that 
 \begin{eqnarray}\label{sect2:massghostrelation}
\sum_{m} \frac{X_{m}^2(1)}{c_{m}} \to 0, \ \ \ \ \sum_{m} \frac{X_{m}^2(1) m^2}{c_{m}}\to  \frac{1}{\alpha},
\end{eqnarray}
which implies $m_g^2= 0/0$ is not well-defined and can be regarded as fake mode. That is reasonable. Since we have already considered all the modes in the effective action, it can not produce additional mode.  We prove (\ref{sect2:massghostrelation}) in the next subsection and verify it for large $\alpha$ perturbatively. 

As a summary, there is always a ghost for the brane-localized HD scalar with $\alpha\ne 0, \lambda=0$. For $\alpha>0$, there is a tachyon mode with $m^2<0$, also a ghost. For $\alpha<0$, a pair of modes with complex conjugate mass exists. One real combination of such a pair is a ghost. The above results agree with the effective action on the brane, which includes HD terms and implies a ghost for $\alpha\ne 0$. We end this second by commenting on the DGP-like scalar with $\alpha=0, \lambda\ne 0$. We have $m^2<0$ for $\lambda<0$ and $m^2>0$ for $\lambda>0$. Unlike the brane-localized HD scalar, there is no complex $m^2$ for the DGP-like scalar. As a result, there is no ghost for $\lambda>0, \alpha=0$. One can easily observe that the inner product $c_m$ (\ref{cmorthogonal}) is always positive for positive $\lambda$. Thus, the DGP-like scalar with $\lambda\ge 0$ is well-defined; it is ghost-free and tachyon-free.   

\subsection{Spectrum identities}

In this subsection, we prove the following spectrum identities
\begin{eqnarray}\label{toymodel: SI DGP}
&&\sum_m  \frac{X^2_m(1)}{c_m}=\frac{1}{\lambda}, \ \ \ \ \ \ \  \ \ \ \ \ \ \  \ \ \ \ \ \ \   \ \ \ \ \ \ \ \ \ \ \   \text{for}\  \alpha=0,\\
&&\sum_{m} \frac{X_{m}^2(1)}{c_{m}} = 0, \ \ \ \ \sum_{m} \frac{X_{m}^2(1) m^2}{c_{m}}= \frac{1}{\alpha}, \ \ \ \text{for}\  \alpha\ne0, \label{toymodel: SI HD}
\end{eqnarray}
where $c_m=\langle X_m, X_{m} \rangle$ denotes the inner product.  We find they are useful to analyze the ghost modes for brane-located interactions.  We comment on this problem at the end of this subsection. 

To start, we use EOM (\ref{EOMX}) to rewrite the orthogonal relation as
\begin{eqnarray}\label{scalarorthogonal1}
\langle X_m, X_{n} \rangle=\int_0^1 dx X_m(x) X_{n}(x) +\lambda X_m(1) X_{n}(1)- \alpha X''_m(1) X_{n}(1)- \alpha X_m(1) X''_{n}(1).
\end{eqnarray}
An advantage of the above expression is that it does not depend on the mass exactly. As a result, it can be easily generalized to arbitrary functions
\begin{eqnarray}\label{scalarorthogonal2}
\langle f, g \rangle=\int_0^1 dx f(x) g(x) +\lambda f(1) g(1)- \alpha f''(1) g(1)- \alpha f(1) g''(1),
\end{eqnarray}
where $f(x)=\sum_{m} f_m X_m(x) $ and $g(x)=\sum_{m} g_m X_m(x) $ with constants $f_m$ and $g_m$. 

{\bf Proof I: DGP-like case with $\alpha=0$}

Let us first prove the spectrum identity (\ref{toymodel: SI DGP}) for the DGP-like case with $\lambda\ne 0$ and $\alpha=0$. The key technique is making use of the step function 
\begin{eqnarray}\label{toymodel: step function}
\Pi_0(x)=\Pi(x-1)=\begin{cases}
0,\  \ \ \ \ \ \ \text{for } x<1,\\
1,\  \ \ \ \ \ \ \text{for } x\ge 1,
\end{cases}
\end{eqnarray}
where $\Pi(x)$ is the usual step function, which is zero for $x<0$ and one otherwise. We remark that the step function $\Pi_0(x)$ naturally describes the brane-localized interactions at $x=1$. We expand the step function in the power of KK modes
\begin{eqnarray}\label{toymodel: step function1}
\Pi_0(x)=\sum_{m} \frac{\langle \Pi_0(x), X_m(x) \rangle }{\langle X_m, X_m \rangle} \ X_m(x) = \lambda \sum_{m}  \frac{X_m(1) X_m(x)}{c_m}.
\end{eqnarray}
In the above calculations, we have used $c_m=\langle X_m, X_m \rangle$ and
\begin{eqnarray}\label{toymodel: step function3}
\langle \Pi_0(x), X_m(x) \rangle =\int_0^1 dx \Pi_0(x) X_m(x) +\lambda \Pi_0(1) X_m(1)=\lambda X_m(1),
\end{eqnarray} 
which can be derived from the orthogonal relation (\ref{scalarorthogonal2}) with $\alpha=0$ and (\ref{toymodel: step function}). From (\ref{toymodel: step function}) and  (\ref{toymodel: step function1}), we obtain the spectrum identity (\ref{toymodel: SI DGP}) with $\lambda\ne 0$ and $\alpha=0$
\begin{eqnarray}\label{toymodel: SI DGP simple proof}
\Pi_0(1)= \lambda \sum_{m}  \frac{X^2_m(1)}{c_m}=1. 
\end{eqnarray}

There is another way to prove the spectrum identity (\ref{toymodel: SI DGP}).  We show it below as a double check of our proofs. From the second equation of (\ref{toymodel: step function1}) and $\langle X_m, X_{n} \rangle=c_m \delta_{m, n}$, we derive
\begin{eqnarray}\label{toymodel: proof DGP 1}
\langle \Pi_0(x), \Pi_0(x) \rangle =\lambda^2 \sum_{m,n}  \frac{X_m(1) }{c_m}  \frac{X_n(1) }{c_n} \langle X_m, X_n \rangle = \lambda^2 \sum_{m}  \frac{X^2_m(1) }{c_m}.
\end{eqnarray} 
From (\ref{scalarorthogonal2}) with $\alpha=0$ and (\ref{toymodel: step function}), we get
\begin{eqnarray}\label{toymodel: proof DGP 2}
\langle \Pi_0(x), \Pi_0(x) \rangle =\int_0^1 dx \Pi_0^2(x) +\lambda \ \Pi_0^2(1)=\lambda. 
\end{eqnarray} 
Comparing (\ref{toymodel: proof DGP 1}) with (\ref{toymodel: proof DGP 2}), we obtain again the spectrum identity (\ref{toymodel: SI DGP}) for $\alpha=0$. 

Let us make some comments. First, the step function $\Pi_0(x)=\Pi(x-1)$ naturally describes the brane-located interactions such as the DGP-like term, since they take non-zero values only on the brane $x=1$. Thus, it is unsurprising that it plays an important role in proving the corresponding spectrum identity on the brane. Second, one may wonder if one could expand the step function regarding the KK modes. We verify that one can do so in an example where the exact expression of the mass spectrum is known. In the large  $\lambda$ limit, the mass spectrum for $\alpha=0$ reads
\begin{eqnarray}\label{toy model: spectrum}
m^2:\   \frac{1}{\lambda }\Big(1-\frac{1}{3 \lambda }+O(\frac{1}{\lambda^2})\Big), \ \ (\pi  q)^2 \Big(1+\frac{2}{\pi ^2  q^2 \lambda }+O(\frac{1}{\lambda^2})\Big),
\end{eqnarray}
which can be solved from $X_m(x)=\sin(\sqrt{m^2}x)$ and (\ref{massscalar}). Here $q=1,2,...$ are integers.  From the mass spectrum and $X_m(x)=\sin(\sqrt{m^2}x)$, we derive for $0\le x<1$
\begin{eqnarray}\label{toy model: test1}
\lambda \frac{X_m(1) X_m(x)}{c_m}: x-\frac{x \left(x^2+1\right)}{6 \lambda }, \ \frac{2 (-1)^q \sin (\pi  q x)}{\pi  q}+\frac{2 (-1)^q (\pi  q x \cos (\pi  q x)-2 \sin (\pi  q x))}{\pi ^3 \lambda  q^3},
\end{eqnarray}
and for $x=1$
\begin{eqnarray}\label{toy model: test2}
\lambda \frac{X_m(1) X_m(1)}{c_m}: \ \  1-\frac{1}{3 \lambda }+O\left(\frac{1}{\lambda^2 }\right), \ \ \ \ \ \ 
\frac{2}{\pi ^2 \lambda q^2}+O\left(\frac{1}{\lambda^2 }\right).
\end{eqnarray}
Note that we have ignored the $O(1/\lambda^2)$ terms in (\ref{toy model: test1}) for presentation purpose. Note also that (\ref{toy model: test1}) agrees with (\ref{toy model: test2}) in the limit $x\to 1$.  From (\ref{toy model: test1}) and  (\ref{toy model: test2}), we obtain for $0\le x <1$
\begin{eqnarray}\label{toy model: test3}
&&\lambda \sum_{m}\frac{X_m(1) X_m(x)}{c_m}\nonumber\\
&=& x-\frac{x \left(x^2+1\right)}{6 \lambda }+
\sum_{q=1}^{\infty}\Big[ \frac{2 (-1)^q \sin (\pi  q x)}{\pi  q}+\frac{2 (-1)^q (\pi  q x \cos (\pi  q x)-2 \sin (\pi  q x))}{\pi ^3 \lambda  q^3} \Big]+O\left(\frac{1}{\lambda^2 }\right),\nonumber\\
&=&0+O\left(\frac{1}{\lambda^2 }\right),
\end{eqnarray}
and for $x=1$
\begin{eqnarray}\label{toy model: test4}
\lambda \sum_{m}\frac{X_m(1) X_m(1)}{c_m}&=& 1-\frac{1}{3 \lambda }+
\sum_{q=1}^{\infty}\frac{2}{\pi ^2 \lambda q^2}+O\left(\frac{1}{\lambda^2 }\right)=1+O\left(\frac{1}{\lambda^2 }\right).
\end{eqnarray}
In the above calculations, we have used the following sum formulas
\begin{eqnarray}\label{toy model: sum 1}
&&\sum _{q=1}^{\infty } \frac{1}{q^2}=\frac{\pi ^2}{6}, \ \ \ \ \ \ \sum_{q=1}^{\infty } \frac{2 (-1)^q \sin (\pi  q x)}{\pi  q}=-x,\\ \label{toy model: sum 2}
&&\sum _{q=1}^{\infty } \frac{2 (-1)^q (\pi  q x \cos (\pi  q x)-2 \sin (\pi  q x))}{\pi ^3 q^3}=\frac{1}{6} x \left(x^2+1\right), 
\end{eqnarray}
which pass the tests of Mathematica for $0\le x< 1$. It is clear from (\ref{toy model: test3},\ref{toy model: test4}) that 
\begin{eqnarray}\label{toy model: step function proof}
\lambda \sum_{m}\frac{X_m(1) X_m(x)}{c_m}=\Pi_0(x)+O(\frac{1}{\lambda^2}),
\end{eqnarray}
which verifies that the step function (\ref{toymodel: step function}) can indeed be expanded in the power of KK modes. Thus, the above proof of spectrum identity (\ref{toymodel: SI DGP}) with $\alpha=0$ is solid.   Setting $x=1$ for (\ref{toy model: step function proof}), we recover the spectrum identity (\ref{toymodel: SI DGP}) in the large $\lambda$ limit.  

{\bf Proof II: HD case with $\alpha\ne 0$}

Let us go on to prove the spectrum identity (\ref{toymodel: SI HD}) with $\alpha\ne 0$.  Let us define a new function 
\begin{eqnarray}\label{toymodel II: step function dd}
\Pi_2(x)=\begin{cases}
0,\  \ \ \ \ \ \ \ \ \ \ \ \ \ \text{for } x<1,\\
\frac{(x-1)^2}{2},\  \ \ \ \ \ \ \text{for } x\ge 1, 
\end{cases}
\end{eqnarray}
whose second derivative is the step function $\Pi''_2(x)=\Pi_0(x)$.  Expanding $\Pi_2(x)$ in the power of KK modes, we have
\begin{eqnarray}\label{toymodel II: step function dd 1}
\Pi_2(x)=\sum_{m} \frac{\langle \Pi_2(x), X_m(x) \rangle }{\langle X_m, X_m \rangle} \ X_m(x) = -\alpha \sum_{m}  \frac{X_m(1) X_m(x)}{c_m}.
\end{eqnarray}
In the above derivation, we have used 
\begin{eqnarray}\label{toymodel: step function3}
\langle \Pi_2(x), X_m(x) \rangle& =&\int_0^1 dx \Pi_2(x) X_m(x) +\lambda \Pi_2(1) X_m(1)-\alpha \Pi''_2(1) X_m(1)-\alpha \Pi_2(1) X''_m(1)\nonumber\\
&=&-\alpha \Pi''_2(1) X_m(1)=-\alpha X_m(1),
\end{eqnarray} 
which can be derived from the orthogonal relation (\ref{scalarorthogonal2}) and (\ref{toymodel II: step function dd}).  From (\ref{toymodel II: step function dd}), $\Pi''_2(x)=\Pi_0(x)$, EOM $X_m''(x)=- m^2 X_m(x)$ and (\ref{toymodel II: step function dd 1}), we derive
\begin{eqnarray}\label{toymodel: SI HD simple proof1}
&&\Pi_2(1)= -\alpha \sum_{m}  \frac{X^2_m(1)}{c_m}=0,  \\ \label{toymodel: SI HD simple proof2}
&&\Pi''_2(1)=\Pi_0(1)= \alpha \sum_{m}  \frac{X^2_m(1) m^2}{c_m}=1, 
\end{eqnarray}
which are just the spectrum identities (\ref{toymodel: SI HD}) with $\alpha\ne 0$. 

Below, we provide another proof. From (\ref{toymodel II: step function dd 1}) and $\langle X_m, X_{n} \rangle=c_m \delta_{m, n}$, we derive
\begin{eqnarray}\label{toymodel II: proof HD 1}
\langle \Pi_2(x), \Pi_2(x) \rangle =\alpha^2 \sum_{m,n}  \frac{X_m(1) }{c_m}  \frac{X_n(1) }{c_n} \langle X_m, X_n \rangle = \alpha^2 \sum_{m}  \frac{X^2_m(1) }{c_m}.
\end{eqnarray} 
From the orthogonal relation (\ref{scalarorthogonal2}) and (\ref{toymodel II: step function dd}), we get
\begin{eqnarray}\label{toymodel II: proof HD 2}
\langle \Pi_2(x), \Pi_2(x) \rangle =\int_0^1 dx \Pi_2^2(x) +\lambda \Pi_2^2(1)-2 \alpha \Pi''_2(1) \Pi_2(1)=0.
\end{eqnarray} 
Comparing (\ref{toymodel II: proof HD 1}) with (\ref{toymodel II: proof HD 2}), we obtain the first equation of the spectrum identity (\ref{toymodel: SI HD}). 

From (\ref{toymodel II: step function dd 1}), EOM $X_m''(x)=- m^2 X_m(x)$ and $\langle X_m, X_{n} \rangle=c_m \delta_{m, n}$, we derive
\begin{eqnarray}\label{toymodel II: proof HD 1b}
\langle \Pi''_2(x), \Pi_2(x) \rangle =-\alpha^2 \sum_{m,n}  \frac{X_m(1) m^2}{c_m}  \frac{X_n(1) }{c_n} \langle X_m, X_n \rangle = -\alpha^2 \sum_{m}  \frac{X^2_m(1)m^2 }{c_m}.
\end{eqnarray} 
From the orthogonal relation (\ref{scalarorthogonal2}) and (\ref{toymodel II: step function dd}), we get
\begin{eqnarray}\label{toymodel II: proof HD 2b}
\langle \Pi''_2(x), \Pi_2(x) \rangle &=&\int_0^1 dx \Pi''_2(x)\Pi_2(x) +\lambda \Pi''_2(1) \Pi_2(1) -\alpha \Pi''''_2(1) \Pi_2(1)-\alpha \Pi''_2(1) \Pi''_2(1)\nonumber\\
&=&-\alpha. 
\end{eqnarray} 
Above, we have used $\int_0^1 dx \Pi''_2(x)\Pi_2(x)=0, \Pi''_2(1) \Pi_2(1)=\Pi_0(1) \Pi_2(1)=0$, $\Pi''''_2(1) \Pi_2(1)=\lim_{x\to 1}\delta'(x-1) (x-1)^2/2=0$ and $ \Pi''_2(1) \Pi''_2(1)= \Pi_0(1) \Pi_0(1)=1$. Here $\delta'(x-1) (x-1)^2=0$ can be derived from the derivative of $\delta(x-1) (x-1)^2=0$ and $\delta(x-1) (x-1)=0$.  Comparing (\ref{toymodel II: proof HD 1b}) with (\ref{toymodel II: proof HD 2b}), we finally obtain the second equation of the spectrum identity (\ref{toymodel: SI HD}).

Similar to DGP-like case, we can verify the spectrum identity (\ref{toymodel: SI HD}) in the large $\alpha$ limit. To do so, we define 
\begin{eqnarray}\label{toy model: fm}
f_m(x)=-\alpha \frac{X_m(1) X_m(x)}{c_m}. 
\end{eqnarray}
 From the mass spectrum (\ref{sect2:masslargealpha}) in the large $\alpha$ limit, we derive for $0\le x<1$
\begin{eqnarray}\label{toy model: HD test1}
f_m(x):\  \frac{\sqrt{\alpha } x}{2}+\frac{1}{12} x \left(x^2-1\right),\  -\frac{\sqrt{\alpha } x}{2}+\frac{1}{12} x \left(x^2-1\right),\  -\frac{2 \left((-1)^q \sin (\pi  q x)\right)}{\pi ^3 q^3}+O(\frac{1}{\sqrt{\alpha}})
\end{eqnarray}
and 
\begin{eqnarray}\label{toy model: HD test2}
f''_m(x):\  \frac{x}{2}+\frac{1}{12} \sqrt{\frac{1}{\alpha }} \left(x^3+3 \lambda  x\right),\  \frac{x}{2}-\frac{1}{12} \sqrt{\frac{1}{\alpha }} \left(x^3+3 \lambda  x\right),\  \frac{2 (-1)^q \sin (\pi  q x)}{\pi  q}+O(\frac{1}{\alpha}).
\end{eqnarray}
Similarly, we obtain for $x=1$
\begin{eqnarray}\label{toy model: HD test3}
f_m(1):  \ \  \frac{\sqrt{\alpha }}{2}+O(\frac{1}{\sqrt{\alpha}}), \  \  -\frac{\sqrt{\alpha }}{2}+O(\frac{1}{\sqrt{\alpha}}), \ \ 0+O(\frac{1}{\sqrt{\alpha}}),
\end{eqnarray}
and 
\begin{eqnarray}\label{toy model: HD test3}
f''_m(1):  \  \frac{1}{2}+\frac{1}{12} \sqrt{\frac{1}{\alpha }} \left(1+3 \lambda  \right),\  \frac{1}{2}-\frac{1}{12} \sqrt{\frac{1}{\alpha }} \left(1+3 \lambda \right),\  0+O(\frac{1}{\alpha})
\end{eqnarray}
Summing above equations, we obtain for $0\le x<1$
\begin{eqnarray}\label{toy model: HD test4a}
&&\sum_m f_m(x)=\frac{1}{6} x \left(x^2-1\right)+\sum_{q=1}^{\infty}  -\frac{2 \left((-1)^q \sin (\pi  q x)\right)}{\pi ^3 q^3}=O(\frac{1}{\sqrt{\alpha}}),\\ \label{toy model: HD test4b}
&&\sum_m f''_m(x)=x+\sum_{q=1}^{\infty}  \frac{2 (-1)^q \sin (\pi  q x)}{\pi  q}=O(\frac{1}{\alpha})
\end{eqnarray}
and for $x=1$
\begin{eqnarray}\label{toy model: HD test5a}
&&\sum_m f_m(1)= \frac{\sqrt{\alpha }}{2}- \frac{\sqrt{\alpha }}{2}=O(\frac{1}{\sqrt{\alpha}}),\\ \label{toy model: HD test5b}
&&\sum_m f''_m(1)=\frac{1}{2}+\frac{1}{12} \sqrt{\frac{1}{\alpha }} \left(1+3 \lambda  \right)+\frac{1}{2}-\frac{1}{12} \sqrt{\frac{1}{\alpha }} \left(1+3 \lambda \right)=1+O(\frac{1}{\alpha}).
\end{eqnarray}
Eqs.(\ref{toy model: HD test4a},\ref{toy model: HD test4b},\ref{toy model: HD test5a},\ref{toy model: HD test5b}) imply 
\begin{eqnarray}\label{toy model: HD test6}
\sum_m f_m(x)= -\alpha \sum_m \frac{X_m(1) X_m(x)}{c_m}= \Pi_2(x),
\end{eqnarray}
where $\Pi_2(x)$ is defined in (\ref{toymodel II: step function dd}). Expressing (\ref{toy model: HD test5a}) and (\ref{toy model: HD test5b}) exactly in KK modes, we recover the spectrum identities (\ref{toymodel: SI HD}) in the large $\alpha$ limit
\begin{eqnarray}\label{toy model: HD test7a}
&&-\alpha \sum_m \frac{X^2_m(1)}{c_m}= O(\frac{1}{\sqrt{\alpha}}),\\ \label{toy model: HD test7b}
&&\alpha \sum_m \frac{X^2_m(1)m^2}{c_m}=1+O(\frac{1}{\alpha}). 
\end{eqnarray}.

Now we finish the proofs of the spectrum identities (\ref{toymodel: SI DGP}) and (\ref{toymodel: SI HD}). Let us make some comments.  

{\bf 1}. The spectrum identities (\ref{toymodel: SI DGP}) and (\ref{toymodel: SI HD}) are useful for analyzing the existence of ghosts. As discussed in section 2.1, there are always ghosts for complex mass spectrums. We are interested in the ghost-free parameters, so we assume the mass spectrum is real below. We will return to this assumption soon. For real mass spectrum, $X_m(1), c_m, m^2$ are all real numbers. For the brane-localized HD scalar with $\alpha\ne 0$, the first equation of spectrum identity (\ref{toymodel: SI HD}) suggests at least one KK mode is a ghost with negative inner product $c_m=\langle X_m, X_m \rangle$. Then, the ghost-free condition requires $\alpha=0$. Next, for the DGP-like case with $\alpha=0$ and $\lambda\ne 0$, the spectrum identity (\ref{toymodel: SI DGP}) implies at least one negative inner product $c_m=\langle X_m, X_m \rangle$ for negative parameter $\lambda$.  Overall, the necessary condition for a ghost-free theory is 
$\lambda\ge 0$ and $\alpha=0$.  Under this condition, it is clear that all the inner products are positive
\begin{eqnarray}\label{toy model: positive inner product}
\langle X_m, X_m \rangle= \int_0^{1} X^2_m(x)  dx+ \lambda X^2_m(1)>0. 
\end{eqnarray}
Thus, the necessary and sufficient ghost-free condition is
\begin{eqnarray}\label{toy model: ghost-free condition}
\text{ghost-free condition}:\ \ \lambda\ge 0,\ \ \ \a=0.
\end{eqnarray} 
Recall that complex $m^2$ always yields ghosts and we have assumed $m^2$ is real in the above derivations. Denote the parameter space with real $m^2$ by $A$, and the parameter space (\ref{toy model: ghost-free condition}) by $B$. Strictly speaking, the ghost-free parameter space we obtained is $A\cap B$. In the following, we will prove $B \in A$. As a result, we have $A\cap B=B$, which means the strict ghost-free condition is still given by (\ref{toy model: ghost-free condition}). 

 {\bf 2}.  The mass spectrum is real and positive $m^2>0$ under the ghost-free condition (\ref{toy model: ghost-free condition}). Thus, our above assumption of the real mass spectrum is consistent. Let us first prove that $m^2$ is real, that $B \in A$. Suppose that there are complex $m^2$ in the mass spectrum. Since the equation of motion and boundary conditions are real, complex $m^2$ must appear in a complex conjugate pair. By applying the orthogonal condition (\ref{scalarorthogonal}) with $\alpha=0$ for the complex conjugate pair, we get 
\begin{eqnarray}\label{toy model: real mass 1}
\int_0^{1} X_m(x) X_{m^*}(x)  dx+\lambda X_m(1) X_{m^*}(1) =c_m \delta_{m, m^*}=0.
\end{eqnarray}
On the other hand, we have for $\lambda\ge 0$
\begin{eqnarray}\label{toy model: real mass 2}
\int_0^{1} X_m(x) X_{m^*}(x)  dx+\lambda X_m(1) X_{m^*}(1) >0.
\end{eqnarray}
The above contradiction suggests no complex mass exists for $\lambda\ge 0$ and $\alpha=0$. Let us go on to prove $m^2>0$. From EOM $X''_m(x)=-m^2 X_m(x)$ and BCs (\ref{scalarDBC1},\ref{scalarNBC1}), we derive for $\alpha=0$
\begin{eqnarray}\label{toy model: positive mass 1}
\int_0^{1} X'_m(x) X'_{m}(x)  dx&=&X'_m(1) X_m(1)-\int_0^{1} X''_m(x) X_{m}(x)  dx\nonumber\\
&=&m^2 \Big(\int_0^{1} X^2_m(x)  dx+ \lambda X^2_m(1) \Big)=m^2 c_m,
\end{eqnarray}
which yields $m^2>0$ since $\int_0^{1} X'_m(x) X'_{m}(x)  dx$ and $c_m=\int_0^{1} X^2_m(x)  dx+ \lambda X^2_m(1) $ are both positive for $\lambda\ge 0$.

\subsection{AdS space}

\begin{figure}[t]
\centering
\includegraphics[width=7.1cm]{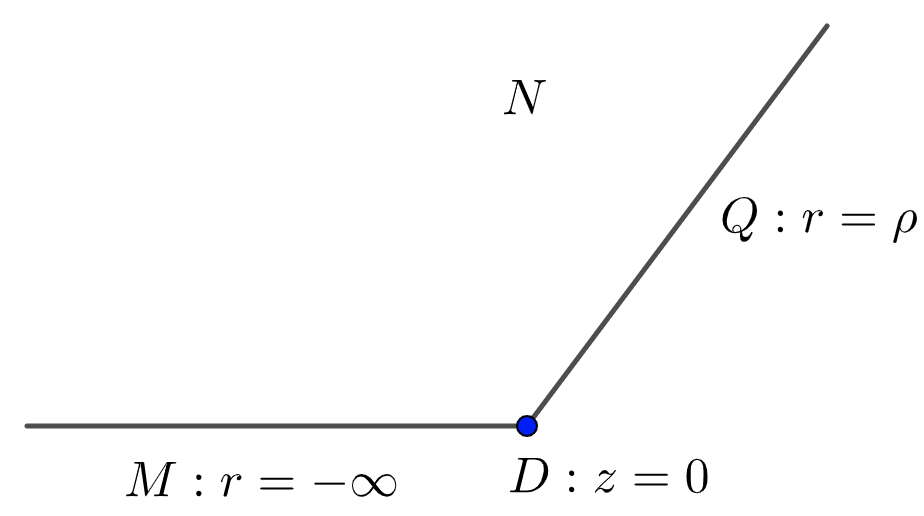}\ \ \ \ \includegraphics[width=7cm]{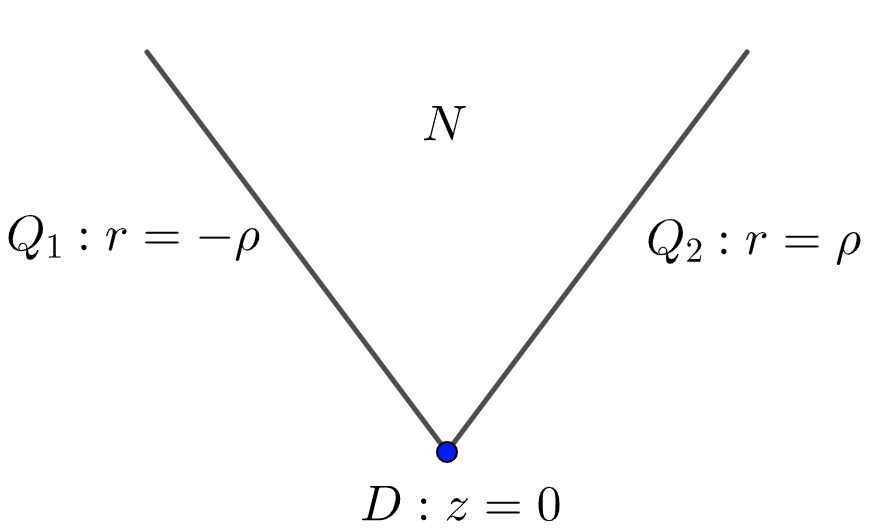}
\caption{Geometry of AdS/BCFT (Left) and wedge holography (Right). Here $N$ is the bulk space, $Q, Q_1, Q_2$ are the end-of-the-world branes, $M$ is the AdS boundary, and $D$ is the defect. }
\label{AdSBCFT}
\end{figure}

The above discussions focus on flat space. 
Let us go on to discuss the brane-localized higher derivative scalars in an AdS space. For simplicity, we focus on AdS/BCFT in the following, and the results can be easily generalized to wedge holography \cite{Akal:2020wfl,Miao:2020oey}. See Fig. \ref{AdSBCFT} for the geometry of AdS/BCFT and wedge holography. Let us focus on the probe limit with the bulk metric
 \begin{eqnarray}\label{metricAdS}
ds^2=g_{\mu\nu} dx^{\mu}dx^{\nu}=dr^2+\cosh^2(r) \bar{h}^{(0)}_{ij}(y) dy^i dy^j, \ \ \ \ \ -\infty < r\le \rho,
\end{eqnarray}
where the AdS boundary $M$ is at $r=-\infty$, the end-of-the-world brane $Q$ locates at $r=\rho$, $ \bar{h}^{(0)}_{ij}=\text{sech}^2(\rho ) h_{ij}$ denotes the AdS$_d$ metric on the brane
 \begin{eqnarray}\label{metricAdS}
ds^2_0= \bar{h}^{(0)}_{ij}(y) dy^i dy^j=\frac{dz^2-dt^2+\sum_{a=1}^{d-2} dw_a^2}{z^2},
\end{eqnarray}
and the defect $D$ is at $z=0$. Below, we list only the main results since the approach is similar to sect.2.1. Besides, we want to leave the more detailed discussions to the case of gravity in the next section.

Substituting the ansatz $\phi=X(r) Y(y)$ into EOM (\ref{EOMscalar}) and separating variables, we derive 
 \begin{eqnarray}\label{EOMscalarAdSY}
&&( \bar{\Box} - m^2) Y(y)=0,\\
&& \cosh^2(r) X''(r)+d \sinh (r) \cosh (r) X'(r) + m^2 X(r)=0, \label{EOMscalarAdSX}
\end{eqnarray}
where $ \bar{\Box}=\bar{D}^i\bar{D}_i$, $\bar{D}_i$ are the covariant derivatives defined by $\bar{h}^{(0)}_{ij}=\text{sech}^2(\rho ) h_{ij}$, and $m$ denotes the mass of scalars.  The ansatz $\phi=X(r) Y(y)$ simplifies DBC $\phi|_M=0$ on the AdS boundary and NBC (\ref{NBCscalar}) on the brane to be
 \begin{eqnarray}\label{DBCscalarAdS1}
&&X(-\infty)=0, \\ \label{NBCscalarAdS1}
&&X'(\rho)=\Big( \lambda\ \text{sech}^2(\rho) m^2 +\alpha\ \text{sech}^4(\rho)  m^2(m^2-d+1)  \Big) X(\rho),
\end{eqnarray}
where we have used (\ref{EOMscalarAdSY}) above.  Similar to the case of above subsection, the BCs (\ref{DBCscalarAdS1}) and (\ref{NBCscalarAdS1}) impose a constraint for the mass of scalar KK modes. 

Following the approach of above subsection, we expand bulk scalar in series of KK modes 
 \begin{eqnarray}\label{scalar-power-AdS}
\phi= \sum_{m}X_m(r) Y_m(y),
\end{eqnarray}
where $X^{(m)}(r)$ obey EOM (\ref{EOMscalarAdSX}) and the orthogonal condition
 \begin{eqnarray}\label{orthogonal-scalar-AdS}
 \langle X_m, X_{m'}\rangle=c_m \delta_{m,m'} &=&\int_{-\infty}^{\rho}\frac{\cosh^{d-2}(r)}{\cosh^{d-2}(\rho)}X_m(r) X_{m'}(r) dr \nonumber\\
&+&\Big(\lambda +\alpha\ \text{sech}^2(\rho )(m^2+m'^2+1-d)  \Big) X_m(\rho) X_{m'}(\rho).
\end{eqnarray}
Note that the orthogonal condition (\ref{orthogonal-scalar-AdS}) can also be read off from the action  (\ref{actionscalar}) and BCs (\ref{DBCscalarAdS1},\ref{NBCscalarAdS1}). One can check the solutions to EOM (\ref{EOMscalarAdSX}) and BCs (\ref{DBCscalarAdS1},\ref{NBCscalarAdS1}) indeed obey the above orthogonal condition. Substituting (\ref{scalar-power-AdS}) into the action (\ref{actionscalar}) and applying the orthogonal condition (\ref{orthogonal-scalar-AdS}), we obtain the effective action on the AdS brane
 \begin{eqnarray}\label{effective-action-scalar-AdS}
I&=&\frac{1}{2}\sum_{m,m'} \int_Q d^dy \sqrt{|\bar{h}^{(0)}|} \cosh^{d-2}(\rho) \nonumber\\
&&\ \times \ Y_m \Big{[} c_m \delta_{m,m'}(\bar{\Box} -m^2) + \alpha\ \text{sech}^2(\rho ) X_m(\rho) X_{m'}(\rho) (\bar{\Box}  -m^2) (\bar{\Box} -m'^2)   \Big{]} Y_{m'}, 
\end{eqnarray}
which takes similar form as the flat case (\ref{action-scalar3}). The effective action on the brane contains higher derivative terms, implying a ghost for $\alpha \ne0$. 

Similar to the case of flat space, the following spectrum identities hold for the brane-localized HD scalar in AdS space
\begin{eqnarray}\label{toy model AdS: SI DGP}
&&\sum_m  \frac{X^2_m(\rho)}{c_m}=\frac{1}{\lambda}, \ \ \ \ \ \ \  \ \ \ \ \ \ \  \ \ \ \ \ \ \   \ \ \ \ \ \ \ \ \ \ \    \ \ \ \ \ \ \ \  \text{for}\  \alpha=0,\\
&&\sum_{m} \frac{X_{m}^2(\rho)}{c_{m}} = 0, \ \ \ \ \sum_{m} \frac{X_{m}^2(\rho) m^2}{c_{m}}= \frac{\cosh^2(\rho)}{\alpha}, \ \ \ \text{for}\  \alpha\ne0, \label{toy model AdS: SI HD}
\end{eqnarray}
where $c_m=\langle X_m, X_{m} \rangle$ labels the inner product. See Appendix B.1 for the proof of the above spectrum identities. Assuming the mass spectrum is real, from the first equation of (\ref{toy model AdS: SI HD}), we observe that at least one inner product $c_m$ is negative for the non-zero HD parameter $\alpha$. To have a ghost-free theory, we require $\alpha=0$. For the case $\alpha=0$, (\ref{toy model AdS: SI DGP}) suggests that at least one inner product $c_m$ is negative for negative DGP parameter $\lambda$. Overall, we get the necessary conditions for a ghost-free theory, i.e., $\lambda\ge 0$ and $\alpha=0$. Under this condition, all the inner products are positive, 
\begin{eqnarray}\label{toy model: positive mass 1 AdS}
\langle X_m, X_{m} \rangle=\int_{-\infty}^{\rho}\frac{\cosh^{d-2}(r)}{\cosh^{d-2}(\rho) }X^2_m(r) dr+\lambda  X^2_m(\rho) >0,
\end{eqnarray}
which implies it is also a sufficient condition. Thus, the necessary and sufficient ghost-free condition is given by (\ref{toy model: ghost-free condition}), i.e., $\lambda\ge 0$ and $\alpha=0$. 

Note that we have assumed real $m^2$ in the above discussions. Label the parameter space of real $m^2$ by $A$, and the constraint (\ref{toy model: ghost-free condition}) by $B$.  The ghost-free parameter space we obtained is $A\cap B$. Below, we will prove $B \in A$, which yields $A\cap B=B$. It means the ghost-free parameter space is indeed given by $B$, i.e., $\lambda\ge 0$ and $\alpha=0$.

Now, let us prove the mass spectrum is real $(B \in A)$ and positive $m^2>0$ for the ghost-free parameters $\lambda\ge 0$ and $\alpha=0$. If there were complex $m^2$, as discussed in sect. 2.1, they must appear in a complex conjugate pair. From the orthogonal condition (\ref{orthogonal-scalar-AdS}) with $\alpha=0$, we derive for the complex conjugate pair
\begin{eqnarray}\label{toy model: real mass 1 AdS}
\int_{-\infty}^{\rho}\frac{\cosh^{d-2}(r)}{\cosh^{d-2}(\rho)}X_m(r) X_{m^*}(r) dr+\lambda  X_m(\rho) X_{m^*}(\rho)=c_m \delta_{m, m^*}=0.
\end{eqnarray}
On the other hand, we have for $\lambda\ge 0$
\begin{eqnarray}\label{toy model: real mass 2 AdS}
\int_{-\infty}^{\rho}\frac{\cosh^{d-2}(r)}{\cosh^{d-2}(\rho)}X_m(r) X_{m^*}(r) dr+\lambda  X_m(\rho) X_{m^*}(\rho) >0.
\end{eqnarray}
The above contradiction rules out complex mass for $\lambda\ge 0$ and $\alpha=0$. Let us go on to prove $m^2>0$. From EOM (\ref{EOMscalarAdSX}) and BCs (\ref{DBCscalarAdS1},\ref{NBCscalarAdS1}), we derive for $\alpha=0$
\begin{eqnarray}\label{toy model: positive mass 1 AdS}
&&\int_{-\infty}^{\rho} \cosh(r)^{d} X'_m(r)X'_m(r)  dr=\cosh(\rho)^{d}  X'_m(\rho) X_m(\rho)-\int_{-\infty}^{\rho} (\cosh^{d}(r) X'_m(r))'  X_{m}(r)  dr\nonumber\\
&=&m^2 \Big(\int_{-\infty}^{\rho}\cosh^{d-2}(r)X^2_m(r) dr+\lambda \cosh^{d-2}(\rho)  X^2_m(\rho) \Big)= m^2 c_m \cosh^{d-2}(\rho).
\end{eqnarray}
Because $\int_{-\infty}^{\rho} \cosh(r)^{d} X'_m(r)X'_m(r)  dr$ and $c_m$ are both positive for $\lambda\ge 0$ and $\alpha=0$, the above equation leads to $m^2>0$.

In summary, we have studied a toy model of a brane-localized scalar in flat and AdS space in this section. First, we derive the effective action on the brane, which includes HD terms and implies a ghost for non-zero HD interaction $\alpha\ne 0$. Furthermore, we obtain novel spectrum identities for KK modes on the brane, which derives the ghost-free condition, i.e., $\lambda\ge 0$ and $\alpha= 0$. Finally, we prove that the mass spectrum is real and positive $m^2>0$ for the ghost-free parameters $\lambda\ge 0$ and $\alpha= 0$. This section's conclusions also work for brane-localized gravity, and we discuss it in detail in the next section.

\section{Brane-localized gravity}

This section investigates some aspects of brane-localized gravity, which includes the spectrum identities for KK modes, the effective action and the brane-bending mode.  We find there is no ghost for positive DGP gravity and brane-localized GB gravity. On the other hand, ghost always appears for one class of brane-localized HD gravity.

For simplicity, we focus on the following action with DGP gravity and curvature-squared gravity on the brane
\begin{eqnarray}\label{gravityaction}
&&I=\int_N d^{d+1}x \sqrt{|g|} \Big(\mathcal{R} +d\left(d-1\right) \Big)\nonumber\\
&&+2\int_Q d^dy \sqrt{|h|} \Big(K-T +\lambda R+ \a_1 L_{\text{GB}}(\hat{R})+ \a_2 (\hat{R}^{ij}\hat{R}_{ij}-\frac{d}{4(d-1)}\hat{R}^2)+ \a_3 \hat{R}^2 \Big)
\end{eqnarray}
where \cite{Miao:2022mdx,Miao:2023unv} \footnote{Note that the brane-localized HD gravity is expressed in terms of $\hat{R}_{ijkl}$ (\ref{background curvature3}) instead of the usual curvature $R_{ijkl}$. Note also that $\hat{R}_{ijkl}$ and thus the HD terms vanish in the background AdS geometry. As a result, the tension parameter (\ref{Tension}) is independent of the HD parameters $\a_i$. On the other hand, the tension depends on the HD parameters if the brane-localized HD gravity is expressed in the usual way, such as $L_{\text{GB}}(R)$.}
 \begin{eqnarray}\label{Tension}
T=(d-1) \tanh(\rho)-\lambda (d-1)(d-2)\text{sech}^2(\rho)
 \end{eqnarray} 
 is the brane tension, $\lambda, \alpha_i$ are parameters, $\mathcal{R}$ and $R$ are Ricci scalars in bulk $N$ and on the brane $Q$ respectively,  $L_{\text{GB}}(\hat{R})=\hat{R}_{ijkl} \hat{R}^{ijkl}-4\hat{R}_{ij} \hat{R}^{ij}+\hat{R}^2$ and $\hat{R}_{ijkl}$ are defined by
\begin{eqnarray}\label{background curvature}
&&\hat{R}=R+d(d-1)\text{sech}^2(\rho),\\ \label{background curvature2}
&&\hat{R}_{ij}=R_{ij}+(d-1)\text{sech}^2(\rho) h_{ij},\\ \label{background curvature3}
&&\hat{R}_{ijkl}=R_{ijkl}+ \text{sech}^2(\rho)(h_{ik}h_{jl}-h_{il}h_{jk}),
\end{eqnarray}
which vanish on the AdS brane with radius $l=\cosh(\rho)$. Note that we have set the bulk AdS radius to be $L=1$ and the brane AdS radius to be $l=\cosh(\rho)$. From the action (\ref{gravityaction}), we can derive NBC on the brane
\begin{eqnarray}\label{sect3:NBC}
K^{ij}-(K-T+\lambda R) h^{ij}+2 \lambda R^{ij}+2E^{ij}=0,
\end{eqnarray}
with
\begin{eqnarray}\label{sect3:NBCHij}
 E_{ij}=P_{( i}^{\ mnl} R_{j) mnl}-2 D^m D^n P_{imnj}-\frac{1}{2} L_{\text{HD}} h_{ij},
\end{eqnarray}
where $L_{\text{HD}}=\a_1 L_{\text{GB}}(\hat{R})+ \a_2 (\hat{R}^{ij}\hat{R}_{ij}-\frac{d}{4(d-1)}\hat{R}^2)+ \a_3 \hat{R}^2$, $D_{i}$ denotes covariant derivative with respect to $h_{ij}$, and
\begin{eqnarray}\label{sect3:NBCPijkl}
P^{ijkl}=\frac{\partial L_{\text{HD}} }{\partial R_{ijkl}}&=&2 \a_1 \hat{R}^{ijkl}+ (\a_1+\a_3-\frac{d \a_2}{4(d-1)}) \hat{R}(g^{ik}g^{jl}-g^{il}g^{jk})\nonumber\\
 &&+(\frac{\a_2}{2}-2\a_1) (\hat{R}^{ik} g^{jl}-\hat{R}^{jk} g^{il}+\hat{R}^{jl} g^{ik}-\hat{R}^{il} g^{jk}).
\end{eqnarray}
Since $\hat{R}^{ijkl}$ vanish on the AdS brane with radius $l=\cosh(\rho)$, so do $P^{ijkl}, L_{\text{HD}}$ and $ E_{ij}$ those depending on $\hat{R}^{ijkl}$. Then the NBC (\ref{sect3:NBC}) on the AdS brane imposes a constraint only on $T$ and $\lambda$, which yields (\ref{Tension}). 

Let us make some comments on the brane-localized HD terms. 
\begin{itemize}
  \item $L_{\text{GB}}(\hat{R})=L_{\text{GB}}(R)+2 (d-3) (d-2) R\ \text{sech}^2(\rho)+(d-3) (d-2) (d-1) d\ \text{sech}^4(\rho )$ is a combination of Gauss-Bonnet gravity and Einstein gravity, whose EOM contains only second-order derivatives. 
   \item The specific combination $(\hat{R}^{ij}\hat{R}_{ij}-\frac{d}{4(d-1)}\hat{R}^2)$ appears in many theories such as new massive gravity \cite{Bergshoeff:2009hq}, holographic renormalization \cite{deHaro:2000vlm,Balasubramanian:1999re}, brane effective action \cite{Chen:2020uac} and critical gravity \cite{Lu:2011zk}. Note that  $(\hat{R}^{ij}\hat{R}_{ij}-\frac{d}{4(d-1)}\hat{R}^2)$ is a Weyl density, which is Weyl invariant up to a total derivative for $d=4$. As we will show below, this HD term leads to a ghost generally. 
  \item  By studying the brane bending mode, we find $\hat{R}^2$ induces a potential scalar ghost generally. However, this ghost does not obey EOM and seems to be harmless. The physical conditions considered in this paper do not restrict $\hat{R}^2$. We leave a careful study of this HD term to future work.  
 \end{itemize}

\subsection{Spectrum identities}

We choose the standard ansatz of the perturbation metric and the embedding function of $Q$
 \begin{eqnarray}\label{perturbationmetric}
&&ds^2=dr^2+\cosh^2 (r) \left( \bar{h}^{(0)}_{ij}(y) + \epsilon H(r) \bar{h}^{(1)}_{ij}(y)  \right)dy^i dy^j+O(\epsilon^2),\\
&& Q: \ r=\rho+O(\epsilon^2), \label{perturbationQ}
\end{eqnarray}
where $\bar{h}^{(0)}_{ij}=\text{sech}^2(\rho) h^{(0)}_{ij}$ is the AdS metric with a unit radius and $\bar{h}^{(1)}_{ij}$ denotes the perturbation.  We assume the brane location is unchanged under the first-order perturbations.  We leave the discussions of fluctuations of brane location to sect.3.3. 

Imposing the transverse traceless gauge 
 \begin{eqnarray}\label{hij1gauge}
\bar{D}^i \bar{h}^{(1)}_{ij}=0,\ \ \  \bar{h}^{(0)ij}\bar{h}^{(1)}_{ij}=0,
\end{eqnarray}
and separating variables of Einstein equations, we get
\begin{eqnarray}\label{EOMMBCmassivehij}
&& \left(\bar{\Box}+2-m^2\right)\bar{h}^{(1)}_{ij}(y)=0,\\
&& \cosh^2(r) H''(r)+d \sinh (r) \cosh (r) H'(r) + m^2 H(r)=0, \label{EOMMBCmassiveH}
\end{eqnarray}
where $m$ denotes the mass of gravitons. We remark that EOM (\ref{EOMMBCmassiveH}) takes the same expression as that of scalar (\ref{EOMscalarAdSX}). Solving the above equation, we get
 \begin{eqnarray}\label{massiveHsolution}
H(r)=\text{sech}^{\frac{d}{2}}(r) \left(b_1 P_{l_m}^{\frac{d}{2}}(-\tanh r)+b_2 Q_{l_m}^{\frac{d}{2}}(-\tanh r)\right),
\end{eqnarray} 
where $P_{l_m}^{\frac{d}{2}}$ and $ Q_{l_m}^{\frac{d}{2}}$ are the Legendre polynomials, $b_1$ and $b_2$ are integral constants and $l_m$ is given by
 \begin{eqnarray}\label{aibia1}
l_m=\frac{1}{2} \left(\sqrt{(d-1)^2+4  m^2}-1\right).
\end{eqnarray}
Note that $l_m$ is real if the Breitenlohner-Freedman (BF) bound of massive gravity in $\text{AdS}_d$ holds
 \begin{eqnarray}\label{BFgravity}
 m^2\ge -(\frac{d-1}{2})^2.
\end{eqnarray}

We impose DBC on the AdS boundary $M$ 
 \begin{eqnarray}\label{DBCsgravity}
H(-\infty)=0,
\end{eqnarray}
which yields 
\begin{equation}\label{Htwocase}
H(r)=\begin{cases}
 \ \text{sech}^{\frac{d}{2}}(r)  \ P_{l_m}^{\frac{d}{2}}(-\tanh r),&\
\text{even $d$} ,\\
\  \text{sech}^{\frac{d}{2}}(r) \ Q_{l_m}^{\frac{d}{2}}(-\tanh r),&\
\text{odd $d$}.
\end{cases}
\end{equation}
Without loss of generality, we have set the integral constants $b_1=b_2=1$.  According to \cite{Chu:2021mvq}, (\ref{massiveHsolution}) are no longer general solutions for 
\begin{equation}\label{neg-m2}
 m^2 =\begin{cases}
 -\frac{1}{4}((d-1)^2-1),&\ \text{even $d$},\\
 -\frac{1}{4}(d-1)^2,&\ \text{odd $d$}.
\end{cases}
\end{equation}
One should resolve EOM (\ref{EOMMBCmassiveH}) for these special $m^2$. In general, (\ref{neg-m2}) does not belong to the mass spectrum in a ghost-free theory. As we will show below, the ghost-free condition yields $m^2\ge 0$.

For the ansatz (\ref{perturbationmetric},\ref{perturbationQ}) and gauge (\ref{hij1gauge}), NBC (\ref{sect3:NBC}) on the brane $Q$ becomes
 \begin{eqnarray}\label{NBCsgravity1}
H'(\rho) h^{(1)}_{ij}=2H(\rho)\Big( \left(\lambda \text{sech}^2(\rho )+ 4(d-3)\a_1 \text{sech}^4(\rho ) \right)(\bar{\Box}+2)+ \a_2 \text{sech}^4(\rho)(\bar{\Box}+2)^2 \Big) h^{(1)}_{ij}.
\end{eqnarray}
By using (\ref{EOMMBCmassivehij}), we can simplify (\ref{NBCsgravity1}) as
 \begin{eqnarray}\label{NBCsgravity2}
H'(\rho) =2H(\rho) m^2\Big(\lambda\ \text{sech}^2(\rho )+ 4(d-3)\a_1\ \text{sech}^4(\rho ) + \a_2\ \text{sech}^4(\rho)m^2 \Big). \end{eqnarray}
For $m^2=0$, the NBC (\ref{NBCsgravity2}) reduces to that of standard AdS/BCFT. Since there is no massless mode in the standard AdS/BCFT, so is our case with DGP and HD gravity on the brane. On the other hand, there is a massless mode in wedge holography with and without brane-localized DGP and HD gravity. 

Substituting (\ref{Htwocase}) into (\ref{NBCsgravity2}), we get a constraint for mass spectrum of gravitons
 \begin{eqnarray}\label{sect3:mass spectrum AdSBCFT}
  \begin{cases}
  P_{l _m+1}^{\frac{d}{2}}(-\tanh\rho)=P_{l _m}^{\frac{d}{2}}(-\tanh\rho) \left(\tilde{\lambda}\ \text{sech}^2(\rho ) \left(d+2 l_m\right)-\tanh\rho\right),&\
\text{even $d$} ,\\
 Q_{l_m+1}^{\frac{d}{2}}(-\tanh\rho)=Q_{l_m}^{\frac{d}{2}}(-\tanh\rho) \left(\tilde{\lambda}\ \text{sech}^2(\rho ) \left(d+2 l_m\right)-\tanh\rho\right),&\
\text{odd $d$},
\end{cases}
\end{eqnarray}
where $l_m$ is given by (\ref{aibia1}) and 
 \begin{eqnarray}\label{sect3.1:lambdabar}
\tilde{\lambda}=\Big(\lambda+ 4(d-3)\a_1\ \text{sech}^2(\rho ) + \a_2\ \text{sech}^2(\rho)m^2 \Big).
\end{eqnarray}
Since $H'(\rho)$ and $H(\rho)$ are finite, the BC (\ref{NBCsgravity2}) approaches DBC
 \begin{eqnarray}\label{NBCsgravity3}
\lim_{m^2\to \infty} H(\rho) \to 0,
\end{eqnarray}
in the large mass limit. It is similar to the toy model of sect.2.1, which also reduces to DBC $X(1)=\sin(\sqrt{m^2})\to 0$ for large mass. According to \cite{Chu:2021mvq}, the DBC (\ref{NBCsgravity3}) yields the mass spectrum
\begin{equation}\label{masslargesmallrho}
  m^2 \approx \begin{cases}
  q(q+d-1),&\ \text{for large $\rho$},\\
  2q(2q+d-1),&\ \text{for small $\rho$}.
\end{cases}
\end{equation}
where $q$ are integers. One can check numerically (\ref{masslargesmallrho}) agrees with (\ref{sect3:mass spectrum AdSBCFT}) for large mass.  

Now, we are ready to discuss the spectrum identities of gravitational KK modes on the brane. The orthogonal condition of KK modes reads
 \begin{eqnarray}\label{orthogonal-gravity}
\langle H_m, H_{m'} \rangle&=&c_m\delta_{m, m'}=\int_{-\infty}^{\rho}\frac{\cosh^{d-2}(r)}{\cosh ^{d-2}(\rho )}H_m(r) H_{m'}(r) dr\nonumber\\
&+&2\Big(\lambda +4(d-3) \a_1 \text{sech}^2(\rho )+\a_2 \text{sech}^2(\rho ) (m^2+m'^2) \Big) H_m(\rho) H_{m'}(\rho).
\end{eqnarray}
One can check that solutions to EOM (\ref{EOMMBCmassiveH}) and BCs (\ref{DBCsgravity},\ref{NBCsgravity2}) obey the above orthogonal relation. From the orthogonal condition (\ref{orthogonal-gravity}), we obtain the following spectrum identities 
\begin{eqnarray}\label{gravity: SI DGP}
&&\sum_{m} \frac{H_m^2(\rho)}{\langle H_m, H_m \rangle} = \frac{1}{2\Big(\lambda +4(d-3) \a_1 \text{sech}^2(\rho )\Big)}, \ \ \ \ \ \ \ \ \ \ \ \ \text{for }\alpha_2=0, \\ \label{gravity: SI HD}
&&\sum_{m} \frac{H_m^2(\rho)}{\langle H_m, H_m \rangle} = 0, \ \ \ \ \sum_{m} \frac{H_m^2(\rho) m^2}{\langle H_m, H_m \rangle}=\frac{\cosh^2(\rho)}{2\alpha_2}, \ \ \ \ \ \ \ \text{for } \alpha_2\ne0,
\end{eqnarray} 
Please refer to the Appendix B.2 for comprehensive proofs of the above spectrum identities. We have also taken great care to verify the identities (\ref{gravity: SI DGP},\ref{gravity: SI HD}) through rigorous analytical calculations in the large $\alpha_2$ limit and numerical calculations in general.  Similar to the toy model of sect. 2 \footnote{See discussions around (\ref{toy model: ghost-free condition}) or below (\ref{toy model AdS: SI HD}). }, the spectrum identities  (\ref{gravity: SI DGP},\ref{gravity: SI HD}) show that at least one inner product $\langle H_m, H_m \rangle$ is negative for nonzero $\alpha_2$ or negative $( \lambda +4(d-3) \a_1 \text{sech}^2(\rho) )$. Then, we obtain the ghost-free condition 
\begin{eqnarray}\label{gravity: ghost-free condition}
\text{ghost-free condition}:\ \ \Big(\lambda +4(d-3) \a_1 \text{sech}^2(\rho )\Big)\ge 0, \ \ \ \a_2=0. 
\end{eqnarray} 
Like the toy model discussed in sect. 2, it is the necessary and sufficient condition for a ghost-free theory. 

Under the ghost-free condition (\ref{gravity: ghost-free condition}), we can prove the mass spectrum is real and positive $m^2>0$. Because EOM (\ref{EOMMBCmassiveH}) and BCs (\ref{DBCsgravity},\ref{NBCsgravity2}) are real, the complex $m^2$ could only appear in a complex conjugate pair. Then the orthogonal condition (\ref{orthogonal-gravity}) with $\alpha_2=0$ yields zero $\langle H_m, H_{m^*} \rangle=c_m \delta_{m,m^*}$ for $m\ne m^*$
\begin{eqnarray}\label{gravity: real mass 1}
\int_{-\infty}^{\rho}\frac{\cosh^{d-2}(r)}{\cosh^{d-2}(\rho )}H_m(r) H_{m^*}(r) dr
+2\Big(\lambda +4(d-3) \a_1 \text{sech}^2(\rho ) \Big) H_m(\rho) H_{m^*}(\rho)
=0.
\end{eqnarray}
On the other hand, we have positive inner product $\langle H_m, H_{m^*} \rangle$
\begin{eqnarray}\label{gravity: real mass 2}
\int_{-\infty}^{\rho}\frac{\cosh^{d-2}(r)}{\cosh ^{d-2}(\rho )}H_m(r) H_{m^*}(r) dr
+2\Big(\lambda +4(d-3) \a_1 \text{sech}^2(\rho ) \Big) H_m(\rho) H_{m^*}(\rho)>0,
\end{eqnarray}
provided that the ghost-free condition (\ref{gravity: ghost-free condition}) holds. The above contradiction suggests no complex mass in the mass spectrum. 

Let us go on to prove $m^2>0$. From EOM (\ref{EOMMBCmassiveH}) and BCs (\ref{DBCsgravity},\ref{NBCsgravity2}), we obtain for $\alpha_2=0$
\begin{eqnarray}\label{gravity: positive mass 1}
&&\int_{-\infty}^{\rho} \cosh^{d} (r)H'_m(r)H'_m(r)  dr=\cosh^{d} (\rho)  H'_m(\rho) H_m(\rho)-\int_{-\infty}^{\rho} (\cosh^{d} (r)H'_m(r))'  H_{m}(r)  dr\nonumber\\
&=&m^2 \Big[\int_{-\infty}^{\rho}\cosh^{d-2}(r)H^2_m(r) dr+2\Big(\lambda +4(d-3) \a_1 \text{sech}^2(\rho )\Big) \cosh^{d-2}(\rho)  H^2_m(\rho) \Big]\nonumber\\
&=& m^2 c_m \cosh^{d-2}(\rho).
\end{eqnarray}
In the ghost-free constraint (\ref{gravity: ghost-free condition}), both $\int_{-\infty}^{\rho} \cosh^{d}(r) H'_m(r)H'_m(r)  dr$ and $c_m$ are positive. As a result, the above equation results in 
\begin{eqnarray}\label{gravity: tachyon-free condition}
\text{tachyon-free condition}:\ \ m^2>0. 
\end{eqnarray} 
In AdS, the theory is tachyon-free if it obeys the BF bound $m^2\ge -(\frac{d-1}{2})^2$. Thus, $m^2$ can be negative generally.  Here we get a stronger version of the tachyon-free condition $m^2> 0$.

\subsection{Squared action}

Let us go on study the squared action of gravitons on the brane $Q$. Following the approach of sect.2, we expand the perturbative metric (\ref{perturbationmetric}) in series of KK modes 
 \begin{eqnarray}\label{gravity-power}
\delta g_{r \mu}=0,  \ \ \delta g_{ij}=\cosh(r)^2 \sum_{m}H_m(r)\bar{h}_{(m)ij}(y), 
\end{eqnarray}
where $\sum_{m}$ denotes the sum over the gravitational mass spectrum,  $ \bar{h}_{(m)ij}(y)$ and $H_m(r)$ satisfy EOM (\ref{EOMMBCmassivehij},\ref{EOMMBCmassiveH}) and the orthogonal condition (\ref{orthogonal-gravity}). 

The squared bulk action of (\ref{gravityaction})  is given by  \cite{Hu:2022lxl}
 \begin{eqnarray}\label{gravityIbulk}
&&I_{\text{bulk}}=\int_N d^{d+1}x\sqrt{|g|} \Big{[}-\frac{1}{4} \nabla_{\a} \bar{H}_{\mu\nu} \nabla^{\alpha} \bar{H}^{\mu\nu}+
\frac{1}{2} \nabla_{\alpha} \bar{H}_{\mu\nu} \nabla^{\nu} \bar{H}^{\mu \alpha}-\frac{d}{2}  \bar{H}_{\mu\nu}  \bar{H}^{\mu\nu} \Big{]} \nonumber\\
&&\ \ \ \ \ \ \ + \int_{Q}d^{d}y\sqrt{|h|} n^{\mu}\bar{H}^{\a\b} (\nabla_{\mu} \bar{H}_{\a\b}-\nabla_{\a} \bar{H}_{\b \mu} ),
 \end{eqnarray}
where  $\bar{H}_{\mu\nu} =\delta g_{\mu\nu}$ denote the metric perturbation.  Recall that we focus on transverse traceless gauge (\ref{hij1gauge}), equivalently,
 \begin{eqnarray}\label{gijgauge}
\nabla^{\mu} \delta g_{\mu\nu}=0,\quad  g^{\mu\nu}\delta g_{\mu\nu}=0,
\end{eqnarray}
where $\nabla_{\mu}$ are covariant derivatives related to $g_{\mu\nu}$. Under this gauge, we rewrite the above action as
 \begin{eqnarray}\label{gravityIbulk1}
I_{\text{bulk}}&=&\int_N d^{d+1}x\sqrt{|g|} \Big{[}-\frac{1}{4} \nabla_{\a} \bar{H}_{\mu\nu} \nabla^{\alpha} \bar{H}^{\mu\nu}+\frac{1}{2}  \bar{H}_{\mu\nu}  \bar{H}^{\mu\nu} \Big{]}\nonumber\\
&&+ \int_{Q}d^{d}y\sqrt{|h|} n^{\mu}\bar{H}^{\a\b} (\nabla_{\mu} \bar{H}_{\a\b}-\frac{1}{2}\nabla_{\a} \bar{H}_{\b \mu} ).
 \end{eqnarray}
We find the following formulas are useful
\begin{eqnarray}\label{gravitygoodformula1}
&&\nabla_r \bar{H}_{ij}=\cosh^2(r) \sum_{m} \bar{h}^{(m)}_{ij}(y) \frac{d}{dr}H^{(m)}(r)=\frac{H'(r)}{H(r)} \bar{H}_{ij}, \\ \label{gravitygoodformula2}
&& \nabla_i \bar{H}_{jr}= -\cosh(r)\sinh(r)\sum_{m} \bar{h}^{(m)}_{ij}(y)H^{(m)}(r)=-\tanh(r) \bar{H}_{ij},\\
\label{gravitygoodformula3}
&& \nabla_i \bar{H}_{jk}= \cosh^2(r)\sum_{m} \bar{D}_i\bar{h}^{(m)}_{jk}(y)H^{(m)}(r),
 \end{eqnarray}
 which yields
  \begin{eqnarray}\label{gravityformula}
n^{\mu}\bar{H}^{\a\b} \nabla_{\mu} \bar{H}_{\a\b}|_{Q}=\bar{H}^{\a\b} \Big(\frac{H'(\rho)}{H(\rho)}  \bar{H}_{\a\b} \Big), \ \ n^{\mu}\bar{H}^{\a\b}\nabla_{\a} \bar{H}_{\b \mu}|_{Q} =-\tanh(\rho) \bar{H}_{\a\b} \bar{H}^{\a\b},
 \end{eqnarray}
 where $\Big(\frac{H'(\rho)}{H(\rho)}  \bar{H}_{\a\b} \Big)$ denotes $\sum_m \frac{H'^{(m)}(\rho)}{H^{(m)}(\rho)}  \bar{H}^{(m)}_{\a\b}$. Then the bulk action (\ref{gravityIbulk1}) becomes
 \begin{eqnarray}\label{gravityIbulk2}
&&I_{\text{bulk}}=\int_N d^{d+1}x\sqrt{|g|} \Big{[}-\frac{1}{4} \nabla_{\a} \bar{H}_{\mu\nu} \nabla^{\alpha} \bar{H}^{\mu\nu}+\frac{1}{2}  \bar{H}_{\mu\nu}  \bar{H}^{\mu\nu} \Big{]} \nonumber\\
&&\ \ \ \ \ \ \ + \int_{Q}d^{d}y\sqrt{|h|}\bar{H}^{\a\b}  \left(\left(\frac{H'(\rho)}{H(\rho)} +\frac{1}{2}\tanh(\rho)\right) \bar{H}_{\a\b}\right).
 \end{eqnarray}
The calculations of boundary action of (\ref{gravityaction}) is more complicated.  By using the gauge (\ref{hij1gauge}) and dropping some total derivatives on the brane $Q$, we derive at the second order of $\bar{H}$

 \begin{eqnarray}\label{someformula1}
2\int_{Q} \sqrt{|h|} \Big(K-(d-1) \tanh(\rho)\Big)=\int_{Q} \sqrt{|h|}\bar{H}^{\a\b}  \Big[\left(-\frac{H'(\rho)}{H(\rho)} -\frac{1}{2}\tanh(\rho)\right) \bar{H}_{\a\b}\Big],
 \end{eqnarray}
 
  \begin{eqnarray}\label{someformula2}
\int_{Q} \sqrt{|h|} \Big(R+\frac{(d-1)(d-2)}{\cosh^2(\rho)}\Big)=\frac{1}{4}\int_{Q} \sqrt{|h|}  \bar{H}_{ij} \Big(\Box+2\text{sech}^2(\rho )\Big) \bar{H}^{ij},
 \end{eqnarray}
 
 \begin{eqnarray}\label{someformula3}
\int_{Q} \sqrt{|h|} L_{\text{GB}}(\hat{R})=(d-3)\text{sech}^2(\rho)\int_{Q} \sqrt{|h|}  \bar{H}^{ij} \Big(\Box+2\text{sech}^2(\rho)\Big) \bar{H}_{ij},
 \end{eqnarray}
 
  \begin{eqnarray}\label{someformula4}
\int_{Q} \sqrt{|h|}  (\hat{R}^{ij}\hat{R}_{ij}-\frac{d}{4(d-1)}\hat{R}^2)=\frac{1}{4}\int_{Q} \sqrt{|h|} \bar{H}^{ij} \Big(\Box+2\text{sech}^2(\rho )\Big)^2 \bar{H}_{ij},
 \end{eqnarray}
 
  \begin{eqnarray}\label{someformula4}
\int_{Q} \sqrt{|h|}  \hat{R}^2=0.
 \end{eqnarray}
 
 By applying the above formulas, we derive the total action
  \begin{eqnarray} \label{gravityItotal1}
&&I=\sum_{m,m'}\int_{-\infty}^{\rho} dr \cosh^d(r)\int_{Q}\sqrt{|\bar{h}^{(0)}|} \Big{[}-\frac{1}{4} \text{sech}^2(r )\bar{D}_{k} \bar{h}^{(m)}_{ij} \bar{D}^{k} \bar{h}^{(m')}{}^{ij} H_{m}(r)H_{m'}(r)\nonumber\\
&&\ \ \ \ \ \ \ \ \ \ \ \ \ \  +\Big(-\frac{1}{4} H'_{m}(r)H'_{m'}(r)+\frac{1}{2}\text{sech}^2(r) H_{m}(r)H_{m'}(r) \Big)\bar{h}^{(m)}_{ij}  \bar{h}^{(m')}{}^{ij} \Big{]}\nonumber\\
&&+\sum_{m,m'}\int_{Q}\sqrt{|\bar{h}^{(0)}|} \cosh^{d-2}(\rho)H_{m}(\rho)H_{m'}(\rho)\bar{h}^{(m)}_{ij}  \Big{[} (\frac{\lambda}{2}+2\a_1 (d-3)\text{sech}^2(\rho ) ) (\bar{\Box}+2)  \nonumber\\
&&\ \ \ \ \ \ \ \ \ \ \ \ \ \  \ \ \ \ \ \ \ \ \ \ \ \ \ \  \ \ \ \ \ \ \ \ \ \ \ \ \ \  \ \ \ \ \ \ \ \ \ \ \ \ \ \ \ \ \ \ \ \ \ \ \ +\frac{\a_2}{2} \text{sech}^2(\rho)  (\bar{\Box}+2)^2  \Big{]}  \bar{h}^{(m')}{}^{ij}.
\end{eqnarray}
Integrating by parts for bulk action and using EOM of $H(r)$ (\ref{EOMMBCmassiveH}), we get
 \begin{eqnarray} \label{gravityItotal2}
I&=&\sum_{m,m'}\int_{-\infty}^{\rho} dr \cosh^{d-2}(r)H_{m}(r)H_{m'}(r)\int_{Q}\sqrt{|\bar{h}^{(0)}|}\  \frac{1}{4}  \bar{h}^{(m)}_{ij} (\bar{\Box}+2-m^2)\bar{h}^{(m')}{}^{ij} \nonumber\\
&&-\frac{1}{4}\sum_{m,m'}\int_{Q}\sqrt{|\bar{h}^{(0)}|}  \cosh^{d}(\rho)H'_{m}(\rho)H_{m'}(\rho)\bar{h}^{(m)}_{ij} \bar{h}^{(m')}{}^{ij}\nonumber\\
&&+\sum_{m,m'}\int_{Q}\sqrt{|\bar{h}^{(0)}|} \cosh^{d-2}(\rho)H_{m}(\rho)H_{m'}(\rho)\bar{h}^{(m)}_{ij}  \Big{[} (\frac{\lambda}{2}+2\a_1 (d-3)\text{sech}^2(\rho ) ) (\bar{\Box}+2)  \nonumber\\
&&\ \ \ \ \ \ \ \ \ \ \ \ \ \  \ \ \ \ \ \ \ \ \ \ \ \ \ \  \ \ \ \ \ \ \ \ \ \ \ \ \ \  \ \ \ \ \ \ \ \ \ \ \ \ \ \ \ \ \ \ \ \ +\frac{\a_2}{2} \text{sech}^2(\rho)  (\bar{\Box}+2)^2  \Big{]}  \bar{h}^{(m')}{}^{ij}.
\end{eqnarray}
By applying the orthogonal condition (\ref{orthogonal-gravity}) and BCs ( \ref{DBCsgravity},\ref{NBCsgravity2}), we finally obtain effective action on the brane,
 \begin{eqnarray} \label{gravityItotal3}
I&=&\frac{1}{4}\sum_{m,m'}\int_{Q}d^dy\sqrt{|\bar{h}^{(0)}|}  \cosh^{d-2}(\rho) \bar{h}^{(m)}_{ij} \Big[ c_m \delta_{m,m'} (\bar{\Box}+2-m^2)   \nonumber\\
&&\ \ \ \ \ \ \ \ \ \ \ \ \ \ \ +2 \a_2 \text{sech}^2(\rho ) H_m(\rho)H_{m'}(\rho)  (\bar{\Box}+2-m^2)   (\bar{\Box}+2-m'^2) \Big] \bar{h}^{(m')}{}^{ij},
\end{eqnarray}
which takes similar expression as the scalar action (\ref{action-scalar3},\ref{effective-action-scalar-AdS}).  

Now we take the effective action to study the ghost problem for brane-localized HD gravity.  Similar to the case of scalar in sect.2.1, the HD terms of the effective action (\ref{gravityItotal3}) suggests a ghost with the mass
 \begin{eqnarray}\label{sect3.2:gravityghost}
m_g^2=\frac{-\frac{\cosh^2(\rho)}{2\a_2}+\sum_{m\ne m_g} \frac{H_{m}^2(\rho) m^2}{c_{m}} }{\sum_{m\ne m_g} \frac{H_{m}^2(\rho)}{c_{m}} },
\end{eqnarray}
where $\sum_{m\ne m_g}$ denotes the sum overall all non-ghost modes. 

It is remarkable that the mass of ghost and non-ghost modes has a nice relation (\ref{sect3.2:gravityghost}). Following the approach of sect.2.1, substituting $\bar{h}^{(m)}_{ij}=\frac{H_m(\rho)}{c_m} \bar{h}_{g ij}$ into (\ref{gravityItotal3}), we derive the effective action of the ghost
\begin{eqnarray}\label{action-scalar4}
I_g&=&\frac{1}{2}\sum_{m'\ne m_g} \a_2 \text{sech}^2(\rho) \frac{H^2_{m'}(\rho)}{c_{m'}}\int_Q d^dy \sqrt{|\bar{h}^{(0)}|}  \cosh^{d-2}(\rho)\ \bar{h}_{gij}  \sum_{m\ne m_g}\frac{H^2_m(\rho)}{c_m}(\bar{\Box}+2  -m^2) (\bar{\Box}+2  -m_g^2) \bar{h}_{g}^{ij} \nonumber\\
&\simeq&-\frac{1}{4}\sum_{m'\ne m_g} \frac{H^2_{m'}(\rho)}{c_{m'}}\int_Q d^dy \sqrt{|\bar{h}^{(0)}|}  \cosh^{d-2}(\rho)\ \bar{h}_{gij}   (\bar{\Box} +2-m_g^2) \bar{h}_{g}^{ij}.
\end{eqnarray}
Since we sum over all non-ghost modes with $c_m>0$, the above action has wrong sign as expected. Now we have verified that the mode with mass $m_g$ is indeed a ghost. 

We remark that the ghost mass (\ref{sect3.2:gravityghost}) can be derived from the spectrum identity (\ref{gravity: SI HD}). Below we numerically verify that the ghost mass (\ref{sect3.2:gravityghost}) lies in the mass spectrum fixed by NBC (\ref{NBCsgravity2}).  It can be regarded as a numerical test of the spectrum identity (\ref{gravity: SI HD}). For simplicity, we focus on the case $d=4$ with small tension so that we can make analytical discussions. First, we verify that the sum $\sum_m ...$ of (\ref{sect3.2:gravityghost}) is convergent so that the ghost mass is well-defined in the small tension limit.  To study the convergence, we only need to consider the large mass limit $m^2\to \infty$. From (\ref{NBCsgravity2}, \ref{NBCsgravity3},\ref{orthogonal-gravity}), we derive
 \begin{eqnarray} \label{largemasslimit1}
H_m(\rho)\to \frac{\cosh^4(\rho)}{2 \a_2 m^4} H'_m(\rho), \ \ c_m \to \int_{-\infty}^{\rho}\frac{\cosh(r)^{d-2}}{\cosh ^{d-2}(\rho )}H^2_m(r)dr.
\end{eqnarray}
By using $m^2\to 2 q (2 q+3)$, $H_m'(0) \to 2 q P_{2 q+2}^2(0)$ and the normalization formula of $H_m(r)\to \text{sech}^2(r) P_{2 q+1}^2(-\tanh (r))$ in the small tension limit $\rho\to 0$, we obtain
 \begin{eqnarray} \label{largemasslimit2}
H_m(\rho)\to \frac{\sqrt{\pi } (q+1) (q+2) (2 q+1)}{4 \alpha_2 q (2 q+3) \Gamma \left(\frac{1}{2}-q\right) \Gamma (q+3)}, \ \ c_m \to \frac{\Gamma (2 q+4)}{(4 q+3) \Gamma (2 q)}, 
\end{eqnarray}
which yields 
 \begin{eqnarray} \label{largemasslimit3}
&&\lim_{m\to \infty, \rho\to0}\frac{H^2_m(\rho) m^2}{c_m} \to \frac{\pi  (q+1) (2 q+1) (4 q+3)}{32 \a_2^2 q^2 (2 q+3)^2 \Gamma \left(\frac{1}{2}-q\right)^2 \Gamma (q+2)^2}\to \frac{1}{16\pi \a_2^2\  q^4},\\
&&\lim_{m\to \infty,\rho\to0}\frac{H^2_m(\rho)}{c_m} \to\frac{\pi  (q+1) (2 q+1) (4 q+3)}{64 \a_2^2 q^3 (2 q+3)^3 \Gamma \left(\frac{1}{2}-q\right)^2 \Gamma (q+2)^2} \to \frac{1}{64\pi \a_2^2\ q^6},\label{largemasslimit4}
\end{eqnarray}
where $q$ are large integers related to the mass (\ref{masslargesmallrho}). The above limits clearly show that the sums $\sum_{m} \frac{H^2_m(\rho) m^2}{c_m}$ and $\sum_{m} \frac{H^2_m(\rho)}{c_m}$ are convergent. Following the approach of sect.2.1, we perform numerical calculations for the first few non-ghost modes and the good approximates (\ref{largemasslimit3},\ref{largemasslimit4}) for the following large-mass modes to derive the ghost mass. Take $d=4, \rho=\lambda=\a_1=0$ as an example, we have 
\begin{equation}\label{sect.3.2:massghost}
m_g^2 \approx \begin{cases}
-2.270, &\ \text{for $\lambda=\a_1=0, \a_2=0.2$},\\
0.353\mp 1.945 i, &\ \text{for $\lambda=\a_1=0, \a_2=-0.2$}
\end{cases}
\end{equation}
One can check that the above ghost mass belongs to the mass spectrum fixed by BC (\ref{NBCsgravity2}). See Table. \ref{sect3:spectrum for ghost} below. We also numerically check that $m_g^2$ lies in the mass spectrum for finite $\rho$. For finite $\rho$, there is no approximate sum formula for the large-mass modes, and we just sum over the first dozens of modes. It also produces the ghost mass (\ref{sect3.2:gravityghost}) with good numerical accuracy. Finally, we numerically verify the spectrum identities (\ref{gravity: SI HD})
 \begin{eqnarray}\label{sect3.2:ghostrelation}
\sum_{m} \frac{H_{m}^2(\rho)}{c_{m}}\to 0,\ \ \ \ \sum_{m} \frac{H_{m}^2(\rho) m^2}{c_{m}} \to \frac{\cosh^2(\rho)}{2\a_2},
\end{eqnarray}
where the sum is over all modes.  It is a strong support for our analytical proof of spectrum identities in Appendix B.2.

\begin{table}[ht]
\caption{Mass spectrum for HD gravity with $d=4, \rho=\lambda=\a_1=0$}
\begin{center}
    \begin{tabular}{| c | c | c | c |  c | c | c | c| c| c|c| }
    \hline
     &0&$1$ & $2$ & 3  & 4 & 5  \\ \hline
    $m^2$ for $\alpha_2=0.2$  &{\color{blue}-2.270} &1.592& 10.315 & 28.115 & 54.059 & 88.036 \\ \hline
    $m^2$ for $\alpha_2=-0.2$ & $\times$ &${\color{blue}0.353\pm1.945 i}$ & 9.658 & 27.884 & 53.941 & 87.964 \\ \hline
   large mass limit & $\times$  & $\times$ & 10 & 28 & 54 & 88  \\ \hline
  \end{tabular}
\end{center}
\label{sect3:spectrum for ghost}
\end{table}

To summarize, we derive the effective action of gravitons on the brane, which implies a ghost for non-zero HD parameter $\a_2$. We also obtain powerful spectrum identities (\ref{gravity: SI DGP},\ref{gravity: SI HD}) for gravitational KK modes on the brane, which yield the ghost-free condition (\ref{gravity: ghost-free condition}).  Under the ghost-free condition (\ref{gravity: ghost-free condition}), we prove that the mass spectrum is real and positive $m^2>0$. The above conclusions are similar to those of the toy model in sect. 2.

\subsection{Brane bending mode}

This subsection investigates the brane bending modes \cite{Garriga:1999yh, Izumi:2022opi}, which originates in the fluctuations of the brane location
 \begin{eqnarray}
Q: r=\rho+\epsilon\ d_1 \phi(y), \label{bendingmodeQ}
\end{eqnarray}
where $d_1$ is a constant to be determined. We take the following ansatz of bulk metric \cite{Charmousis:1999rg, Kanno:2002ia}
\begin{eqnarray}\label{newbendingmetric}
ds^2=\Big(1+\epsilon H_1(r)\phi(y) \Big) dr^2+ \Big(1+\epsilon H_2(r)\phi(y) \Big) \cosh^2(r) \bar{h}^{(0)}_{ij}(y) dy^i dy^j,
 \end{eqnarray}
 where $\bar{h}^{(0)}_{ij}$ is the AdS metric with unit radius. Solving Einstein equations of order $O(\epsilon)$, we get \cite{Miao:2023unv}
\begin{eqnarray}\label{newbendingH1H2}
H_1(r)=-d_2 (d-2) \text{sech}^{d-2}(r), \  H_2(r)=d_2 \text{sech}^{d-2}(r),
 \end{eqnarray}
 and 
 \begin{eqnarray}\label{newbendingphi}
(\bar{\Box} -d )\phi(y)=0,
 \end{eqnarray}
 where $d_2$ is an integral constant and $\bar{\Box}$ is the d'Alembert operator with respect to $ \bar{h}^{(0)}_{ij}$.  Imposing the NBC (\ref{sect3:NBC}) on the brane, we fix the embedding function (\ref{bendingmodeQ}) to be
   \begin{eqnarray}\label{newbendingphi1phi2}
r=\rho-\epsilon \frac{d_2 (d-2) \hat{\lambda} \ \text{sech}^{d-2}\left(\rho\right)}{1+2 (d-2) \hat{\lambda} \tanh \left(\rho\right)} \phi(y),
 \end{eqnarray}
 where 
 \begin{eqnarray}\label{bending-hatlambda}
\hat{\lambda}=\lambda + 4\a_1(d-3) \text{sech}^2(\rho) - \a_2 (d-2) \text{sech}^2(\rho).
 \end{eqnarray}

 Substituting the metric (\ref{newbendingmetric}) together with the embedding function (\ref{newbendingphi1phi2}) into the action (\ref{gravityaction}) and integrating along $r$, we can obtain the squared effective action of the brane bending mode.  For AdS/BCFT, the integration is divergent near the AdS boundary $r=-\infty$. To get a finite effective action, we perform the standard holographic renormalization \cite{deHaro:2000vlm,Balasubramanian:1999re} on the AdS boundary. Interestingly, we find only the integration of the range $0\le r \le \rho$ contributes to the renormalized action. In this way, we derive the squared action of brane bending mode
 \begin{eqnarray}\label{action of  bending mode}
I_{\phi}= d_2^2 \epsilon^2 \int d^dy\sqrt{|\bar{h}^{(0)}|} \Big[\frac{B_1}{2}\phi (\bar{\Box}-d) \phi+ \frac{B_2}{2}\phi (\bar{\Box}-d)^2 \phi  \Big] ,
 \end{eqnarray}
 with
 \begin{eqnarray}\label{bendmodeB1}
B_1=\frac{(d-1) (d-2) }{2} \int_0^{\rho} dr \ \text{sech}^{d-2}(r)+ \frac{(d-1) (d-2) \text{sech}^{d-2}\left(\rho\right)}{1+2 (d-2) \hat{\lambda} \tanh \left(\rho\right)}\hat{\lambda},
 \end{eqnarray}
and 
 \begin{eqnarray}\label{bendmodeB2}
B_2=\frac{4(d-1)^2 \text{sech}^d(\rho)\ \alpha_3}{\left(1+2(d-2) \hat{\lambda} \tanh(\rho)\right)^2}. 
 \end{eqnarray} 
Recall that we have $-\rho\le r\le \rho$ for wedge holography. As a result, the effective action of brane bending mode for wedge holography is just double the one (\ref{action of  bending mode}) of AdS/BCFT. 
 
 Note that only the HD gravity $\alpha_3 \hat{R}^2$ yields HD interaction in the effective action (\ref{action of  bending mode}) of the brane bending mode.  To remove the potential ghost from this HD term, we can simply set $\alpha_3=0$.  In fact, similar to the case of sect.3.2, the potential ghost is not allowed by bulk EOM (\ref{newbendingphi}) even for $\alpha_3\ne 0$. Let us explain more.  The effective action (\ref{action of  bending mode}) implies two independent modes obeying 
  \begin{eqnarray}\label{bendmode1}
(\bar{\Box}-d) \phi=0,\ \  (\bar{\Box}-d+\frac{B_1}{B_2}) \phi=0. 
 \end{eqnarray} 
However, only the first mode is consistent with the bulk EOM (\ref{newbendingphi}) determined by Einstein equations. Thus the potential ghost satisfying $(\bar{\Box}-d+\frac{B_1}{B_2}) \phi=0$ is ruled out. To have the positive kinetic energy for the brane bending mode obeying $(\bar{\Box}-d) \phi=0$, we require $B_1\ge 0$, which yields a constraint on the parameter
 \begin{eqnarray}\label{sect3.3:condition of kinetic energy}
\hat{\lambda}\ge -\frac{\int_0^{\rho} dr \ \text{sech}^{d-2}(r)}{2 \left(\text{sech}^{d-2}(\rho )+(d-2)  \tanh (\rho ) \int_0^{\rho} dr \ \text{sech}^{d-2}(r)\right)}.
 \end{eqnarray}
We shall discuss this constraint carefully in the following sections.

\section{Constraints from AdS/BCFT}

In this section, we discuss various constraints of the parameters $(\lambda,\a_1, \a_2, \a_3)$ in AdS/BCFT. We first study boundary central charges and entanglement entropy and then summarize all the constraints discussed in this paper. We find the ghost-free condition sets the most substantial parameter constraints for pure DGP gravity with $\a_i=0$ and brane-localized GB gravity with $\lambda=\a_2=\a_3=0$. 

\subsection{Boundary central charge}

There are boundary contributions to the Weyl anomaly when there is a boundary. Take 3d BCFT as an example; the Weyl anomaly takes the form 
 \begin{eqnarray}\label{3dWeylanomaly}
\mathcal{A}=\int_{\partial M} d^2y\sqrt{|\sigma|} \Big( a R_{\partial M}- b \bar{k}_{ab} \bar{k}^{ab} \Big),
 \end{eqnarray}
 where $a,b$ are boundary central charges, $R_{\partial M}$ and $\bar{k}_{ab}$ denote the intrinsic Ricci scalar and traceless parts of extrinsic curvatures on the boundary $\partial M$. $a$ regarding the Euler density is the A-type boundary central charge, which obeys the g-theorem. $b$ is the B-type boundary central charge, which is related to the norm of the displacement operator. Both $a$ and $b$ should be positive, which imposes constraints on the parameters. 
 
Let us first study the A-type boundary central charges in general odd dimensions, i.e., $d=2m+1$.  Consider a BCFT living in $d-$dimensional ball, where the dual bulk metric is given by
\begin{eqnarray}\label{BCFTball}
ds^2=\frac{dz^2+dr^2+r^2 d^2{\Omega_{d-1}}}{z^2},
\end{eqnarray}
and the brane $Q$ locates at
\begin{eqnarray}\label{BCFTballQ}
r(z)^2+\left(z-r_b \sinh(\rho)\right)^2= r_b^2 \cosh^2(\rho),
\end{eqnarray}
where $r_b$ labels the ball radius. Note that (\ref{BCFTballQ}) obeys NBC (\ref{sect3:NBC}) for arbitrary parameters $(\lambda, \alpha_1, \alpha_2, \alpha_3)$.  Recall that the holographic Weyl anomaly can be calculated by the UV logarithmic divergent term of the gravitational action \cite{Henningson:1998gx}. 
Substituting (\ref{BCFTball},\ref{BCFTballQ}) into the gravitational action (\ref{gravityaction}), we derive the Weyl anomaly
\begin{eqnarray}\label{WeylanomalyA}
\mathcal{A}=I_{\log \frac{1}{\epsilon}}=-2 (-2)^{\frac{d-1}{2}} a_{\text{bdy}} \int_{\partial M} d^{d-1}y \sqrt{|\sigma|}  E_{d-1},
\end{eqnarray}
where $\epsilon$ is the UV cutoff, $E_{d-1}$ is the Euler density on the boundary, and $a_{\text{bdy}}$ is the A-type boundary central charge given by
\begin{eqnarray}\label{acbdy}
a_{\text{bdy}}=S(d-2) \Big( \int_0^{\rho } \cosh ^{d-2}(r) \, dr+ 2 \lambda  \cosh ^{d-2}(\rho) \Big),
\end{eqnarray}
where $S(d-2)=2\pi^{\frac{d-2}{2}}/\Gamma(\frac{d-1}{2})$ is the volume of $(d-2)-$dimensional unit sphere. Readers who are interested in the calculation details can refer to the appendix of \cite{Hu:2022ymx} \footnote{\cite{Hu:2022ymx} studied the A-type boundary central charge for Gauss-Bonnet gravity in bulk, which can be straightforwardly generalized to the case of this paper. }. The A-type boundary central charge defined by Weyl anomaly works only for odd $d$. To get the central charge for general $d$, one can study the universal terms of the boundary entropy of half sphere, which yields the same expression as (\ref{acbdy}) \cite{Hu:2022ymx}. Thus, (\ref{acbdy}) applies to general dimensions.  Requiring $a_{\text{bdy}}\ge 0$, we derive a lower bound of DGP parameter
\begin{eqnarray}\label{sect4.1:boundoflambda}
\lambda \ge -\frac{1}{2} \int_0^{\rho } \frac{\cosh ^{d-2}(r)}{\cosh ^{d-2}(\rho)} \, dr
\end{eqnarray}

Let us go on to study the B-type boundary central charge. Near the boundary, the renormalized stress tensor takes the following universal form \cite{Miao:2017aba} 
\begin{eqnarray}\label{Tijuniversal}
T_{ab}=-2 b_{\text{bdy}} \frac{\bar{k}_{ab}}{x^{d-1}}+O(\frac{1}{x^{d-2}}),
\end{eqnarray}
where $ b_{\text{bdy}} $ is the B-type boundary central charge and $x$ is the distance to the boundary.  Note that $ b_{\text{bdy}} $ is related to the norm of displacement operator \cite{Miao:2018dvm,Herzog:2017kkj,Herzog:2017xha}, and thus must be non-negative.

Following \cite{Miao:2017aba}, we take the following ansatz of bulk metric and brane embedding function
\begin{eqnarray}\label{HDmetric}
&&\text{bulk metric}:\ ds^2=\frac{dz^2+dx^2+(\delta_{ab}-2x \epsilon \bar{k}_{ab} f(\frac{z}{x}) )dy^a dy^b}{z^2}+O(\epsilon^2), \\ \label{GBQ} 
&&\text{brane Q}: \ \ \ \ \ x=-\sinh(\rho) z+ O(\epsilon^2),
\end{eqnarray}
where $\e$ denotes the order of perturbations.  Substituting (\ref{HDmetric}) into Einstein equations, we derive one independent equation at order $O(\e)$ \cite{Miao:2017aba}
\begin{eqnarray}\label{HDeq}
s(s^2+1)f''(s)-(d-1) f'(s)=0,
\end{eqnarray}
where $s=z/x$. Solving (\ref{HDeq}), we get
\begin{eqnarray}\label{HDsolution}
f(s)=1+b_{\text{bdy}} \ s^d \, _2F_1\left(\frac{d-1}{2},\frac{d}{2};\frac{d+2}{2};-s^2\right),
\end{eqnarray}
where we have used the DBC $f(0)=1$ on the AdS boundary $s=z/x=0$ \cite{Miao:2017aba}.  
Note that we have assumed $s=z/x\ge 0$ (equivalently $x\ge 0$) in the above derivations. Suitable analytic continuation should be performed in order to get a smooth function $f(z/x)$ for $-\sinh(\rho) z\le x < 0$. Please see  \cite{Miao:2017aba} and sect.2.1 of \cite{Hu:2022ymx} for more discussions. 

From NBC (\ref{sect3:NBC}) on the brane (\ref{GBQ}), we can fix the integral constant 
\begin{eqnarray}\label{HDBcharge}
b_{\text{bdy}}=\frac{d (-\text{csch}(\rho ))^{-d}}{\frac{d \sinh (\rho ) \cosh (\rho ) (-\coth (\rho ))^{3-d}}{2 (d-2) \hat{\lambda}+\coth (\rho )}-\, _2F_1\left(\frac{d-1}{2},\frac{d}{2};\frac{d+2}{2};-\text{csch}^2(\rho )\right)},
\end{eqnarray}
where $\hat{\lambda}$ is defined by (\ref{bending-hatlambda}).  Recall that the solution (\ref{HDsolution}) works for only $z/x>0$ before the analytic continuation. As a result, the integral constant (\ref{HDBcharge}) applies to only $z/x=-1/\sinh(\rho) \ge 0$, or equivalently, $\rho\le 0$.  To get the expression for $\rho>0$, a suitable analytic continuation of the hypergeometric function of (\ref{HDBcharge}) should be taken. One can first simplify $b_{\text{bdy}}$ (\ref{HDBcharge})  under the assumption $\rho<0$, and then analytically extend the result to $\rho>0$.  In this way, we obtain for $d=4$ and $d=5$,
\begin{equation}\label{HDBcharge4d5d}
b_{\text{bdy}}=\begin{cases}
\frac{\cosh (\rho ) (\cosh (\rho )+4 \hat{\lambda} \sinh (\rho ))}{e^{2 \rho } (4 \hat{\lambda}+1)+1},&\ \text{for } d=4,\\
\frac{4 \cosh ^2(\rho )}{6 \sinh (\rho )+3 \cosh ^2(\rho ) \left(4 \tan ^{-1}\left(\tanh \left(\frac{\rho }{2}\right)\right)+\pi \right)+\frac{24 \hat{\lambda}}{\cosh (\rho )+6 \hat{\lambda} \sinh (\rho )}}, &\ \text{for } d=5.
\end{cases}
\end{equation}
Requiring $b_{\text{bdy}}\ge 0$, we derive a lower bound of $\hat{\lambda}$ from (\ref{HDBcharge})
\begin{eqnarray}\label{sect4.1:boundofhatlambda}
\hat{\lambda} \ge \frac{-\coth (\rho )}{2(d-2)}. 
\end{eqnarray}

The renormalized stress tensor can be calculated by the holographic formula
\begin{eqnarray}\label{holoTijHD}
T_{ij}=d h^{(d)}_{ij},
\end{eqnarray}
where $h^{(d)}_{ij}$ is defined in the Fefferman-Graham expansion of the asymptotically AdS metric
\begin{eqnarray}\label{FG}
ds^2=\frac{dz^2+(g^{(0)}_{ij}+z^2g^{(1)}_{ij}+...+z^dh^{(d)}_{ij}+...) dy^i dy^j}{z^2}.
\end{eqnarray}
From (\ref{HDmetric},\ref{HDsolution},\ref{holoTijHD},\ref{FG}), we obtain the holographic stress tensor
\begin{eqnarray}\label{TijGB1}
T_{ab}=-2 \e \ b_{\text{bdy}} \frac{\bar{k}_{ab}}{x^{d-1}}+O(\e^2),
\end{eqnarray}
which takes the expected expression (\ref{Tijuniversal}). 

To summarize, we have obtained the A-type and B-type boundary central charges (\ref{acbdy},\ref{HDBcharge},\ref{HDBcharge4d5d}) in this subsection. The positive nature of these boundary central charges imposes constraints (\ref{sect4.1:boundoflambda},\ref{sect4.1:boundofhatlambda}) on the parameters $(\lambda,\alpha_1,\alpha_2)$.  We discuss this constraint together with others at the end of this section. 

\subsection{Entanglement entropy}

In this subsection, we study the constraints of parameters $(\lambda,\alpha_1, \alpha_2, \alpha_3)$ from entanglement entropy.  From the gravitational action (\ref{gravityaction}), we derive the holographic entanglement entropy (HEE)
\cite{Dong:2013qoa,Camps:2013zua}

\begin{eqnarray}\label{sect4:HEE}
S_{\text{HEE}}&=& 4\pi \int_{\Gamma} d^{d-1}x \sqrt{\gamma}+8\pi  \int_{\partial \Gamma} d^{d-2}x \sqrt{\sigma}\Big( (\lambda+2\a_1(d-2)(d-3)\text{sech}^2(\rho ))\nonumber\\
&&\ \ \ \ \ \ \ \ \ \ + 2\a_1 R_{\partial \Gamma} + \a_2(\hat{R}^{\b}_{\b}-\frac{1}{2} K_{\b} K^{\b}-\frac{d}{2(d-1)}\hat{R} ) +2\a_3 \hat{R}\Big),
\end{eqnarray}
where $\Gamma$ denotes the bulk RT surface, $\partial \Gamma=\Gamma\cap Q$ is the intersection of the RT surface and the brane, $K_{a}$ denote the trace of extrinsic curvatures on $\partial \Gamma$, as viewed from the brane geometry, $R_{\partial \Gamma} $ is the intrinsic Ricci scalar on $\partial \Gamma$, $\hat{R}$ is defined by (\ref{background curvature},\ref{background curvature2},\ref{background curvature3}), and $\b$ are the directions normal to $\partial \Gamma$ on the branes.  

For simplicity, we consider the AdS metric
\begin{eqnarray}\label{sect4:AdS}
ds^2=dr^2+\cosh^2(r)\frac{dz^2-dt^2+\sum_{a=1}^{d-2} dw_a^2}{z^2}, \ \ \ -\infty<r\le \rho,
\end{eqnarray}
where the brane is at $r=\rho$. One can check that the above metric satisfies NBC (\ref{sect3:NBC}) for arbitrary parameters $(\lambda, \a_1, \a_2, \a_3)$ provided the brane tension is given by (\ref{Tension}).  Note that the black string does not obey the NBC (\ref{sect3:NBC}) for non-vanishing GB gravity on the brane, i.e., $ \a_1 \ne 0$.

From (\ref{sect4:HEE}) together with the embedding functions $z=z(r), t=\text{constant}$, we derive the area functional of the RT surface 
\begin{eqnarray}\label{sect4:areaisland}
A=\frac{S_{\text{HEE}}}{4\pi}=\int_{-\infty}^{\rho} dr\frac{\cosh^{d-2}(r)}{z(r)^{d-2}} \sqrt{1+\frac{\cosh^{2}(r) z'(r)^2}{z(r)^{2}}}
+\frac{2\cosh^{d-2}(\rho)}{z^{d-2}_{\rho}} \bar{\lambda},
\end{eqnarray}
where $z_{\rho}=z(\rho)$ denotes the endpoint of RT surface on the brane, and
\begin{eqnarray}\label{sect4:barlambda}
\bar{\lambda}= \lambda+2\a_1(d-2)(d-3)\text{sech}^2(\rho )- \frac{\a_2}{2} (d-2)^2 \text{sech}^2(\rho). 
\end{eqnarray}
Interestingly, $\bar{\lambda}$ is equal to $\hat{\lambda}$ (\ref{bending-hatlambda}) for $d=4$, while different generally. In the above derivations, we have set the tangential volume $V=\int d^{d-2}w=1$ and have used $R_{\partial \Gamma} =\bar{R}^{\b}_{\b}=\bar{R}=0$ and 
\begin{eqnarray}\label{sect4:KK}
 K_{\beta} K^{\beta}=(d-2)^2\text{sech}^2(\rho). 
\end{eqnarray}
Taking the variation of the area functional (\ref{sect4:areaisland}), we derive the NBC on the brane
\begin{eqnarray}\label{sect4:NBCisland}
\frac{z'_{\rho}}{\sqrt{1+\frac{\cosh^{2}(\rho) z_{\rho}'^2}{z_{\rho}^{2}}}}=\frac{2(d-2)z_{\rho}}{\cosh^2(\rho)} \bar{\lambda }.
\end{eqnarray}
On the other hand, we impose the standard DBC $z(-\infty)=z_M$ on the AdS boundary $M$, where $z_M$ is a positive constant.  

\begin{figure}[t]
\centering
\includegraphics[width=10cm]{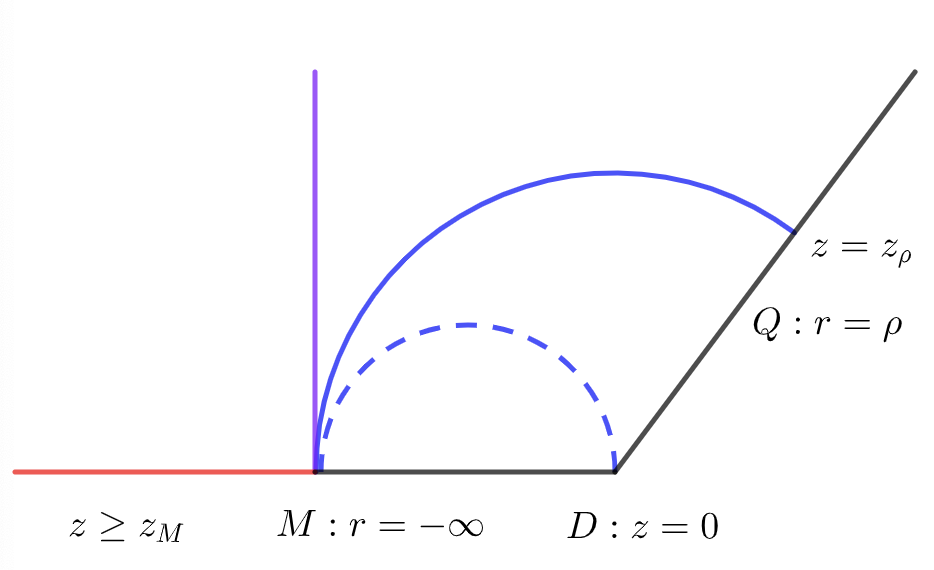}
\caption{RT surfaces of AdS/BCFT. We aim to calculate entanglement entropy of the red region $z\ge z_M$ on the AdS boundary $M$, which is given by the area of RT surfaces. There are three kinds of possible RT surfaces depending on the parameters. The first kind (purple curve) does not intersect with the brane $Q$ and defect $D$, the second kind (blue curve) intersects with the brane but not the defect, and the third kind (blue dotted curve) intersects with the defect. To have a finite renormalized entropy, we rule out the third kind of RT surface, which imposes a lower bound (\ref{sect4:EEbound}) on the parameter $\bar{\lambda}$ (\ref{sect4:barlambda}).}
\label{HEEAdSBCFT}
\end{figure}

Let us calculate the entanglement entropy of a half space on the AdS boundary, i.e., $z\ge z_M$ (red region of Fig. \ref{HEEAdSBCFT}). Depending on the parameters, there are three kinds of RT surfaces. See purple, blue, and blue dotted curves of Fig. \ref{HEEAdSBCFT}. For non-negative $\bar{\lambda}$ and large enough $\rho$, the NBC (\ref{sect4:NBCisland}) cannot be satisfied, and only the first kind of RT surface (purple curve of Fig. \ref{HEEAdSBCFT} is allowed \cite{Chu:2017aab} \footnote{Although \cite{Chu:2017aab} focuses on zero $\bar{\lambda}$, the conclusion can be generalized to positive $\bar{\lambda}$.}. For negative enough $\bar{\lambda}$, one can recover the NBC (\ref{sect4:NBCisland}) on the brane \cite{Miao:2023unv} and thus allows the existence of the second kind of RT surface (blue curve of Fig. \ref{HEEAdSBCFT}). Note that the bulk term of area functional (\ref{sect4:areaisland}) decreases with $z(r)$, while the boundary term of (\ref{sect4:areaisland}) increases with $z_{\rho}$ for negative $\bar{\lambda}$. These two terms compete and minimize the area functional (\ref{sect4:areaisland}) for suitable $z_{\rho}$. For overly negative $\bar{\lambda}$, the boundary area functional $2 \bar{\lambda} \cosh^{d-2}(\rho)/z^{d-2}_{\rho}$ dominates and the minimization approach leads to $z_{\rho}\to 0$. The third kind of RT surface (blue dotted curves of Fig. \ref{HEEAdSBCFT}) appears in this case.   To compare the three cases, we calculate the area difference $\Delta A_i= A_i-A_1$, where $A_i$ denotes the area of i-th kind of RT surface. Note that $4\pi \Delta A_i$ can be regarded as a renormalized entanglement entropy, since it remove the UV divergence at $z=z_M$ on the AdS boundary.  As a result, $\Delta A_1$ and $\Delta A_2$ are finite. However, there is another UV divergence at $z=0$ for the third case, and $\Delta A_3\to -\infty$ for overly negative $\bar{\lambda}$. We require the renormalized area (entanglement entropy) to be finite and thus should rule out the third case. In this way, we get a lower bound of $\bar{\lambda}$. 

Now let us calculate the lower bound of $\bar{\lambda}$.  We recall two facts. First, all the RT surfaces are perpendicular to the AdS boundary. Thus, the included angle between the third kind of RT surface  (blue dotted curve of Fig. \ref{HEEAdSBCFT}) and the AdS boundary is $\frac{\pi}{2}$. Second, the angle between the brane and the AdS boundary is $\frac{\pi}{2}+2 \tan ^{-1}\left(\tanh \left(\frac{\rho }{2}\right)\right)$ \cite{Miao:2017aba}. As a result, the angle between the brane $Q$ and the blue dotted curve $\Gamma_3$ is 
\begin{eqnarray}\label{sect4:angle1}
 \theta=2 \tan ^{-1}\left(\tanh \left(\frac{\rho }{2}\right)\right).
\end{eqnarray}
The angle can be obtained in another way.  We consider the critical parameter $\bar{\lambda}$ that the blue curve approaches the blue dotted curve $\Gamma_2 \to \Gamma_3$ ($z_{\rho}\to 0$). The angle between the blue curve $\Gamma_2$ and the brane $Q$ obeys
\begin{eqnarray}\label{sect4:angle2}
\cos( \theta) =\lim_{z_{\rho}\to 0}n_{\Gamma_2}^{\mu} n_{Q\ \mu}=\lim_{z_{\rho}\to 0}\frac{-z'_{\rho}}{\sqrt{z'^2_{\rho}+\frac{z_{\rho}^2}{\cosh^2(\rho)}}},
\end{eqnarray}
where $n_{\Gamma_2\ \mu}=(-z'_{\rho},1,0,...,0)/\sqrt{z'^2_{\rho}+\frac{z_{\rho}^2}{\cosh^2(\rho)}}$ and $n_{Q\ \mu}=(1,0,0,...,0)$ are normal vectors for the second kind of RT surface and the brane $Q$, respectively.  We take the limit $z_{\rho}\to 0$ at the end of calculations so that $\Gamma_2 \to \Gamma_3$.  From the NBC (\ref{sect4:NBCisland}) and (\ref{sect4:angle1},\ref{sect4:angle2}), we derive the critical parameter
\begin{eqnarray}\label{sect4:angle3}
\bar{\lambda}_{\text{cri}}=-\frac{\cosh(\rho)\cos(\theta)}{2(d-2)}=-\frac{1}{2(d-2)},
\end{eqnarray}
which yields a lower bound of the parameter
\begin{eqnarray}\label{sect4:EEbound}
\bar{\lambda}\ge \bar{\lambda}_{\text{cri}}=-\frac{1}{2(d-2)}.
\end{eqnarray}
Interestingly, the lower bound is independent of the position of brane $r=\rho$. It should be mentioned that \cite{Chen:2020uac} derives the same bound (\ref{sect4:EEbound}) for DGP gravity in the large tension limit $\rho\to \infty$. Here we find the bound (\ref{sect4:EEbound}) works for general $\rho$ and generalize it to brane-localized HD gravity. Finally, we remark that the renormalized area $\Delta A_3=A_3-A_1$ becomes negative infinity if the bound (\ref{sect4:EEbound}) is violated. Please see the appendix B for more discussions.

\subsection{Summary of constraints}

\begin{table}[ht]
\caption{Constraints on parameters of AdS/BCFT}
\begin{center}
\begin{tabular}{| c | c |  }
\hline
 ghost-free condition & $\lambda+ 4(d-3)\a_1\ \text{sech}^2(\rho )\ge0$,\  $\a_2=0$ \\ \hline
 tachyon-free condition & $m^2>0$, \\ \hline
    bending mode  & $ \hat{\lambda}\ge -\frac{\int_0^{\rho} dr \ \text{sech}^{d-2}(r)}{2 \left(\text{sech}^{d-2}(\rho )+(d-2)  \tanh (\rho ) \int_0^{\rho} dr \ \text{sech}^{d-2}(r)\right)}$,  \\ \hline
   A-type charge  & $\lambda \ge -\int_0^{\rho } \frac{\cosh ^{d-2}(r)}{2\cosh ^{d-2}(\rho)} \, dr$,  \\ \hline
   B-type charge  & $ \hat{\lambda}\ge  -\frac{\coth(\rho)}{2(d-2)}$,  \\ \hline
HEE & $\bar{\lambda}\ge -\frac{1}{2(d-2)}$, \\ \hline
notations& $\hat{\lambda}=\lambda + 4\a_1(d-3) \text{sech}^2(\rho) - \a_2 (d-2) \text{sech}^2(\rho)$, \\
& $\bar{\lambda}=\lambda+2\a_1(d-2)(d-3)\text{sech}^2(\rho )- \frac{\a_2}{2} (d-2)^2 \text{sech}^2(\rho)$. \\ \hline
    \end{tabular}
\end{center}
\label{sect4:Constraint in AdSBCFT}
\end{table}

Let us summarize various constraints of parameters in Table.\ref{sect4:Constraint in AdSBCFT}. Recall some notations. The DGP and HD gravity parameters are labeled by $\lambda$ and $\alpha_i$, respectively. The ghost-free condition (\ref{gravity: ghost-free condition}) is derived from the spectrum identities (\ref{gravity: SI DGP},\ref{gravity: SI HD}) for gravitational KK modes. It rules out one type of brane-localized HD gravity, i.e., $\alpha_2=0$.  As we have proved in sect. 3.1, the ghost-free condition automatically results in the tachyon-free condition $m^2> 0$. In AdS, the theory is tachyon-free if it obeys the BF bound $m^2\ge -(\frac{d-1}{2})^2$. Thus, $m^2$ can be negative generally. Here, we get a stronger version of the tachyon-free condition $m^2> 0$. In the bending-mode constraint and B-type-charge constraint, we have $ 
\hat{\lambda}=\lambda + 4\a_1(d-3) \text{sech}^2(\rho) + \a_2 (2-d) \text{sech}^2(\rho)$. In the HEE constraint, we have $\bar{\lambda}=\lambda+2\a_1(d-2)(d-3)\text{sech}^2(\rho )- \frac{\a_2}{2} (d-2)^2 \text{sech}^2(\rho)$. Note that $\bar{\lambda}$ (\ref{sect4:barlambda}) equals to $\hat{\lambda}$ (\ref{bending-hatlambda}) for $d=4$, while different generally.  Note also that the HD term $\alpha_3 \hat{R}^2$ is not restricted by Table.\ref{sect4:Constraint in AdSBCFT}. We leave the study of the constraint of $\alpha_3$ to future work. 

Let us discuss some general characteristics. 

\begin{itemize}
  \item The DGP coupling $\lambda$ dominates in various constraints in the large $\rho$ limit. That is because the HD couplings $\a_1, \a_2$ are always associated with $\text{sech}^2(\rho)$; thus, they are suppressed in the large $\rho$ limit for fixed $\a_1, \a_2$. Remarkably, in such limit, except the ghost-free and tachyon-free conditions, all the other constraints of Table.\ref{sect4:Constraint in AdSBCFT} yield the same lower bound 
\begin{eqnarray}\label{sect4:largerhoconstrain}
\lambda \ge -\frac{1}{2(d-2)}, \ \ \text{for } \rho \to \infty.
\end{eqnarray}
Similar characters are also observed in \cite{Miao:2023unv, Li:2023fly} for DGP gravity. We remark that the above universal lower bound coincides with the counter term $R$ coefficient in the holographic renormalization \cite{deHaro:2000vlm, Balasubramanian:1999re}.

  \item  The ghost-free condition imposes the most robust constraint  $\lambda\ge 0$ for DGP gravity with $\a_i=0$. The other constraints of Table. \ref{sect4:Constraint in AdSBCFT} are automatically obeyed under the ghost-free condition.
  
  \item  The most substantial constraint is $\a_1\ge0$ for brane-localized GB gravity with $\lambda=\a_2=\a_3=0$. 
Interestingly, string theory also predicts a positive GB gravity \cite{Boulware:1985wk}, consistent with our results. According to \cite{Buchel:2009sk, Hofman:2008ar}, there is an upper bound of the GB parameter to avoid the negative energy flux of CFTs. It is interesting to generalize the discussions to our case on branes. We leave it to future works. 

\item In general, the strongest constraints set by Table.\ref{sect4:Constraint in AdSBCFT}  are as follows
\begin{eqnarray}\label{sect4: constraint HD0}
&& \a_2=0,\\ \label{sect4: constraint HD1}
&& \lambda+ 4(d-3)\a_1\ \text{sech}^2(\rho )\ge0,\\ \label{sect4: constraint HD2}
&& \lambda \ge -\int_0^{\rho } \frac{\cosh ^{d-2}(r)}{2\cosh ^{d-2}(\rho)} \, dr,\\ \label{sect4: constraint HD3}
&&\lambda+2\a_1(d-2)(d-3)\text{sech}^2(\rho )\ge -\frac{1}{2(d-2)}.
\end{eqnarray} 
See Fig. \ref{boundAdSBCFT} for the parameter space fixed by the above inequalities. 

\begin{figure}[t]
\centering
\includegraphics[width=10cm]{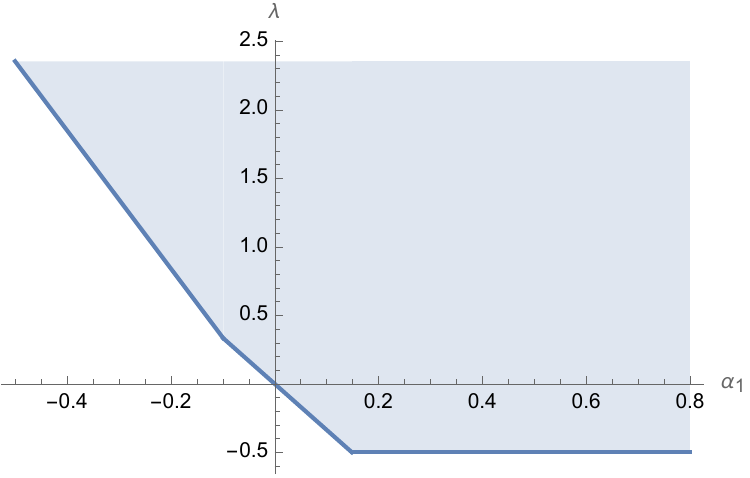}
\caption{ The gray region denotes the parameter space of AdS/BCFT fixed by eqs.(\ref{sect4: constraint HD1},\ref{sect4: constraint HD2},\ref{sect4: constraint HD3}) with $d=5$ and $\rho=1$. }
\label{boundAdSBCFT}
\end{figure}

\end{itemize}

In summary, we have derived the boundary central charges and holographic entanglement entropy in this section. Besides, we discuss various constraints on the parameters of AdS/BCFT, which are summarized in Table.\ref{sect4:Constraint in AdSBCFT} and eqs. (\ref{sect4: constraint HD0},\ref{sect4: constraint HD1},\ref{sect4: constraint HD2},\ref{sect4: constraint HD3}). The ghost-free condition imposes the most substantial constraint. It requires a non-negative DGP gravity ($\a_i=0$) $\lambda\ge0$, non-negative GB gravity ($\lambda=\a_2=\a_3=0$) $\alpha_1\ge 0$ and rules out the HD gravity $\alpha_2=0$.

\section{Constraints from wedge holography}

In the above sections, we mainly focus on AdS/BCFT. This section generalizes the discussions to wedge holography. Recall the geometry of wedge holography in Fig. \ref{AdSBCFT}, where two branes are located at $r=\rho$ and $r=-\rho$, respectively. Wedge holography proposes the classical gravity in bulk $N$ is dual to ``quantum gravity" on the branes $Q=Q_1\cup Q_2$ and is dual to CFTs on the defect $D$. The bulk action is still given by (\ref{gravityaction}). The only difference from AdS/BCFT is that $Q=Q_1\cup Q_2$ denotes two branes, and we impose NBC (\ref{sect3:NBC}) on both branes. Unlike AdS/BCFT, there is a massless mode on the branes in wedge holography \cite{Hu:2022lxl}. For simplicity, we focus on the symmetric case with the same parameters $(T, \lambda, \a_i)$ on the two branes. Since the calculations are similar to those of AdS/BCFT, we do not repeat the derivations; we list the main results below.

The NBCs on the branes read
 \begin{eqnarray}\label{wedge: NBC1}
&&H'(\rho) =2H(\rho) m^2\Big(\lambda\ \text{sech}^2(\rho )+ 4(d-3)\a_1\ \text{sech}^4(\rho ) + \a_2\ \text{sech}^4(\rho)m^2 \Big),\\ \label{wedge: NBC2}
&&H'(-\rho) =-2H(-\rho) m^2\Big(\lambda\ \text{sech}^2(\rho )+ 4(d-3)\a_1\ \text{sech}^4(\rho ) + \a_2\ \text{sech}^4(\rho)m^2 \Big). 
\end{eqnarray}
Following the approach of sect. 3.2 and \cite{Miao:2023unv,Hu:2022lxl}, we can derive graviton mass spectrum on the branes
\begin{eqnarray}\label{sect5:spectrum}
m^2\big( M_{00} + 2M_{10} \tilde{\lambda}+ M_{11} \tilde{\lambda}^2 \big)=0,
\end{eqnarray}
with $\tilde{\lambda}=\lambda+ 4(d-3)\a_1\ \text{sech}^2(\rho ) + \a_2\ \text{sech}^2(\rho)m^2$ and
 \begin{eqnarray}\label{M00}
&&M_{00}=\text{sech}^2(\rho) \left(P_{l _m}^{\frac{d}{2}-1}\left(x\right) Q_{l _m}^{\frac{d}{2}-1}\left(-x\right)-P_{l _m}^{\frac{d}{2}-1}\left(-x\right) Q_{l_m}^{\frac{d}{2}-1}\left(x\right)\right),\\ \label{M10}
&&M_{10}=-2 \text{sech}^3(\rho) \left(P_{l_m}^{\frac{d}{2}}\left(-x\right) Q_{l_m}^{\frac{d}{2}-1}\left(x\right)-P_{l_m}^{\frac{d}{2}-1}\left(x\right) Q_{l_m}^{\frac{d}{2}}\left(-x\right)\right),\\ \label{M01}
&&M_{11}=-4 \text{sech}^4(\rho) \left(P_{l_m}^{\frac{d}{2}}\left(x\right) Q_{l_m}^{\frac{d}{2}}\left(-x\right)-P_{l _m}^{\frac{d}{2}}\left(-x\right) Q_{l_m}^{\frac{d}{2}}\left(x\right)\right),\label{M11}
\end{eqnarray}
where $x=\tanh\rho$ and $l_m$ is given by (\ref{aibia1}). In fact, we only need to replace $(\lambda_1, \lambda_2)$ with $\tilde{\lambda}$ and $(x_1, x_2)$ with $x$ in eq.(2.24) of \cite{Miao:2023unv}.

The orthogonal relations of gravitational KK modes become
 \begin{eqnarray}\label{sect5:orthogonal-gravity}
 \langle H_m, H_{m'}\rangle&=&c_m\delta_{m, m'}=\int_{-\rho}^{\rho}\frac{\cosh(r)^{d-2}}{\cosh ^{d-2}(\rho )}H_m(r) H_{m'}(r) dr \\
&&+2\Big(\lambda +4(d-3) \a_1 \text{sech}^2(\rho )+\a_2 \text{sech}^2(\rho ) (m^2+m'^2) \Big) H_m(\rho) H_{m'}(\rho)\nonumber\\
&&+2\Big(\lambda +4(d-3) \a_1 \text{sech}^2(\rho )+\a_2 \text{sech}^2(\rho ) (m^2+m'^2) \Big) H_m(-\rho) H_{m'}(-\rho)\nonumber
\end{eqnarray}
The gravitational effective action on the branes takes the same form as (\ref{gravityItotal3}), where the mass obeys (\ref{sect5:spectrum}), and the inner product $c_m=\langle H_m, H_m\rangle$ is given by
 \begin{eqnarray}\label{sect5:cm}
c_m=2\int_{0}^{\rho}\frac{\cosh(r)^{d-2}}{\cosh ^{d-2}(\rho )}H^2_m(r)dr+4\Big(\lambda +4(d-3) \a_1 \text{sech}^2(\rho )+2\a_2 \text{sech}^2(\rho ) m^2 \Big) H^2_m(\rho),
\end{eqnarray}
where we have used $H_m^2(r)=H_m^2(-r)$ for the symmetric wedge holography. 

The gravitational spectrum identities take the same form as those of AdS/BCFT on each brane.  Since we focus on the symmetric case of wedge holography, we have
\begin{eqnarray}\label{wedge: SI DGP}
&&\sum_{m} \frac{H_m^2(\rho)}{\langle H_m, H_m \rangle}=\sum_{m} \frac{H_m^2(-\rho)}{\langle H_m, H_m \rangle}  = \frac{1}{2\Big(\lambda +4(d-3) \a_1 \text{sech}^2(\rho )\Big)}, \ \ \ \text{for }\alpha_2=0, \\ \label{wedge: SI HD1}
&&\sum_{m} \frac{H_m^2(\rho)}{\langle H_m, H_m \rangle} =\sum_{m} \frac{H_m^2(-\rho)}{\langle H_m, H_m \rangle}= 0, \ \ \ \ \ \ \ \ \ \ \ \ \ \ \ \ \ \ \  \ \ \ \ \ \ \ \ \ \ \ \  \ \ \ \ \ \ \text{for } \alpha_2\ne0,\\
&&\sum_{m} \frac{H_m^2(\rho) m^2}{\langle H_m, H_m \rangle}=\sum_{m} \frac{H_m^2(-\rho) m^2}{\langle H_m, H_m \rangle}=\frac{\cosh^2(\rho)}{2\alpha_2}, \ \ \ \ \ \ \  \ \ \ \ \ \ \ \ \ \ \ \  \ \ \ \ \ \ \text{for } \alpha_2\ne0.  \label{wedge: SI HD2}
\end{eqnarray} 
For the asymmetric case and the corresponding proofs of spectrum identities, please see the Appendix B.3.  Following the approach of AdS/BCFT, we can derive the ghost-free condition (\ref{gravity: ghost-free condition}) and prove that the mass spectrum is non-negative $m^2\ge 0$. Note that wedge holography includes a massless mode. The massless mode with $m^2=0$ and $H(r)=1$ obeys the consistent equation (\ref{gravity: positive mass 1}) trivially.

The discussions of brane bending mode are similar to AdS/BCFT, and $B_1$ is still given by (\ref{bendmodeB1}) for half wedge space. 
 The A-type boundary central charges (\ref{acbdy}) of BCFT$_d$ become the A-type central charges of CFT$_{d-1}$ in wedge holography. The B-type boundary central charges (\ref{HDBcharge},\ref{HDBcharge4d5d}) are more subtle. The solution discussed in sect.4.1 does not satisfy NBC on both branes. So far, it is unclear if other ansatz of bulk metric can solve this problem. Thus we do not discuss the B-type boundary central charge in wedge holography. Please do not confuse B-type boundary central charge of BCFT$_d$ and B-type bulk central charge of CFT$_{d-1}$, they are different.

\begin{figure}[t]
\centering
\includegraphics[width=10cm]{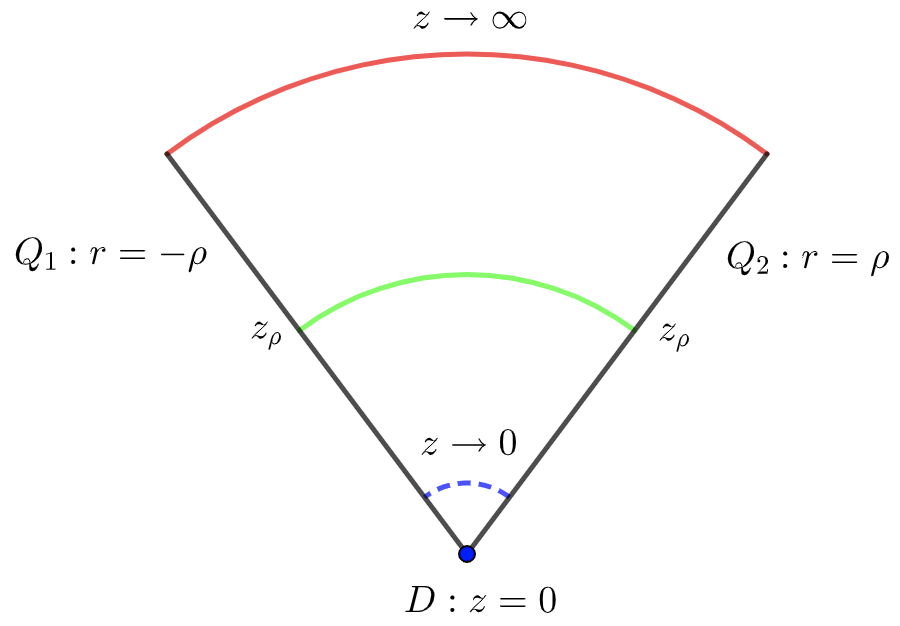}
\caption{HEE in wedge holography.  HEE of the whole defect $D$ is given by the RT surface ending on the branes. Depending on the parameters, there are three kinds of RT surfaces labelled by red, green and blue dotted curves. The blue dotted curve yields negative infinity HEE and should be ruled out. This imposes a lower bound on the parameter. }
\label{HEEwedge}
\end{figure}
 
Now let us discuss the HEE in wedge holography. Note that the black string is not a solution if there exists brane-localized GB gravity \footnote{We focus on the most straightforward case in which the brane is located at $r=\pm \rho$. In this case, the black string does not obey NBC if brane-localized GB gravity exists. Whether the black string could satisfy NBC for the more general embedding function is an open question, and we do not discuss it in this paper.}. To discuss the most general HD gravity on branes, we focus on AdS space (\ref{sect4:AdS}) with $-\rho \le r\le \rho$ in bulk. We are interested in the HEE of the whole defect $D$, which can be calculated by the RT surface ending on the branes. 
There are three kinds of RT surfaces depending on the parameters. See Fig. \ref{HEEwedge}.  Now we follow the approach of  \cite{Miao:2023unv,Li:2023fly}. The area of RT surface is given by 
\begin{eqnarray}\label{sect5:areawedge}
A=\int_{0}^{\rho} dr\frac{\cosh^{d-2}(r)}{z(r)^{d-2}} \sqrt{1+\frac{\cosh^{2}(r) z'(r)^2}{z(r)^{2}}}
+\frac{2\cosh^{d-2}(\rho)}{z^{d-2}_{\rho}} \bar{\lambda},
\end{eqnarray}
where $z_{\rho}=z(\rho)$ denotes the endpoint of RT surface and $\bar{\lambda}$ is defined by (\ref{sect4:barlambda}). Again we focus on the half wedge space.  Since the AdS space (\ref{sect4:AdS}) is invariant under a constant rescaling $z\to c z$, if $z=z_0(r)$ is an extremal surface so does $z= c z_0(r)$. And the area (\ref{sect5:areawedge}) transforms as $A\to A/c^{d-2}$. 
Consider an extremal surface $z=z_0(r)$ with an arbitrary endpoint $z_{\rho}$.  Note that the extremal surface needs not to satisfy the NBC (\ref{sect4:NBCisland}). If the area $A_0$ is positive, we can choose $c\to \infty$ and thus $z\to c z\to \infty$ to get the RT surface with minimal area, i.e., $A=A_0/c^{d-2}\to 0$. On the other hand, if $A_0$ is negative, we choose $c\to 0$ and thus $z\to c z\to 0$ to minimize the RT surface, i.e., $A=A_0/c^{d-2}\to -\infty$.  The later case yields negative infinity area and should be ruled out. The critical case has zero area $A_0=0$ and imposes a lower bound of $\bar{\lambda}$
\begin{eqnarray}\label{sect5:lambdacri}
\bar{\lambda}\ge \bar{\lambda}_{\text{cri}}.
\end{eqnarray}
 For the critical case, we have $A\to A_0/c^{d-2}=0$ for arbitrary $c$, which means the extremal surface minimizes at arbitrary endpoint $z_{\rho}$ and is actually the RT surface.  As a RT surface, it should obey the NBC (\ref{sect4:NBCisland}), which yields
\begin{eqnarray}\label{sect5:lambdacri}
\bar{\lambda}_{\text{cri}}=\frac{\cosh^2(\rho)z'_{\rho} }{2(d-2)\sqrt{z_{\rho}^{2}+\cosh^{2}(\rho) z_{\rho}'^2}}.
\end{eqnarray}
The above equation can be calculated numerically by any symmetrical extremal surface with the same endpoints $z_{\rho}$ on the two branes. The choice of $z_{\rho}$ does not affect the result. 

\begin{table}[ht]
\caption{Constraints on parameters of wedge holography}
\begin{center}
\begin{tabular}{| c | c |  }
\hline
 ghost-free condition & $\lambda+ 4(d-3)\a_1\ \text{sech}^2(\rho )\ge0$,\   $\a_2=0$, \\ \hline
  tachyon-free condition   &  $m^2\ge0 $, \\ \hline
    bending mode  & $ \hat{\lambda}\ge -\frac{\int_0^{\rho} dr \ \text{sech}^{d-2}(r)}{2 \left(\text{sech}^{d-2}(\rho )+(d-2)  \tanh (\rho ) \int_0^{\rho} dr \ \text{sech}^{d-2}(r)\right)}$,  \\ \hline
   A-type charge  & $\lambda \ge -\int_0^{\rho } \frac{\cosh ^{d-2}(r)}{2\cosh ^{d-2}(\rho)} \, dr$,  \\ \hline
HEE & $\bar{\lambda}\ge \bar{\lambda}_{\text{cri}}=\frac{\cosh^2(\rho)z'_{\rho} }{2(d-2)\sqrt{z_{\rho}^{2}+\cosh^{2}(\rho) z_{\rho}'^2}}$ , \\ \hline
notations& $\hat{\lambda}=\lambda + 4\a_1(d-3) \text{sech}^2(\rho) - \a_2 (d-2) \text{sech}^2(\rho)$, \\
& $\bar{\lambda}=\lambda+2\a_1(d-2)(d-3)\text{sech}^2(\rho )- \frac{\a_2}{2} (d-2)^2 \text{sech}^2(\rho)$. \\ \hline
    \end{tabular}
\end{center}
\label{table:Constraint in wedge holography}
\end{table}

\begin{figure}[t]
\centering
\includegraphics[width=10cm]{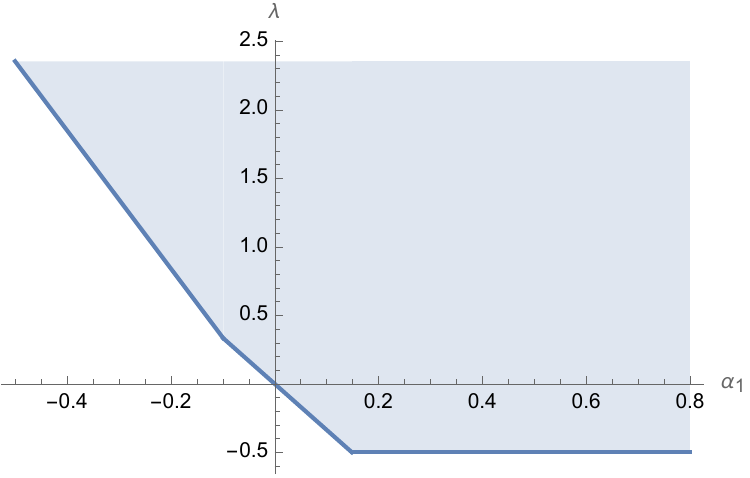}
\caption{ The gray region denotes the parameter space of wedge holography fixed by eqs.(\ref{sect5: constraint HD1},\ref{sect5: constraint HD2},\ref{sect5: constraint HD3}) with $d=5$ and $\rho=1$. It is smaller than that of AdS/BCFT.}
\label{boundwedge}
\end{figure}

Let us summarize all the above constraints in Table. \ref{table:Constraint in wedge holography}. Like AdS/BCFT, the ghost-free condition rules out brane-localized HD gravity related to $\a_2$. Besides, it automatically yields the `strong' tachyon-free condition $m^2\ge 0$. In total, Table. \ref{table:Constraint in wedge holography} set the following independent constraints 

\begin{eqnarray}\label{sect5: constraint HD0}
&& \a_2=0,\\ \label{sect5: constraint HD1}
&& \lambda+ 4(d-3)\a_1\ \text{sech}^2(\rho )\ge0,\\ \label{sect5: constraint HD2}
&& \lambda \ge -\int_0^{\rho } \frac{\cosh ^{d-2}(r)}{2\cosh ^{d-2}(\rho)} \, dr,\\ \label{sect5: constraint HD3}
&&\lambda+2\a_1(d-2)(d-3)\text{sech}^2(\rho )\ge  \bar{\lambda}_{\text{cri}}=\frac{\cosh^2(\rho)z'_{\rho} }{2(d-2)\sqrt{z_{\rho}^{2}+\cosh^{2}(\rho) z_{\rho}'^2}}.
\end{eqnarray} 
For pure DGP gravity with $\a_i=0$, we have $\lambda\ge 0$. For pure brane-localized GB gravity with $\lambda=\a_2=\a_3=0$, we have $\a_1\ge 0$. In general, see Fig. \ref{boundwedge} for an example. We numerically observe that $\bar{\lambda}_{\text{cri}}\ge -1/2(d-2)$ and $\lim_{\rho\to \infty} \bar{\lambda}_{\text{cri}}\to -1/2(d-2)$.  Comparing (\ref{sect5: constraint HD3}) of wedge holography and  (\ref{sect4: constraint HD3}) of AdS/BCFT, we see that the parameter space becomes smaller in wedge holography due to $\bar{\lambda}_{\text{cri}}\ge-1/2(d-2)$.

\section{Conclusions and Discussions}

This paper studies the brane-localized interactions, including DGP and curvature-square gravity on the brane. To warm up, we first study a toy model of a scalar in flat space and AdS space, respectively, and then generalize the results to gravity. We work out the effective action on the brane, implying the brane-localized HD gravity suffers the ghost problem generally. Next, we prove novel algebraic identities of the mass spectrum on the brane. These spectrum identities can describe the phase transformations of the mass spectrum and yield strict ghost-free conditions. Under the ghost-free conditions, we prove the mass spectrum is real and non-negative $m^2\ge 0$.  

This paper also discusses various constraints on DGP gravity and brane-localized HD gravity in AdS/BCFT and wedge holography. See Table. \ref{sect4:Constraint in AdSBCFT} and Table. \ref{table:Constraint in wedge holography}, which include the ghost-free conditions of KK and brane-bending modes, the positive definiteness of boundary central charges, and the finiteness of renormalized entanglement entropy. The ghost-free condition imposes the strongest constraints for pure DGP gravity ($\a_i=0$) and pure brane-localized GB gravity ($\lambda=\a_2=\a_3=0$). It requires non-negative DGP gravity $\lambda\ge 0$, non-negative GB gravity $\alpha_1\ge 0$, and rules out one kind of brane-localized HD gravity, i.e., $\alpha_2=0$. Inspired by the case in bulk \cite{Buchel:2009sk, Hofman:2008ar}, it is expected that there is an upper bound of brane GB parameter $\a_1$ to avoid the negative energy flux of CFTs. We leave this interesting problem to future work. 

Let us discuss possible applications and generalizations of our results below. 

First, there are several possible ways to remove the ghost of HD gravity, which includes suitable boundary conditions \cite{Maldacena:2011mk,Lu:2011ks,Hell:2023rbf}, fine-turning the couplings of infinite towers of HD terms \cite{Biswas:2011ar,Modesto:2017sdr}, non-Hermitian \cite{Mannheim:2021oat}, and so on. Whether these mechanisms can remove the ghost for brane-localized HD gravity is an interesting problem, and we leave it to future works. 

Second, our result suggests that the DGP parameter $\lambda$ should be non-negative to be ghost-free. Recent works \cite{Miao:2023unv,Li:2023fly} show that negative DGP gravity can recover entanglement islands in wedge/cone holography with massless gravity. Unfortunately, according to the above discussions, this model includes a ghost and thus is not well-defined. It should be mentioned that, recently, \cite{Geng:2023qwm,Geng:2023iqd} imposed a no-island condition that the island surface should be within the horizon in bulk. This condition is related to the island constraint of \cite{Li:2023fly}, which sets a negative lower bound of the DGP parameter and is weaker than the constraint in this paper. Whether entanglement islands exist in massless gravity is an interesting and important problem. According to \cite{Geng:2023qwm}, if one divides the black hole and radiation regions by minimizing the entanglement entropy, one always gets zero radiation region. There is no information loss for a zero radiation region and, thus, no need for islands. However, we are not interested in a zero radiation region. It raises the question of whether there are other physical partitions of the black holes and radiation regions. The resolution may lie in the observer \cite{Witten:2023qsv}, who will see a non-zero radiation region in his lifetime. We leave this interesting problem to future work. 

Third, the ghost problem does not matter if one takes HD gravity as an effective instead of a fundamental theory. Take wedge holography as an example; there is always a massless mode with $m^2=0$. For suitable HD parameter $\alpha_2$, all the other modes (including the ghost mode) have $m^2>0$ and are frozen at low energy. Then, the ghost mode can be ignored since it is not excited at low energy. It is interesting to take our HD model of AdS/BCFT as an effective theory to study the boundary quantum effects such as the Casimir effect, anomalous transports, etc. It is also interesting to generalize the discussions of this paper to cone holography \cite{Miao:2021ual}, which is the holographic dual of edge modes on codim-n defects.

\section*{Acknowledgements}

We thank Yi Pang, Jie Ren, Run-Qiu Yang, Yu-Xiao Liu, Lixin Xu, Zheng-quan Cui and Yu Guo for valuable comments and discussions. This work is supported by the National Natural Science Foundation of China (No.12275366 and No.11905297).

\appendix

\section{Ghost problem of bulk HD gravity}
This appendix shows that bulk HD gravity suffers the ghost problem generally. For simplicity, we take curvature-squared gravity as an example. The squared action expanded around a flat background is given by
 \begin{eqnarray}\label{Ibulk}
I\sim \int_N d^{d+1}x \sqrt{|g|} \bar{g}^{\mu\nu} \Box_N(\Box_N- m^2) \bar{g}_{\mu\nu},
\end{eqnarray}
where $ \bar{g}_{\mu\nu} $ denotes the metric perturbation obeying transverse traceless gauges $\nabla^{\mu} \bar{g}_{\mu\nu}= \bar{g}^{\mu}_{\ \mu}=0$, $\Box_N=\nabla_{\mu}\nabla^{\mu}$ is the d'Alembert operator in bulk $N$, $m$ is the graviton mass depending on the parameters of the theory. In general, curvature-squared gravity includes a massless graviton, a massive graviton, and a scalar mode \cite{Stelle:1977ry}. We are interested in the gravitons, so we 
focus on transverse traceless gauges, and the scalar mode does not appear in (\ref{Ibulk}). From the action (\ref{Ibulk}), we read off the graviton propagator
 \begin{eqnarray}\label{bulk-propagator}
G\sim \frac{1}{p^2(p^2+m^2)} \sim \frac{1}{m^2}\Big(\frac{1}{p^2} -\frac{1}{p^2+m^2}\Big),
\end{eqnarray}
where we have ignored the tensor structures for simplicity. Since $G\sim 1/p^4$, the curvature-squared gravity has a better momentum integral convergence than Einstein gravity, which results in a renormalizable theory \cite{Stelle:1976gc}. However, the propagator (\ref{bulk-propagator}) does not have the correct signs for both the massless and massive modes, which leads to a ghost. There is an equivalent way to see it through the action. For the massless mode $\Box_N \bar{g}_{\mu\nu}=0$, we can replace $(\Box_N- m^2)$ with $(0- m^2)$ and rewrite the fourth-order action (\ref{Ibulk}) into an effective second-order action
 \begin{eqnarray}\label{IbulkM1}
I_{\text{massless}}\sim - m^2\int_N d^{d+1}x \sqrt{|g|} \bar{g}^{\mu\nu} \Box_N\bar{g}_{\mu\nu},
\end{eqnarray}
and similar for the massive mode $(\Box_N- m^2)\bar{g}_{\mu\nu}=0$, 
 \begin{eqnarray}\label{IbulkM2}
I_{\text{massive}}\sim m^2\int_N d^{d+1}x \sqrt{|g|} \bar{g}^{\mu\nu} (\Box_N- m^2) \bar{g}_{\mu\nu}.
\end{eqnarray}
We can re-derive the propagator (\ref{bulk-propagator}) from the above two actions for each mode. Since actions (\ref{IbulkM1},\ref{IbulkM2}) have different signs, one of them must be a ghost. 

\section{Proofs of spectrum identities}

\subsection{Brane-localized scalar in AdS/BCFT}

This subsection proves the spectrum identities (\ref{toy model AdS: SI DGP},\ref{toy model AdS: SI HD}) of KK modes for brane-localized scalar in AdS/BCFT. 

To start, we use EOM (\ref{EOMscalarAdSX}) to rewrite the orthogonal relationship (\ref{orthogonal-scalar-AdS}) into a mass independent form
 \begin{eqnarray}\label{appA1: orthogonal}
 \langle X_m, X_{n}\rangle &=&\int_{-\infty}^{\rho}\frac{\cosh(r)^{d-2}}{\cosh(\rho)^{d-2}}X_m(r) X_{n}(r) dr+\Big(\lambda +\a (1-d)\text{sech}^2(\rho )  \Big) X_m(\rho) X_{n}(\rho)\nonumber\\
&-&\alpha\ \text{sech}^2(\rho )\Big(  \cosh^2(\rho) X_m''(\rho)+d \sinh (\rho) \cosh (\rho) X_m'(\rho) \Big) X_{n}(\rho) \nonumber\\
&-&\alpha\ \text{sech}^2(\rho ) X_{m}(\rho)\Big(  \cosh^2(\rho) X_n''(\rho)+d \sinh (\rho) \cosh (\rho) X_n'(\rho) \Big).
\end{eqnarray}
So we have for general functions $f(r)=\sum_m f_m X_m(r)$ and $g(r)=\sum_m g_m X_m(r)$ that 
 \begin{eqnarray}\label{appA1: orthogonal 1}
 \langle f, g\rangle &=&\int_{-\infty}^{\rho}\frac{\cosh(r)^{d-2}}{\cosh(\rho)^{d-2}}f(r) g(r) dr+\Big(\lambda +\a (1-d)\text{sech}^2(\rho )  \Big) f(\rho) g(\rho)\nonumber\\
&-&\alpha\ \text{sech}^2(\rho )\Big(  \cosh^2(\rho) f''(\rho)+d \sinh (\rho) \cosh (\rho) f'(\rho) \Big) g(\rho) \nonumber\\
&-&\alpha\ \text{sech}^2(\rho ) f(\rho)\Big(  \cosh^2(\rho) g''(\rho)+d \sinh (\rho) \cosh (\rho) g'(\rho) \Big).
\end{eqnarray}
Similar to the toy model of sect.2, we define the following step function 
\begin{eqnarray}\label{appA1: step function 0}
\Pi_0(r)=\Pi(r-\rho)=\begin{cases}
0,\  \ \ \ \ \ \ \text{for } r<\rho,\\
1,\  \ \ \ \ \ \ \text{for } r\ge \rho,
\end{cases}
\end{eqnarray}
and its integral function
\begin{eqnarray}\label{appA1: step function 1}
\Pi_1(r)=\int_{\rho}^{r}  ds_1 \Pi_0(s_1) =
\begin{cases}
0,\  \ \ \ \ \ \ \ \ \ \ \ \ \ \text{for } r<\rho,\\
(r-\rho),\  \ \ \ \ \ \ \text{for } r\ge \rho.
\end{cases}
\end{eqnarray}
and 
\begin{eqnarray}\label{appA1: step function 2}
\Pi_2(r)=\int_{\rho}^{r} ds_1 \Pi_1(s_1) =
\begin{cases}
0,\  \ \ \ \ \ \ \ \ \ \ \ \ \text{for } r<\rho,\\
\frac{(r-\rho)^2}{2},\  \ \ \ \ \ \ \text{for } r\ge \rho.
\end{cases}
\end{eqnarray}
By definition, we have $\Pi''_2(r)=\Pi'_1(r)=\Pi_0(r)$. 

Expanding $\Pi_2(r)$ in the power of KK modes, we have 
\begin{eqnarray}\label{appA1: Pi2 expand}
\Pi_2(r)=\sum_m \frac{\langle \Pi_2, X_m \rangle}{\langle X_m, X_m \rangle } X_m(r)=- \a \sum_m  \frac{X_m(\rho) X_m(r)}{c_m} 
\end{eqnarray}
where we have used $c_m=\langle X_m, X_m \rangle$ and $\langle \Pi_2, X_m \rangle=-\a X_m(\rho)$ which can de derived from  (\ref{appA1: orthogonal 1},\ref{appA1: step function 2}) and $\Pi''_2(\rho)=\Pi_0(\rho)=1$.  From (\ref{appA1: step function 2}, \ref{appA1: Pi2 expand}), we get
\begin{eqnarray}\label{appA1: formula 1}
&& \Pi_2(\rho)=- \a \sum_m  \frac{X_m(\rho) X_m(\rho)}{c_m} =0,\\ \label{appA1: formula 2}
&& \Pi'_2(\rho)=- \a \sum_m  \frac{X_m(\rho) X'_m(\rho)}{c_m} =0,\\ \label{appA1: formula 3}
&& \Pi''_2(\rho)=- \a \sum_m  \frac{X_m(\rho) X''_m(\rho)}{c_m} =1. 
\end{eqnarray}
Note that (\ref{appA1: formula 1}) is just the first equation of spectrum identities (\ref{toy model AdS: SI HD}). Substituting EOM (\ref{EOMscalarAdSX}) into (\ref{appA1: formula 3}) and using (\ref{appA1: formula 2}), we derive
\begin{eqnarray}\label{appA1: proof 2}
1&=&- \a \sum_m  \frac{X_m(\rho) X''_m(\rho)}{c_m} \nonumber\\
&=& \a  \sum_m  \frac{X_m(\rho)}{c_m}\Big( \frac{d \sinh(\rho)\cosh(\rho) X'_m(\rho)+m^2 X_m(\rho)}{\cosh^2(\rho)}\Big)\nonumber\\
&=& \frac{\a}{\cosh^2(\rho)} \sum_m \frac{X^2_m(\rho) m^2}{c_m},
\end{eqnarray}
which proves the second equation of spectrum identities (\ref{toy model AdS: SI HD}).  Substituting NBC (\ref{NBCscalarAdS1}) into (\ref{appA1: formula 2}) and using (\ref{toy model AdS: SI HD}), we derive another spectrum identity for $\a \ne0$
\begin{eqnarray}\label{appA1: new spectrum identity}
\sum_m  \frac{X^2_m(\rho) m^4}{c_m} =-\frac{\lambda+(1-d) \a \text{sech}^2(\rho)}{\a^2 \text{sech}^4(\rho)}.
\end{eqnarray}

Let us go on to prove the spectrum identity (\ref{toy model AdS: SI DGP}) with $\a=0$. Expanding $\Pi_0(r)$ in the power of KK modes with $\a=0$, we have 
\begin{eqnarray}\label{appA1: Pi0 expand}
\Pi_0(r)=\sum_m \frac{\langle \Pi_0, X_m \rangle}{\langle X_m, X_m \rangle } X_m(r)= \lambda \sum_m  \frac{X_m(\rho) X_m(r)}{c_m} ,
\end{eqnarray}
where we have use $c_m=\langle X_m, X_m \rangle$ and $\langle \Pi_0, X_m \rangle=\lambda X_m(\rho)$ which can de derived from  (\ref{appA1: orthogonal 1},\ref{appA1: step function 0}) with $\a=0$.  From (\ref{appA1: step function 0},\ref{appA1: Pi0 expand}), we obtain the spectrum identity (\ref{toy model AdS: SI DGP}) with $\a=0$
\begin{eqnarray}\label{appA1: Pi0 proof}
\Pi_0(\rho)= \lambda \sum_m  \frac{X^2_m(\rho)}{c_m}=1.
\end{eqnarray}

Now we finish the proofs of spectrum identities (\ref{toy model AdS: SI DGP},\ref{toy model AdS: SI HD}) for the brane-localized scalar in AdS/BCFT. As a by-product, we get a new spectrum identity (\ref{appA1: new spectrum identity}) for $\a\ne0$.

\subsection{Brane-localized gravity in AdS/BCFT}

This subsection proves the spectrum identities (\ref{gravity: SI DGP},\ref{gravity: SI DGP}) of KK modes for brane-localized gravity in AdS/BCFT. 

By EOM (\ref{EOMMBCmassiveH}), we rewrite the orthogonal relationship (\ref{orthogonal-gravity}) into a mass-independent expression
\begin{eqnarray}\label{appA2: orthogonal}
 \langle f, g\rangle &=&\int_{-\infty}^{\rho}\frac{\cosh(r)^{d-2}}{\cosh(\rho)^{d-2}}f(r) g(r) dr+2\Big(\lambda +4(d-3) \a_1 \text{sech}^2(\rho ) \Big) f(\rho) g(\rho)\nonumber\\
&-&2\alpha_2\ \text{sech}^2(\rho )\Big(  \cosh^2(\rho) f''(\rho)+d \sinh (\rho) \cosh (\rho) f'(\rho) \Big) g(\rho) \nonumber\\
&-&2\alpha_2\ \text{sech}^2(\rho ) f(\rho)\Big(  \cosh^2(\rho) g''(\rho)+d \sinh (\rho) \cosh (\rho) g'(\rho) \Big),
\end{eqnarray}
where  $f(r)=\sum_m f_m H_m(r)$ and $g(r)=\sum_m g_m H_m(r)$. 

Expanding $\Pi_2(r)$ in terms of gravitational KK modes, we get 
\begin{eqnarray}\label{appA2: Pi2 expand}
\Pi_2(r)=\sum_m \frac{\langle \Pi_2, H_m \rangle}{\langle H_m, H_m \rangle } H_m(r)=- 2\a_2 \sum_m  \frac{H_m(\rho) H_m(r)}{c_m} 
\end{eqnarray}
where we have used $c_m=\langle H_m, H_m \rangle$ and $\langle \Pi_2, H_m \rangle=-2\a_2 H_m(\rho)$ derived from  (\ref{appA2: orthogonal},\ref{appA1: step function 2}) and $\Pi''_2(\rho)=\Pi_0(\rho)=1$.  From (\ref{appA1: step function 2}, \ref{appA2: Pi2 expand}), we obtain
\begin{eqnarray}\label{appA2: formula 1}
&& \Pi_2(\rho)=- 2\a_2 \sum_m  \frac{H_m(\rho) H_m(\rho)}{c_m} =0,\\ \label{appA2: formula 2}
&& \Pi'_2(\rho)=- 2\a_2 \sum_m  \frac{H_m(\rho) H'_m(\rho)}{c_m} =0,\\ \label{appA2: formula 3}
&& \Pi''_2(\rho)=- 2\a_2 \sum_m  \frac{H_m(\rho) H''_m(\rho)}{c_m} =1. 
\end{eqnarray}
Eq.(\ref{appA2: formula 1}) gives the first equation of spectrum identities (\ref{gravity: SI HD}). Substituting EOM (\ref{EOMMBCmassiveH}) into (\ref{appA2: formula 3}) and using (\ref{appA2: formula 2}), we derive
\begin{eqnarray}\label{appA1: proof 2}
1&=&- 2\a_2 \sum_m  \frac{H_m(\rho) H''_m(\rho)}{c_m} \nonumber\\
&=& 2\a_2  \sum_m  \frac{H_m(\rho)}{c_m}\Big( \frac{d \sinh(\rho)\cosh(\rho) H'_m(\rho)+m^2 H_m(\rho)}{\cosh^2(\rho)}\Big)\nonumber\\
&=& \frac{2\a_2}{\cosh^2(\rho)} \sum_m \frac{H^2_m(\rho) m^2}{c_m},
\end{eqnarray}
which yields the second equation of spectrum identities (\ref{gravity: SI HD}).  Substituting NBC (\ref{NBCsgravity2}) into (\ref{appA2: formula 2}) and using (\ref{gravity: SI HD}), we get a new spectrum identity for $\a_2 \ne0$
\begin{eqnarray}\label{appA2: new spectrum identity}
\sum_m  \frac{H^2_m(\rho) m^4}{c_m} =-\frac{\lambda+4(d-3) \a_1 \text{sech}^2(\rho)}{2\a_2^2 \text{sech}^4(\rho)}.
\end{eqnarray}

Let us go on to prove the spectrum identity (\ref{gravity: SI DGP}) with $\a_2=0$. Expanding $\Pi_0(r)$ in the power of gravitational KK modes with $\a_2=0$, we have 
\begin{eqnarray}\label{appA2: Pi0 expand}
\Pi_0(r)=\sum_m \frac{\langle \Pi_0, H_m \rangle}{\langle H_m, H_m \rangle } H_m(r)=2\Big(\lambda+4(d-3) \a_1 \text{sech}^2(\rho)\Big) \sum_m  \frac{H_m(\rho) H_m(r)}{c_m} ,
\end{eqnarray}
where we have used $c_m=\langle H_m, H_m \rangle$ and $\langle \Pi_0, H_m \rangle=2\Big(\lambda+4(d-3) \a_1 \text{sech}^2(\rho)\Big) H_m(\rho)$ which can de derived from  (\ref{appA2: orthogonal},\ref{appA1: step function 0}) with $\a_2=0$.  From (\ref{appA1: step function 0},\ref{appA2: Pi0 expand}), we obtain the spectrum identity (\ref{gravity: SI DGP}) with $\a_2=0$
\begin{eqnarray}\label{appA2: Pi0 proof}
\Pi_0(\rho)= 2\Big(\lambda+4(d-3) \a_1 \text{sech}^2(\rho)\Big) \sum_m  \frac{H^2_m(\rho)}{c_m}=1.
\end{eqnarray}

Now we finish the proofs of spectrum identities (\ref{gravity: SI DGP},\ref{gravity: SI HD}) for the brane-localized gravity in AdS/BCFT. As a by-product, we get a new spectrum identity (\ref{appA2: new spectrum identity}) for $\a_2\ne0$. 

To end this section, we verify the spectrum identities (\ref{gravity: SI DGP},\ref{gravity: SI HD}) in large parameter limits, where the perturbative mass spectrum can be obtained. For simplicity, we focus on the case $d=4, \rho=0$ so that we can make analytical discussions. Solving (\ref{sect3:mass spectrum AdSBCFT}) perturbatively, we derive the mass spectrum $m^2$ for large $\a_2$
\begin{eqnarray}\label{app: spectrum2}
-\frac{\sqrt{3}}{2 \sqrt{\a_2}}\Big( 1+\frac{b_1}{\sqrt{\a_2}}+O(\frac{1}{\a_2}) \Big), \ \frac{\sqrt{3}}{2 \sqrt{\a_2}}\Big( 1- \frac{b_1}{\sqrt{\a_2}} +O(\frac{1}{\a_2})\Big), \ 2q(2q+3)\Big(1+O(\frac{1}{\a_2})\Big)
\end{eqnarray}
where $q=1,2,...$ are integers and 
\begin{eqnarray}\label{app: gravity b1}
b_1=\frac{-4 \text{LegendreP}^{(1,0,0)}(2,2,0)+3 \text{LegendreP}^{(2,0,0)}(1,2,0)+192 \alpha _1+48 \lambda }{48 \sqrt{3}}.
\end{eqnarray} 
From (\ref{Htwocase},\ref{orthogonal-gravity}) with $d=4, \rho=0$ and the above mass spectrum, we obtain at order $O(1/\alpha_2)$: 
\begin{eqnarray}\label{app: gravity relation1}
\frac{H^2_m(0)}{c_m}:  && -\frac{\sqrt{\frac{1}{\alpha _2}}}{2 \sqrt{3}}+\frac{-32 \alpha _1+8 \sqrt{3} b_1-\text{Int}-8 \lambda }{48 \alpha _2}+O\left(\frac{1}{\alpha _2^{3/2}}\right), \nonumber\\
&& \frac{\sqrt{\frac{1}{\alpha _2}}}{2 \sqrt{3}}+\frac{-32 \alpha _1+8 \sqrt{3} b_1-\text{Int}-8 \lambda }{48 \alpha _2}+O\left(\frac{1}{\alpha _2^{3/2}}\right), \nonumber\\
&& O\left(\frac{1}{\alpha _2^2}\right),
\end{eqnarray}
where 
\begin{eqnarray}\label{app: gravity Int}
\text{Int}&=&\int_{-\infty }^0 \text{sech}^2(r) \text{LegendreP}^{(1,0,0)}(1,2,-\tanh (r))^2 \, dr\nonumber\\
&=&\frac{1}{2} \text{LegendreP}^{(2,0,0)}(1,2,0)-\frac{2}{3} \text{LegendreP}^{(1,0,0)}(2,2,0).
\end{eqnarray}
From (\ref{app: gravity b1}, \ref{app: gravity relation1},\ref{app: gravity Int}), we derive
\begin{eqnarray}\label{app: gravity relation1a}
\sum_m \frac{H^2_m(0)}{c_m}=\frac{-32 \alpha _1+8 \sqrt{3} b_1-\text{Int}-8 \lambda }{24 \alpha _2}+O\left(\frac{1}{\alpha _2^{3/2}}\right)=O\left(\frac{1}{\alpha _2^{3/2}}\right).
\end{eqnarray}
Similarly, we have 
\begin{eqnarray}\label{app: gravity relation2}
\frac{H^2_m(0)m^2}{c_m}:  &&\frac{1}{4 \alpha _2}+\frac{\left(\frac{1}{\alpha _2}\right){}^{3/2} \left(32 \alpha _1+\text{Int}+8 \lambda \right)}{32 \sqrt{3}}+O\left(\frac{1}{\alpha _2^2}\right), \nonumber\\
&&\frac{1}{4 \alpha _2}-\frac{\left(\frac{1}{\alpha _2}\right){}^{3/2} \left(32 \alpha _1+\text{Int}+8 \lambda \right)}{32 \sqrt{3}}+O\left(\frac{1}{\alpha _2^2}\right),  \nonumber\\
&&O\left(\frac{1}{\alpha _2^2}\right),
\end{eqnarray}
which gives
\begin{eqnarray}\label{app: gravity relation2a}
\sum_m \frac{H^2_m(0) m^2}{c_m}=\frac{1}{2 \alpha _2}+O\left(\frac{1}{\alpha _2^2}\right).
\end{eqnarray}
We finish the perturbative proof of spectrum identity (\ref{gravity: SI HD}) for $d=4, \rho=0$. 

Let us go on to verify the spectrum identity (\ref{gravity: SI DGP}) with $\a_i=0$ and $d=4, \rho=0$. For DGP gravity with $\alpha_i=0$, we get the mass spectrum in large $\lambda$ limit
\begin{eqnarray}\label{app a2: spectrum2}
m^2: \ \frac{3}{4 \lambda }\Big(1+O(\frac{1}{\lambda})\Big), \ \ \ 2q(2q+3)\Big(1+O(\frac{1}{\lambda})\Big),
\end{eqnarray}
where $q=1,2,...$ are integers. From (\ref{Htwocase},\ref{orthogonal-gravity}) with $d=4, \rho=0, \a_i=0$ and the above mass spectrum, we obtain
\begin{eqnarray}\label{app a2: gravity relation1}
\frac{H^2_m(0)}{c_m}: \ \frac{1}{2 \lambda }-\frac{\text{Int}}{16 \lambda ^2}+O\left(\frac{1}{\lambda^3 }\right), \ \ \frac{\pi  (q+1) (2 q+1) (4 q+3)}{16  q (2 q+3) \Gamma \left(\frac{1}{2}-q\right)^2 \Gamma (q+2)^2}\frac{1}{\lambda^2}+O\left(\frac{1}{\lambda^3 }\right)
\end{eqnarray}
where $\text{Int}$ is defined by (\ref{app: gravity Int}). 
Summing the above equation, we derive
\begin{eqnarray}\label{app a2: gravity relation1a}
\sum_m \frac{H^2_m(0)}{c_m}=\ \frac{1}{2 \lambda }+O\left(\frac{1}{\lambda^3}\right).
\end{eqnarray}
In the above derivation, we have used the sum formula 
\begin{eqnarray}\label{app a2: sum formula}
&&\sum_{q=1}^{\infty} \frac{\pi  (q+1) (2 q+1) (4 q+3)}{16  q (2 q+3) \Gamma \left(\frac{1}{2}-q\right)^2 \Gamma (q+2)^2}=\frac{2 C-3}{4 \pi }+\frac{7}{48}
+\frac{3}{256} \, _4F_3\left(1,1,\frac{3}{2},\frac{5}{2};2,3,3;1\right)\nonumber\\
&=&\frac{1}{16} \left(\frac{3}{6} \text{LegendreP}^{(2,0,0)}(1,2,0)-\frac{4}{6} \text{LegendreP}^{(1,0,0)}(2,2,0)\right)=\frac{\text{Int}}{16}
\end{eqnarray}
where $C\simeq 0.915966$ denotes the Catalan's constant, $_4F_3$ is the generalized hypergeometric function. We do not find an analytical derivation of the second line from the first line of (\ref{app a2: sum formula}). Instead, we find they agree with each other at high numerical accuracy. We finish the perturbative proof of spectrum identity (\ref{gravity: SI DGP}) with $\a_i=0$ and $d=4, \rho=0$. 

In addition to the above perturbative proofs for large parameters, we also check numerically all of the spectrum identities for finite parameters and find they holds with high numerical accuracy.  These perturbative and numerical verifications are strong supports to our general proofs of the spectrum identities in this appendix.

\subsection{Brane-localized gravity in wedge holography}

This subsection studies the spectrum identities of KK modes for brane-localized gravity in wedge holography. We consider general bulk space given by $-\rho_L\le r\le \rho_R$. The parameters on the right and left branes are labelled by $(\lambda_R, \a_{R\ i})$ and $(\lambda_L, \a_{L\ i})$, respectively. The NBCs on the two branes read

 \begin{eqnarray}\label{app A3: wedge NBC1}
&&H'(\rho_R) =2H(\rho_R) m^2\Big(\lambda_R\ \text{sech}^2(\rho_R )+ 4(d-3)\a_{R\ 1}\ \text{sech}^4(\rho_R ) + \a_{R\ 2}\ \text{sech}^4(\rho_R)m^2 \Big)\nonumber\\ \label{app A3: wedge NBC2}
&& \\
&&H'(-\rho_L) =-2H(-\rho_L) m^2\Big(\lambda_L\ \text{sech}^2(\rho_L )+ 4(d-3)\a_{L\ 1}\ \text{sech}^4(\rho_L ) + \a_{R\ 2}\ \text{sech}^4(\rho_L)m^2 \Big).\nonumber\\
\end{eqnarray}

To make terms on the left and right branes symmetrical, we slightly change the expression of orthogonal relation as follows
\begin{eqnarray}\label{appA3: orthogonal}
&& \langle f, g\rangle_{W} =\int_{-\rho_L}^{\rho_R}\cosh(r)^{d-2}f(r) g(r) dr \nonumber\\
 &+&2\cosh(\rho_R)^{d-2}\Big(\lambda_R +4(d-3) \a_{R\ 1}\ \text{sech}^2(\rho_R ) \Big) f(\rho_R) g(\rho_R)\nonumber\\
  &+&2\cosh(\rho_L)^{d-2}\Big(\lambda_L +4(d-3) \a_{L\ 1}\ \text{sech}^2(\rho_L ) \Big) f(-\rho_L) g(-\rho_L)\nonumber\\
&-&2\alpha_{R\ 2}\ \cosh(\rho_R)^{d-4}\Big(  \cosh^2(\rho_R) f''(\rho_R)+d \sinh (\rho_R) \cosh (\rho_R) f'(\rho_R) \Big) g(\rho_R) \nonumber\\
&-&2\alpha_{R\ 2}\ \cosh(\rho_R)^{d-4}f(\rho_R)\Big(  \cosh^2(\rho_R) g''(\rho_R)+d \sinh (\rho_R) \cosh (\rho_R) g'(\rho_R) \Big)  \nonumber\\
&-&2\alpha_{L\ 2}\ \cosh(\rho_L)^{d-4}\Big(  \cosh^2(\rho_L) f''(-\rho_L)-d \sinh (\rho_L) \cosh (\rho_L) f'(-\rho_L) \Big)  g(-\rho_L), \nonumber\\
&-&2\alpha_{L\ 2}\ \cosh(\rho_L)^{d-4} f(-\rho_L)\Big(  \cosh^2(\rho_L) g''(-\rho_L)-d \sinh (\rho_L) \cosh (\rho_L) g'(-\rho_L) \Big), \nonumber\\
\end{eqnarray}
where  $f(r)=\sum_m f_m H_m(r)$ and $g(r)=\sum_m g_m H_m(r)$.  Note that the above inner product is different from the one of sect.5 up to a constant factor. For the symmetric case with $\rho_R=\rho_L=\rho$,  we have $\langle f, g\rangle_{W}=\cosh^{d-2}(\rho) \langle f, g\rangle$.

We define the following step functions 
\begin{eqnarray}\label{appA3: step function 0 R}
&&\Pi_{R\ 0}(r)=\begin{cases}
0,\  \ \ \ \ \ \ \text{for } r<\rho_R,\\
1,\  \ \ \ \ \ \ \text{for } r\ge \rho_R,
\end{cases}\\
\label{appA3: step function 0 L}
&&\Pi_{L\ 0}(r)=\begin{cases}
0,\  \ \ \ \ \ \ \text{for } r>-\rho_L,\\
1,\  \ \ \ \ \ \ \text{for } r\le -\rho_L,
\end{cases}
\end{eqnarray}
and its integral functions
\begin{eqnarray}\label{appA3: step function 2 R}
&&\Pi_{R\ 2}(r)=
\begin{cases}
0,\  \ \ \ \ \ \ \ \ \ \ \ \ \text{for } r<\rho_R,\\
\frac{(r-\rho_R)^2}{2},\  \ \ \ \ \text{for } r\ge \rho_R.
\end{cases}\\
\label{appA3: step function 2 L}
&&\Pi_{L\ 2}(r)=
\begin{cases}
0,\  \ \ \ \ \ \ \ \ \ \ \ \ \text{for } r>-\rho_L,\\
\frac{(r+\rho_L)^2}{2},\  \ \ \ \ \text{for } r\le -\rho_L.
\end{cases}
\end{eqnarray}
Following the approaches of B.2 and using $\Big( \Pi_{R\ 0}(r), \Pi_{R\ 2}(r) \Big)$ and $\Big( \Pi_{L\ 0}(r), \Pi_{L\ 2}(r) \Big)$, we derive the spectrum identities on the right brane 
\begin{eqnarray}\label{app A3: SI DGP R}
&&\sum_{m} \frac{H_m^2(\rho_R)}{\langle H_m, H_m \rangle_W} = \frac{\text{sech}^{d-2}(\rho_R)}{2\Big(\lambda_R +4(d-3) \a_{R \ 1} \ \text{sech}^2(\rho_R )\Big)}, \ \ \ \ \ \ \ \ \ \ \ \ \text{for }\alpha_{R\ 2}=0, \\ \label{app A3: SI HD R}
&&\sum_{m} \frac{H_m^2(\rho_R)}{\langle H_m, H_m \rangle_W} = 0, \ \ \ \ \sum_{m} \frac{H_m^2(\rho_R) m^2}{\langle H_m, H_m \rangle_W}=\frac{\text{sech}^{d-4}(\rho_R)}{2\alpha_{R\ 2}}, \ \ \ \ \ \ \ \text{for } \alpha_{R\ 2}\ne0,
\end{eqnarray} 
and the left brane
\begin{eqnarray}\label{app A3: SI DGP L}
&&\sum_{m} \frac{H_m^2(-\rho_L)}{\langle H_m, H_m \rangle_W} = \frac{\text{sech}^{d-2}(\rho_L)}{2\Big(\lambda_L +4(d-3) \a_{L \ 1} \ \text{sech}^2(\rho_L )\Big)}, \ \ \ \ \ \ \ \ \ \ \ \ \text{for }\alpha_{L\ 2}=0, \\ \label{app A3: SI HD L}
&&\sum_{m} \frac{H_m^2(-\rho_L)}{\langle H_m, H_m \rangle_W} = 0, \ \ \ \ \sum_{m} \frac{H_m^2(-\rho_L) m^2}{\langle H_m, H_m \rangle_W}=\frac{\text{sech}^{d-4}(\rho_L)}{2\alpha_{L\ 2}}, \ \ \ \ \ \  \text{for } \alpha_{L\ 2}\ne0.
\end{eqnarray} 
Taking into account the factor difference of inner product, the above spectrum identities reduce to (\ref{wedge: SI DGP},\ref{wedge: SI HD1},\ref{wedge: SI HD2}) of sect.5 for the symmetric case with $\rho_R=\rho_L=\rho$ and $(\lambda_R, \a_{R\ i})=(\lambda_L, \a_{L\ i})=(\lambda, \a_{i})$. Since the derivations are almost the same as those of Appendix B.2, we do not repeat them. Following the arguments of sect.3.1, from the spectrum identities (\ref{app A3: SI DGP R}, \ref{app A3: SI HD R}, \ref{app A3: SI DGP L},\ref{app A3: SI HD L}), we obtain the necessary and sufficient ghost-free conditions
\begin{eqnarray}\label{appA3: ghost-free R}
&&\lambda_R +4(d-3) \a_{R \ 1} \ \text{sech}^2(\rho_R )\ge 0, \ \ \a_{R\ 2}=0,\\
&&\lambda_L +4(d-3) \a_{L \ 1} \ \text{sech}^2(\rho_L )\ge 0, \ \ \ \a_{L\ 2}=0. \label{appA3: ghost-free L}
\end{eqnarray}
By applying the methods of sect.3.1, we can prove that the mass spectrum is real and negative $m^2\ge 0$ under the above ghost-free conditions. For DGP gravity with $\a_{R\ i}=\a_{L\ i}=0$, we must have non-negative DGP couplings on both branes to be ghost-free, i.e., $\lambda_R \ge 0\  \& \ \lambda_L \ge 0$.

\section{ HEE in AdS/BCFT }

\begin{figure}[t]
\centering
\includegraphics[width=10cm]{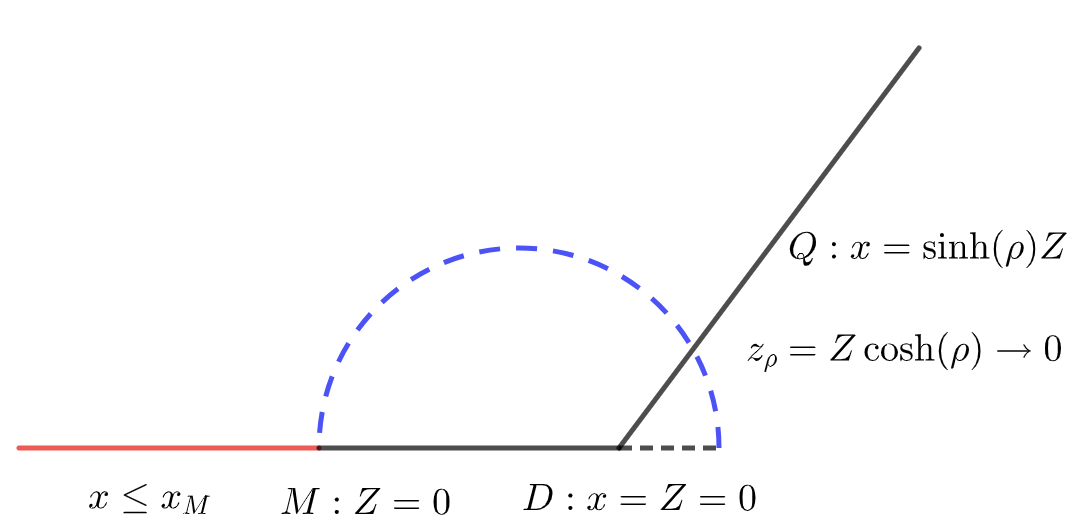}
\caption{The third kind of RT surface in AdS/BCFT. The RT surface is labelled by the blue dotted curve, and its intersection with the brane is $z_{\rho}\to 0$.}
\label{RT3AdSBCFT}
\end{figure}

This appendix shows that the renormalized area $\Delta A_3=A_3-A_1$ becomes negative infinity when the bound (\ref{sect4:EEbound}) is violated. We focus on the region near the defect $D$, where the potential divergence appears. See Fig. \ref{RT3AdSBCFT}.  By performing the coordinate transformation $Z=z/\cosh(r), x=z \tanh(r)$, we rewrite the bulk metric (\ref{sect4:AdS}) as
\begin{eqnarray}\label{app:AdSmetric}
ds^2=\frac{dZ^2+dx^2-dt^2+\sum_{a=1}^{d-2} dw_a^2}{Z^2},
\end{eqnarray}
with the brane located at
\begin{eqnarray}\label{app:brane}
Q:\ x=\sinh(\rho) Z. 
\end{eqnarray}
We are interested in the leading divergence of the area of the RT surface (blue dotted curve of Fig. \ref{RT3AdSBCFT}). Assuming the embedding function $x=x(Z), t=\text{constant}$ for the RT surface, we derive the bulk area functional 
\begin{eqnarray}\label{app:areabulk1}
A_{\text{bulk}}=\int_{\epsilon} dZ L=\int_{\epsilon} dZ \frac{\sqrt{1+ x'(Z)^2}}{Z^{d-1}},
\end{eqnarray}
where $\epsilon$ is the UV cutoff of $Z$. Since the area functional contains no $x(Z)$, we can define a conserved quantity
\begin{eqnarray}\label{app: E}
E=\frac{\partial L}{\partial x'(Z)}=\frac{Z^{1-d} x'(Z)}{\sqrt{x'(Z)^2+1}},
\end{eqnarray}
which yields 
\begin{eqnarray}\label{app:bx}
x'(Z)\to\pm \frac{E Z^d}{\sqrt{Z^2-E^2 Z^{2 d}}}. 
 \end{eqnarray}
 Substituting (\ref{app:bx}) into (\ref{app:areabulk1}), we obtain 
 \begin{eqnarray}\label{app:areabulk2}
A_{\text{bulk}}&=&\int_{\epsilon} dZ Z^{2-d} \sqrt{\frac{1}{Z^2-E^2 Z^{2 d}}}=\frac{1}{(d-2) \epsilon^{d-2}}+...\nonumber\\
&=&\frac{\cosh^{d-2}(\rho)}{(d-2) z_{\rho}^{d-2}}+...
\end{eqnarray}
where dots label the sub-leading divergence. Recall that $Z=z/\cosh(r)$, thus we have $\epsilon=z_{\rho}/\cosh(\rho)$ on the intersection
of the RT surface and brane. Recall the boundary area of RT surface is given by (\ref{sect4:areaisland})
\begin{eqnarray}\label{app:areabdy}
A_{\text{bdy}}=\frac{2\cosh^{d-2}(\rho)}{z^{d-2}_{\rho}} \bar{\lambda}.
\end{eqnarray}
From the above two equations, we obtain the total area of the third kind of RT surface
\begin{eqnarray}\label{app:areatotal}
A=A_{\text{bulk}}+A_{\text{bdy}}=\lim_{z_{\rho}\to0}2\frac{\cosh^{d-2}(\rho)}{z^{d-2}_{\rho}} (\bar{\lambda}+\frac{1}{2(d-2)})+...
\end{eqnarray}
which becomes negative infinity for $\bar{\lambda}<-\frac{1}{2(d-2)}.$ Note that the area (\ref{app:areatotal}) approaches positive infinity for $\bar{\lambda}>-\frac{1}{2(d-2)}$. It means the RT surface does not minimize at the defect $z_{\rho}\to 0$ for $\bar{\lambda}>-\frac{1}{2(d-2)}$. In other words, for $\bar{\lambda}>-\frac{1}{2(d-2)}$, the RT surface is given by the first or second kinds with $z_{\rho}>0$ rather than the third kind with $z_{\rho}\to 0$. Now we finish the proof that the renormalized area of the third kind of RT surface becomes negative infinity when the bound (\ref{sect4:EEbound}) is violated. One should impose the bound (\ref{sect4:EEbound}) to rule out this unphysical case.

\begin{thebibliography}{00}



\bibitem{Gross:1986mw}
D.~J.~Gross and J.~H.~Sloan,
Nucl. Phys. B \textbf{291} (1987), 41-89

\bibitem{Gross:1986iv}
D.~J.~Gross and E.~Witten,
Nucl. Phys. B \textbf{277} (1986), 1

\bibitem{Fradkin:1984pq}
E.~S.~Fradkin and A.~A.~Tseytlin,
Phys. Lett. B \textbf{158} (1985), 316-322



\bibitem{Stelle:1976gc}
K.~S.~Stelle,
Phys. Rev. D \textbf{16}, 953-969 (1977)

\bibitem{Buchbinder:1992rb}
I.~L.~Buchbinder, S.~D.~Odintsov and I.~L.~Shapiro,
``Effective action in quantum gravity,''

\bibitem{Buchel:2009sk}
A.~Buchel, J.~Escobedo, R.~C.~Myers, M.~F.~Paulos, A.~Sinha and M.~Smolkin,
JHEP \textbf{03}, 111 (2010)

\bibitem{Gregory:2009fj}
R.~Gregory, S.~Kanno and J.~Soda,
JHEP \textbf{10}, 010 (2009)

\bibitem{Nojiri:1999mh}
S.~Nojiri and S.~D.~Odintsov,
Int. J. Mod. Phys. A \textbf{15} (2000), 413-428

\bibitem{DeFelice:2010aj}
A.~De Felice and S.~Tsujikawa,
Living Rev. Rel. \textbf{13}, 3 (2010)

\bibitem{Nojiri:2010wj}
S.~Nojiri and S.~D.~Odintsov,
Phys. Rept. \textbf{505} (2011), 59-144

\bibitem{Nojiri:2017ncd}
S.~Nojiri, S.~D.~Odintsov and V.~K.~Oikonomou,
Phys. Rept. \textbf{692} (2017), 1-104

\bibitem{Stelle:1977ry}
K.~S.~Stelle,
Gen. Rel. Grav. \textbf{9}, 353-371 (1978)

\bibitem{Dvali:2000hr}
G.~R.~Dvali, G.~Gabadadze and M.~Porrati,
Phys. Lett. B \textbf{485}, 208-214 (2000)



\bibitem{Takayanagi:2011zk}
  T.~Takayanagi,
  Phys.\ Rev.\ Lett.\  {\bf 107} (2011) 101602
  [arXiv:1105.5165 [hep-th]].
  
\bibitem{Fujita:2011fp}
M.~Fujita, T.~Takayanagi and E.~Tonni,
JHEP \textbf{11}, 043 (2011)
[arXiv:1108.5152 [hep-th]].

\bibitem{Nozaki:2012qd}
M.~Nozaki, T.~Takayanagi and T.~Ugajin,
JHEP \textbf{06}, 066 (2012)
[arXiv:1205.1573 [hep-th]].

\bibitem{Miao:2018qkc}
R.~X.~Miao,
JHEP \textbf{02}, 025 (2019)
[arXiv:1806.10777 [hep-th]].

\bibitem{Miao:2017gyt}
R.~X.~Miao, C.~S.~Chu and W.~Z.~Guo,
Phys. Rev. D \textbf{96}, no.4, 046005 (2017)
[arXiv:1701.04275 [hep-th]].

\bibitem{Chu:2017aab}
C.~S.~Chu, R.~X.~Miao and W.~Z.~Guo,
JHEP \textbf{04}, 089 (2017)
[arXiv:1701.07202 [hep-th]].


\bibitem{Chu:2021mvq}
C.~S.~Chu and R.~X.~Miao,
JHEP \textbf{01}, 084 (2022)
[arXiv:2110.03159 [hep-th]].

\bibitem{Randall:1999ee}
L.~Randall and R.~Sundrum,
Phys. Rev. Lett. \textbf{83}, 3370-3373 (1999)

\bibitem{Randall:1999vf}
L.~Randall and R.~Sundrum,
Phys. Rev. Lett. \textbf{83}, 4690-4693 (1999)

\bibitem{Karch:2000ct}
A.~Karch and L.~Randall,
JHEP \textbf{05}, 008 (2001)



\bibitem{Penington:2019npb}
G.~Penington,
JHEP \textbf{09}, 002 (2020)

\bibitem{Almheiri:2019psf}
A.~Almheiri, N.~Engelhardt, D.~Marolf and H.~Maxfield,
JHEP \textbf{12}, 063 (2019)

\bibitem{Almheiri:2019hni}
A.~Almheiri, R.~Mahajan, J.~Maldacena and Y.~Zhao,
JHEP \textbf{03}, 149 (2020)

\bibitem{Almheiri:2019yqk}
A.~Almheiri, R.~Mahajan and J.~Maldacena,
[arXiv:1910.11077 [hep-th]].

\bibitem{Almheiri:2019psy}
A.~Almheiri, R.~Mahajan and J.~E.~Santos,
SciPost Phys. \textbf{9}, no.1, 001 (2020)




\bibitem{Chen:2020uac}
H.~Z.~Chen, R.~C.~Myers, D.~Neuenfeld, I.~A.~Reyes and J.~Sandor,
JHEP \textbf{10}, 166 (2020)

\bibitem{Chen:2020hmv}
H.~Z.~Chen, R.~C.~Myers, D.~Neuenfeld, I.~A.~Reyes and J.~Sandor,
JHEP \textbf{12}, 025 (2020)


\bibitem{Ling:2020laa}
Y.~Ling, Y.~Liu and Z.~Y.~Xian,
JHEP \textbf{03}, 251 (2021)

\bibitem{Geng:2020qvw}
H.~Geng and A.~Karch,
JHEP \textbf{09} (2020), 121


\bibitem{Krishnan:2020fer}
C.~Krishnan,
JHEP \textbf{01}, 179 (2021)


\bibitem{Yadav:2022mnv}
G.~Yadav and A.~Misra,
Phys. Rev. D \textbf{107}, no.10, 106015 (2023)


\bibitem{Emparan:2023dxm}
R.~Emparan, R.~Luna, R.~Suzuki, M.~Toma\v{s}evi\'c and B.~Way,
JHEP \textbf{05}, 182 (2023)



\bibitem{Kawabata:2021hac}
K.~Kawabata, T.~Nishioka, Y.~Okuyama and K.~Watanabe,
JHEP \textbf{05}, 062 (2021)
[arXiv:2102.02425 [hep-th]].





\bibitem{Chou:2021boq}
C.~J.~Chou, H.~B.~Lao and Y.~Yang,
Phys. Rev. D \textbf{106}, no.6, 066008 (2022)
[arXiv:2111.14551 [hep-th]].




\bibitem{Alishahiha:2020qza}
M.~Alishahiha, A.~Faraji Astaneh and A.~Naseh,
JHEP \textbf{02}, 035 (2021)
[arXiv:2005.08715 [hep-th]].



\bibitem{Hu:2022ymx}
Q.~L.~Hu, D.~Li, R.~X.~Miao and Y.~Q.~Zeng,
JHEP \textbf{09}, 037 (2022)
[arXiv:2202.03304 [hep-th]].

\bibitem{Hu:2022zgy}
P.~J.~Hu, D.~Li and R.~X.~Miao,
JHEP \textbf{11}, 008 (2022)

\bibitem{Miao:2022mdx}
R.~X.~Miao,
[arXiv:2212.07645 [hep-th]].

\bibitem{Miao:2023unv}
R.~X.~Miao,
JHEP \textbf{03}, 214 (2023)

\bibitem{Li:2023fly}
D.~Li and R.~X.~Miao,
JHEP \textbf{06}, 056 (2023)
[arXiv:2303.10958 [hep-th]].

\bibitem{Jeong:2023hrb}
H.~S.~Jeong, K.~Y.~Kim and Y.~W.~Sun,
[arXiv:2305.18122 [hep-th]].

\bibitem{Yu:2023whl}
M.~H.~Yu, X.~H.~Ge and C.~Y.~Lu,
[arXiv:2306.11407 [hep-th]].

\bibitem{Chang:2023gkt}
J.~C.~Chang, S.~He, Y.~X.~Liu and L.~Zhao,
[arXiv:2308.03645 [hep-th]].

\bibitem{Tong:2023nvi}
C.~W.~Tong, D.~H.~Du and J.~R.~Sun,
[arXiv:2306.06682 [hep-th]].

\bibitem{Ghodrati:2022hbb}
M.~Ghodrati,
JHEP \textbf{08}, 059 (2023)

\bibitem{Lee:2022efh}
J.~H.~Lee, D.~Neuenfeld and A.~Shukla,
JHEP \textbf{10}, 139 (2022)

\bibitem{Aguilar-Gutierrez:2023kfn}
S.~E.~Aguilar-Gutierrez, P.~Bueno, P.~A.~Cano, R.~A.~Hennigar and Q.~Llorens,
[arXiv:2310.09333 [hep-th]].

\bibitem{Aguilar-Gutierrez:2023tic}
S.~E.~Aguilar-Gutierrez, A.~K.~Patra and J.~F.~Pedraza,
JHEP \textbf{10}, 156 (2023)

\bibitem{dS wedge}
S.~E.~Aguilar-Gutierrez, Filip.~Landgren,
[arXiv:2311.02074[hep-th]].



\bibitem{Miao:2017aba}
R.~X.~Miao and C.~S.~Chu,
JHEP \textbf{03}, 046 (2018)


\bibitem{Chu:2018ntx}
C.~S.~Chu and R.~X.~Miao,
JHEP \textbf{07}, 005 (2018)

\bibitem{Miao:2022oas}
R.~X.~Miao and Y.~Q.~Zeng,
Phys. Lett. B \textbf{838}, 137700 (2023)




\bibitem{Kanda:2023zse}
H.~Kanda, M.~Sato, Y.~k.~Suzuki, T.~Takayanagi and Z.~Wei,
JHEP \textbf{03}, 105 (2023)




\bibitem{Akal:2020wfl}
I.~Akal, Y.~Kusuki, T.~Takayanagi and Z.~Wei,
Phys. Rev. D \textbf{102}, no.12, 126007 (2020)


\bibitem{Miao:2020oey}
R.~X.~Miao,
JHEP \textbf{01}, 150 (2021)

\bibitem{Lu:2011zk}
H.~Lu and C.~N.~Pope,
Phys. Rev. Lett. \textbf{106}, 181302 (2011)

\bibitem{Bergshoeff:2009hq}
E.~A.~Bergshoeff, O.~Hohm and P.~K.~Townsend,
Phys. Rev. Lett. \textbf{102}, 201301 (2009)

\bibitem{deHaro:2000vlm}
S.~de Haro, S.~N.~Solodukhin and K.~Skenderis,
Commun. Math. Phys. \textbf{217}, 595-622 (2001)

\bibitem{Balasubramanian:1999re}
V.~Balasubramanian and P.~Kraus,
Commun. Math. Phys. \textbf{208}, 413-428 (1999)






\bibitem{Garriga:1999yh}
J.~Garriga and T.~Tanaka,
Phys. Rev. Lett. \textbf{84}, 2778-2781 (2000)

\bibitem{Izumi:2022opi}
K.~Izumi, T.~Shiromizu, K.~Suzuki, T.~Takayanagi and N.~Tanahashi,
JHEP \textbf{10}, 050 (2022)


\bibitem{Charmousis:1999rg}
C.~Charmousis, R.~Gregory and V.~A.~Rubakov,
Phys. Rev. D \textbf{62}, 067505 (2000)

\bibitem{Kanno:2002ia}
S.~Kanno and J.~Soda,
Phys. Rev. D \textbf{66}, 083506 (2002)


\bibitem{Henningson:1998gx}
M.~Henningson and K.~Skenderis,
JHEP \textbf{07}, 023 (1998)




\bibitem{Miao:2018dvm}
R.~X.~Miao,
JHEP \textbf{07}, 098 (2019)
[arXiv:1808.05783 [hep-th]].

\bibitem{Herzog:2017kkj} 
  C.~Herzog, K.~W.~Huang and K.~Jensen,
  Phys.\ Rev.\ Lett.\  {\bf 120}, no. 2, 021601 (2018)
  
\bibitem{Herzog:2017xha} 
  C.~P.~Herzog and K.~W.~Huang,
  JHEP {\bf 1710}, 189 (2017)
  
\bibitem{Dong:2013qoa}
X.~Dong,
JHEP \textbf{01}, 044 (2014)

\bibitem{Camps:2013zua}
J.~Camps,
JHEP \textbf{03}, 070 (2014)




\bibitem{Boulware:1985wk}
D.~G.~Boulware and S.~Deser,
Phys. Rev. Lett. \textbf{55}, 2656 (1985)


\bibitem{Hofman:2008ar}
D.~M.~Hofman and J.~Maldacena,
JHEP \textbf{05}, 012 (2008)



\bibitem{Hu:2022lxl}
P.~J.~Hu and R.~X.~Miao,
JHEP \textbf{03}, 145 (2022)
[arXiv:2201.02014 [hep-th]].


\bibitem{Maldacena:2011mk}
J.~Maldacena,
[arXiv:1105.5632 [hep-th]].


\bibitem{Lu:2011ks}
H.~Lu, Y.~Pang and C.~N.~Pope,
Phys. Rev. D \textbf{84}, 064001 (2011)

\bibitem{Hell:2023rbf}
A.~Hell, D.~Lust and G.~Zoupanos,
JHEP \textbf{08}, 168 (2023)

\bibitem{Biswas:2011ar}
T.~Biswas, E.~Gerwick, T.~Koivisto and A.~Mazumdar,
Phys. Rev. Lett. \textbf{108}, 031101 (2012)
[arXiv:1110.5249 [gr-qc]].

\bibitem{Modesto:2017sdr}
L.~Modesto and L.~Rachwal,
Int. J. Mod. Phys. D \textbf{26}, no.11, 1730020 (2017)

\bibitem{Mannheim:2021oat}
P.~D.~Mannheim,
Nuovo Cim. C \textbf{45}, no.2, 27 (2022)

\bibitem{Geng:2023qwm}
H.~Geng,
[arXiv:2306.15671 [hep-th]].

\bibitem{Geng:2023iqd}
H.~Geng, A.~Karch, C.~Perez-Pardavila, L.~Randall, M.~Riojas, S.~Shashi and M.~Youssef,
[arXiv:2306.15672 [hep-th]].

\bibitem{Witten:2023qsv}
E.~Witten,
[arXiv:2303.02837 [hep-th]].



\bibitem{Miao:2021ual}
R.~X.~Miao,
Phys. Rev. D \textbf{104} (2021) no.8, 086031


\end{thebibliography}
\end{document}